%% file: arxiv.tex
\documentclass[a4paper,USenglish]{eurocg26}

\usepackage[style=numeric,sorting=nyt,sortcites,backend=biber,maxbibnames=99,
maxcitenames=2,
mincitenames=1,
maxalphanames=4,
minalphanames=3,
doi=true,
url=true,
giveninits=true,
backref=false,
]{biblatex}

\def\nobreakbefore{\relax
  \ifvmode\else
    \ifhmode
      \ifdim\lastskip > 0pt\relax
        \unskip\nobreakspace
      \fi
    \fi
  \fi
}
\let\oldcite\cite
\renewcommand\cite{\nobreakbefore\oldcite}

\defcounter{abbrvpenalty}{9000}

\DeclareFieldFormat{sentencecase}{#1}

\DeclareSourcemap{
  \maps[datatype=bibtex]{
    \map{
      \step[fieldsource=url,
            match=doi,
            final]
      \step[fieldset=url, null]
    }
  }
}

\AtEveryBibitem{%
 \ifentrytype{inproceedings}{%
  \clearfield{series}
  \clearfield{number}
  \clearfield{volume}
 }{}%
 \clearlist{address}%
 \clearfield{issn}%
 \clearlist{location}%
 \clearfield{month}%
 \ifentrytype{book}{}{%
  \clearlist{publisher}%
  \clearname{editor}%
 }%
}

\DefineBibliographyStrings{english}{%
    andothers = {et\addabbrvspace al\adddot}
}

\makeatletter
\defbibenvironment{bibliography}
  {\thebibliography{99}}
  {\endthebibliography}
  {\item}
\makeatother

\usepackage{amsmath}
\usepackage{mathtools}
\def\norm#1{\left\lVert#1\right\rVert}

\usepackage{microtype}
\usepackage[ruled,vlined,linesnumbered]{algorithm2e}
\makeatletter
\patchcmd{\algocf@makecaption@ruled}{\hsize}{\textwidth}{}{}
\patchcmd{\@algocf@start}{-1.5em}{0em}{}{}
\makeatother

\usepackage[noabbrev,capitalize]{cleveref}

\usepackage{float}
\usepackage{pgfplots}
\pgfplotsset{compat=1.18}

\let\paragraph\subparagraph

\title{Finding Patient Zero via Low-Dimensional Geometric Embeddings\footnote{The financial support by the Austrian Federal Ministry of Economy, Energy and Tourism, the National Foundation for Research, Technology and Development and the Christian Doppler Research Association is gratefully acknowledged.}
}

\author[1]{Stefan Huber}
\author[2]{Dominik Kaaser}
\affil[1]{Josef Ressel Centre for Intelligent and Secure Industrial Automation,\\ Salzburg University of Applied Sciences, Austria \\
  \texttt{stefan.huber@fh-salzburg.ac.at}}
\affil[2]{TU Hamburg, Germany\\
  \texttt{dominik.kaaser@tuhh.de}}
\authorrunning{S.\ Huber and D.\ Kaaser}
\def\?#1{}
\ArticleNo{\?}
\def\copyrightline{}

\makeatletter
\gdef\@EventLongTitle{}
\gdef\@EventShortTitle{}
\makeatother

\begin{document}

\maketitle

\begin{abstract}
We study the patient zero problem in epidemic spreading processes in the independent cascade model and propose a geometric approach for source reconstruction. Using Johnson-Lindenstrauss projections, we embed the contact network into a low-dimensional Euclidean space and estimate the infection source as the node closest to the center of gravity of infected nodes. Simulations on Erdős-Rényi graphs demonstrate that our estimator achieves meaningful reconstruction accuracy despite operating on compressed observations.
\end{abstract}

\section{Introduction}

Epidemic spreading processes are omnipresent in our modern world, and the COVID-19 pandemic has led to a surge in research to model and analyze their fundamental properties. 
Interestingly, the underlying dynamics of epidemic spreading processes are analyzed (often independently) in seemingly unrelated research areas:
In epidemiology, they model the spread of diseases.
In sociology, they model opinion formation in social networks.
In computer science, they are used, e.g., to update distributed databases \cite{DBLP:journals/sigops/DemersGHILSSST88},
model the spread of cyber attacks in large networks \cite{DBLP:conf/IEEEares/WilkensHKK019}, and build routing tables in the internet \cite{DBLP:conf/stoc/LenzenP13}.
Surprisingly, less is known about the reverse direction: given the state of an epidemic spreading process, can we find its source ``\emph{patient zero}''?

\paragraph{Contribution.} In this paper we take a geometric perspective on the patient zero problem. 
We consider a simple yet expressive model of epidemic spreading based on the prominent SIR model \cite{Kermack_1927,DBLP:journals/amc/HarkoLM14,ross1916application} and simulate a simple reconstruction algorithm.
In our algorithm we embed the contact network into a geometric space.
Our main idea is to estimate patient zero by selecting the node whose embedded position is closest to the center of gravity of all infected nodes.
To make this approach scalable, we apply the \emph{Johnson-Lindenstrauss} (JL) lemma \cite{jl} to project the network onto a substantially lower-dimensional space while approximately preserving pairwise distances. Our empirical results show that despite significantly reduced dimensions this heavily compressed representation retains sufficient geometric information so that our center-of-gravity estimator achieves accurate reconstruction.

\section{Related Work}

Global parameters of epidemic spreading processes are well understood, with known results modeling the spread of diseases in epidemiology \cite{Brauer_2017,Keeling_2011,Kermack_1927} and
the spread of rumors in social networks \cite{DBLP:conf/icalp/KempeKT05,DBLP:journals/toc/KempeKT15,DBLP:journals/talg/DoerrFS14,DBLP:journals/cacm/DoerrFF12,DBLP:journals/endm/DoerrFF11,DBLP:journals/algorithmica/BerenbrinkES15,Murayama_2021,Goffman_1964,DBLP:conf/icwsm/LermanG10,Nekovee_2007,DBLP:conf/icwsm/SadilekKS12,DBLP:journals/sigkdd/GruhlLGT04,DBLP:journals/toc/KempeKT15,DBLP:journals/kais/ZhaoWFXX12}. They model blog post propagation \cite{DBLP:conf/sdm/LeskovecMFGH07} and fake news \cite{Saxena_2022,DBLP:journals/tweb/RaponiKOP22}, marketing strategies \cite{DBLP:conf/aaai/BeckerCDG20,DBLP:journals/toc/KempeKT15}, attacker lateral movement in corporate networks \cite{DBLP:conf/IEEEares/WilkensHKK019},
updates in replicated databases and information spreading \cite{DBLP:journals/sigops/DemersGHILSSST88,DBLP:conf/focs/KarpSSV00,DBLP:conf/soda/ChierichettiLP10,DBLP:journals/jacm/ChierichettiGLP18,DBLP:conf/podc/GiakkoupisMS19},
or attacks on large enterprise networks and their countermeasures by hardening certain components at an additional cost \cite{DBLP:journals/jcss/AspnesCY06,DBLP:conf/wine/GoyalJKKM16,DBLP:conf/sagt/FriedrichIKLNS17}.
Early work on opinion spreading dates back to \textcite{downs1957economic}, and there is a vast body of research that models opinion spreading via autonomous agents
\cite{abelson1963computer,DeGroot74,FJ90,HK02,DBLP:conf/atal/Berenbrink0KLRS22}.
For more details we refer to the surveys by \textcite{lorenz2007continuous,pastor2015epidemic} and the work by \textcite{castellano2009statistical}.

Regarding the reverse direction, only a small number of works analyze the patient zero problem in a rigorous way \cite{Kazemitabar_2019, DBLP:conf/sigmetrics/ShahZ10, DBLP:conf/sigmetrics/ShahZ12, DBLP:conf/icml/DawkinsLX21, zhuchenying_sir, zhuying_ic}.
\Textcite{DBLP:conf/sigmetrics/ShahZ10, DBLP:conf/sigmetrics/ShahZ12} consider infinite acyclic networks and analyze a special case of the SIR model with parameters SIR$(p, 0)$.
They use a fundamental connection between generalized Pólya urns and infection spreading on acyclic networks to prove that approximate inference of patient zero is possible, provided the networks satisfy certain expansion properties.
In addition, they bound the distance between the reported patient zero and the true source of the epidemic spreading process.
\Textcite{DBLP:conf/icml/DawkinsLX21} introduce a statistical framework to analyze the patient zero problem that covers multiple epidemic spreading models.
\Textcite{zhuchenying_sir, zhuying_ic} consider the more general SIR model and study an inference algorithm for $d$-regular trees based on Jordan centrality.
\Textcite{berenbrink_icm} deepen some of the results from \cite{zhuchenying_sir, zhuying_ic} w.r.t.\ the independent cascade model. They first prove the Bayes optimality of the model and then  establish a phase transition for $d$-regular trees and Galton-Watson trees in terms of the spreading probability $\beta$.
In addition, the authors bound the probability that the reported patient zero is far away from the true source of the epidemic spreading process.
\Textcite{DBLP:conf/icdcs/Hahn-KlimrothK23} incorporate a monitoring strategy for the epidemic spreading process.
They place sensors on certain edges in the network which raise an alarm when the infection spreads across that edge.
The authors prove a lower bound on the required number of sensors, and they simulate heuristics on real-world networks.

In all of these results, complete knowledge of all infected individuals is necessary. This poses a major limitation: in practice, monitoring strategies only deliver incomplete information.
Without complete information, multiple works aim to infer an underlying ground truth from a compressed signal.
However, to the best of our knowledge, none of these works can be applied directly to the patient zero problem.
Prominent examples in this respect are compressed sensing \cite{DBLP:journals/tit/DonohoJM13a} in information theory, the Stochastic Block Model \cite{DBLP:journals/jmlr/Abbe17} in network science, and Group Testing \cite{DBLP:journals/cpc/Coja-OghlanGHL21,GEBHARD2023104718,DBLP:conf/icdcs/HahnKlimrothKR24,DBLP:journals/cpc/CojaOghlanHHKKRS25} in information theory and distributed computing.

\section{Model}

The \emph{independent cascade model} \cite{DBLP:journals/toc/KempeKT15} is a simple model of epidemic spreading.
It is a special case of the classical SIR model \cite{Kermack_1927,DBLP:journals/amc/HarkoLM14,ross1916application}, where agents are either \emph{susceptible}, \emph{infected}, or \emph{removed}.
Agents are modeled as nodes in a social interaction graph $G$.
Time progresses in discrete rounds.
Initially, a single agent---the \emph{patient zero} $\omega$---is \emph{infected}, while all other agents are \emph{susceptible}.
In each round $t$, every infected agent $u$ spreads the infection to each susceptible neighbor $v$ independently with probability $\beta$.
In case of a success, agent $v$ will be infected at the beginning of round $t+1$.
At the end of each round, every infected agent is removed.
Removed agents cannot ever be infected again.
The process concludes when either all nodes are removed, or no further infections are possible.

\paragraph{Patient Zero Problem.}
We are given a graph $G = (V,E)$ and a node $\omega \in V$ chosen uniformly at random from $V$.
Starting from patient zero~$\omega$, we run the infection spreading process according to the independent cascade model with infection probability $\beta \in (0,1)$ as a  \emph{forward process}.
Given a snapshot consisting of the set of infected nodes after an unknown number of $t$ rounds, the task is to infer the identity of patient zero $\omega$.
Formally, the forward process on $G$ with source $\omega$ defines a distribution over observable outcomes.  
Let $\mathcal{O}$ denote the observed final infection state.  
The patient zero problem is to design an estimator $\widehat{\omega} = \widehat{\omega}(G,\beta,\mathcal{O})$ that maximizes the posterior probability $\Pr[\widehat{\omega} = \omega \mid \mathcal{O} ]$ or, more generally, minimizes the distance $\operatorname{dist}(\widehat{\omega}, \omega)$ in~$G$.

\section{Reconstruction via Geometric Embeddings}

In our algorithm we first embed the contact network into a Euclidean space in a way that preserves the relevant structural properties of the graph.
To achieve this, we rely on the JL lemma \cite{jl} which guarantees low-distortion dimensionality reduction.
It states that any set of $n$ points in a high-dimensional Euclidean space can be linearly embedded into a Euclidean space with $\operatorname{O}(\ln n / \varepsilon^2)$ dimensions such that
all pairwise distances are preserved up to a factor of $1 \pm \varepsilon$. In practice, this mapping is typically realized using random projections via sparse random matrices $A \in \mathbb{R}^{k \times d}$.
The lemma is formalized as follows.

\begin{lemma}[Johnson-Lindenstrauss lemma \cite{jl}]
Let $X = {x_1, \dots , x_n} \in \mathbb{E}^d$ be a set of $n$ points in $d$-dimensional Euclidean space.
There exists a linear function $f\colon \mathbb{R}^d \rightarrow \mathbb{R}^k$ with $k \geq 8\ln(n) / \varepsilon^2 $ such that for any pair of points $u,v \in X$
\[
(1-\varepsilon)\cdot \norm{u-v}_2^2 \leq \norm{f(u)-f(v)}_2^2 \leq (1+\varepsilon) \cdot \norm{u-v}_2^2.
\]
\end{lemma}

The spreading process induces a spatial structure on the set of infected nodes.
We estimate patient zero by selecting the node whose embedded position is closest to the center of gravity of all infected nodes.
To this end, our reconstruction algorithm operates in three stages: extraction of distance signatures, low-distortion dimensionality reduction, and geometric scoring.
Our algorithm is formalized in \cref{alg:reconstruction}.

\begin{algorithm}[t]
\DontPrintSemicolon
\SetKwComment{tcp}{$\triangleright$\ }{}
\newcommand\mycommfont[1]{\footnotesize\ttfamily\textcolor{gray}{#1}}
\SetCommentSty{mycommfont}

\caption{Patient Zero Scoring via JL-Embedded Distance Signatures\label{alg:reconstruction}}

\KwIn{graph $G=(V,E)$, infected set $I \subseteq V$, JL dimension $k$}
\KwOut{score array $D$}

\ForEach{$i \in I$}{
    $F[\cdot, i] \gets \mathrm{BFS}(G,i)$ \tcp*{compute distance signatures $F \in \mathbb{R}^{|I|\times |V|}$ for infected nodes}
}

sample $R \in \mathbb{R}^{k \times |I|}$ with $R_{ij} \sim \mathcal{N}(0,1/k)$ \tcp*{sample matrix for JL projection}
$X \gets R F$ \tcp*{compute embedded coordinates}

$\displaystyle c \gets \frac{1}{|I|}\sum_{i \in I} X[i, \cdot]$ \tcp*{compute epidemic center of gravity $c$}

\ForEach{$u \in V$}{
    $D(u) \gets \left\lVert X[\cdot,u] - c\right\rVert_2$ \tcp*{compute distance of $u$ to $c$}
}

\Return{$D$} \tcp*{patient-zero likelihood scores}

\end{algorithm}

\paragraph{Distance Signatures.}
For every infected node $i \in I$, the algorithm computes a \emph{distance signature} $F[i, \cdot]$, defined as the array of shortest-path distances from~$i$ to all nodes in $V$.
These signatures contain information about the spatial structure of the spread.
The first step produces a matrix $F \in \mathbb{R}^{|I|\times|V|}$, where the $i$th row contains the graph distances from infected node~$i$ to all nodes in $V$.

\paragraph{Geometric Embedding.}
To obtain a compact geometric representation, the algorithm projects the high-dimensional matrix $F$ onto a $k$-dimensional Euclidean space using a random Gaussian JL matrix $R \in \mathbb{R}^{k \times |I|}$ with $R_{ij} \sim \mathcal{N}(0, 1/k)$.
The embedded coordinates are computed as $X = R  F$.

\paragraph{Epidemic Center Estimation.}
Given the embedded distance signatures $\{X[\cdot, i] \colon i \in I\}$ of all infected nodes, the algorithm computes the center of gravity $c \in \mathbb{R}^k$ by taking their mean.

\paragraph{Patient-Zero Scoring.}
Finally, the algorithm computes for each node $u \in V$ a score $D(u)$ as the Euclidean distance between the embedded position of node $u$ and the epidemic center of gravity $c$, that is,
$
    D(u) = \left(\sum_{\ell=1}^k \left( X[\ell, u] - c_{\ell} \right)^2\right)^{1/2}
$.
Sorting the nodes by $D$ provides a \emph{ranking} of nodes, and the node with rank one is estimated to be patient zero.

\section{Evaluation}

To evaluate our reconstruction algorithm we implement a custom simulation framework for epidemic spreading on large synthetic networks.
The core of our framework is written in Python, with separate modules for network generation, infection propagation, reconstruction analysis, and visualization.

While our results apply to a large number of models of social interactions, we focus in this extended abstract only on Erdős-Rényi random graphs $G(n,p)$.
Unless noted otherwise, we use for our evaluation a $G(n,p)$ with $n = 10^4$ nodes and expected node degree $n\cdot p = 10$.
We initialize a single known patient zero as infected and then run the epidemic spreading process with transmission probability $\beta = 0.25$.
After $t = 4$ rounds we record the infection states of all nodes and apply our reconstruction algorithm to evaluate how well our geometric embedding locates the original source.

\begin{figure}[t]
\parbox[t]{6.5cm}{\vspace{-0.5cm}\hspace{-0.5cm}\input{figures/n.pgf}\vspace{-0.5cm}\caption{Rank of true patient zero over $n$.}\label{fig:n}}\hfill
\parbox[t]{6.5cm}{\vspace{-0.5cm}\hspace{-0.5cm}\input{figures/p.pgf}\vspace{-0.5cm}\caption{Rank of true patient zero over $p$.}\label{fig:p}}
\end{figure}

First, we show in \cref{fig:n,fig:p} the impact of the underlying graph $G(n,p)$ on our reconstruction algorithm.
Each data point is based on $100$ independent simulation runs.
We report the rank of the true patient zero normalized over the graph size $n$, so values lie in $[0,1]$ with lower being better.
Our data in \cref{fig:n} indicate that the rank improves with $n$, with some variance across multiple runs.
In \cref{fig:p} we show the rank over the graph density, spanning subcritical to supercritical connectivity regimes of the $G(n,p)$.
For small values of $p$, outbreaks rarely spread far, yielding little structural information and thus high ranks. 
Near the critical threshold, a giant component emerges and infection trees retain topology.
This enables strong inference, and the normalized rank reaches its minimum.
As $p$ continues to increase, the graph becomes dense and diffusion homogenizes.
In this regime, the patient-zero signatures blur, and the reconstruction performance worsens again.

\begin{figure}[t]
\parbox[t]{6.5cm}{\vspace{-0.5cm}\hspace{-0.5cm}\input{figures/beta.pgf}\vspace{-0.5cm}\caption{Rank of true patient zero over $\beta$.}\label{fig:beta}}\hfill
\parbox[t]{6.5cm}{\vspace{-0.5cm}\hspace{-0.5cm}\input{figures/k.pgf}\vspace{-0.5cm}\caption{Rank of true patient zero over $k$.}\label{fig:k}}
\end{figure}

Next, we show in \cref{fig:beta} the impact of the spreading process on our reconstruction algorithm.
Our data confirm that the reconstruction quality depends on the infectiousness of the epidemic: low infection probabilities lead to sparse or quickly extinguished outbreaks, making accurate source inference difficult.
Starting with $\beta \approx 0.25$, the epidemic spreading generates enough infected nodes to allow reliable localization of patient zero.

Finally, we show in \cref{fig:k} the impact of the dimension $k$ of the JL embedding.
For small values of $k$, the geometric distortion becomes large, so the center-of-gravity estimator is weak and patient zero ranks poorly.
As soon as $k$ approaches $\log n$, the embedding preserves distance signatures faithfully, and the median rank drops.
The curve decreases and then plateaus close to zero, indicating that additional dimensions provide diminishing returns.

\section{Conclusion}

Our geometric approach to the patient zero problem is based on low-dimensional embeddings of epidemic distance signatures. We propose a simple center-of-gravity estimator applied to observations that are heavily compressed via JL projections.
Our empirical data indicate that
much of the information required for source reconstruction is preserved under strong dimensionality reduction. This suggests that patient zero inference is feasible even when only limited or compressed data is available, a situation that is commonly encountered in real-world epidemic spreading scenarios.
In the full version of this paper we will extend our approach to more realistic network models, real-world contact networks, and partially observed infection states.

\printbibliography[heading=none]

\end{document}

%% file: figures/n.pgf
%% Creator: Matplotlib, PGF backend
%%
%% To include the figure in your LaTeX document, write
%%   \input{<filename>.pgf}
%%
%% Make sure the required packages are loaded in your preamble
%%   \usepackage{pgf}
%%
%% Also ensure that all the required font packages are loaded; for instance,
%% the lmodern package is sometimes necessary when using math font.
%%   \usepackage{lmodern}
%%
%% Figures using additional raster images can only be included by \input if
%% they are in the same directory as the main LaTeX file. For loading figures
%% from other directories you can use the `import` package
%%   \usepackage{import}
%%
%% and then include the figures with
%%   \import{<path to file>}{<filename>.pgf}
%%
%% Matplotlib used the following preamble
%%   \def\mathdefault#1{#1}
%%   \everymath=\expandafter{\the\everymath\displaystyle}
%%   \IfFileExists{scrextend.sty}{
%%     \usepackage[fontsize=10.000000pt]{scrextend}
%%   }{
%%     \renewcommand{\normalsize}{\fontsize{10.000000}{12.000000}\selectfont}
%%     \normalsize
%%   }
%%   
%%   \makeatletter\@ifpackageloaded{underscore}{}{\usepackage[strings]{underscore}}\makeatother
%%
\begingroup%
\makeatletter%
\begin{pgfpicture}%
\pgfpathrectangle{\pgfpointorigin}{\pgfqpoint{2.952756in}{2.952756in}}%
\pgfusepath{use as bounding box, clip}%
\begin{pgfscope}%
\pgfsetbuttcap%
\pgfsetmiterjoin%
\pgfsetlinewidth{0.000000pt}%
\definecolor{currentstroke}{rgb}{1.000000,1.000000,1.000000}%
\pgfsetstrokecolor{currentstroke}%
\pgfsetdash{}{0pt}%
\pgfpathmoveto{\pgfqpoint{0.000000in}{0.000000in}}%
\pgfpathlineto{\pgfqpoint{2.952756in}{0.000000in}}%
\pgfpathlineto{\pgfqpoint{2.952756in}{2.952756in}}%
\pgfpathlineto{\pgfqpoint{0.000000in}{2.952756in}}%
\pgfpathlineto{\pgfqpoint{0.000000in}{0.000000in}}%
\pgfpathclose%
\pgfusepath{}%
\end{pgfscope}%
\begin{pgfscope}%
\pgfsetbuttcap%
\pgfsetmiterjoin%
\pgfsetlinewidth{0.000000pt}%
\definecolor{currentstroke}{rgb}{0.000000,0.000000,0.000000}%
\pgfsetstrokecolor{currentstroke}%
\pgfsetstrokeopacity{0.000000}%
\pgfsetdash{}{0pt}%
\pgfpathmoveto{\pgfqpoint{0.590551in}{0.590551in}}%
\pgfpathlineto{\pgfqpoint{2.755906in}{0.590551in}}%
\pgfpathlineto{\pgfqpoint{2.755906in}{2.755906in}}%
\pgfpathlineto{\pgfqpoint{0.590551in}{2.755906in}}%
\pgfpathlineto{\pgfqpoint{0.590551in}{0.590551in}}%
\pgfpathclose%
\pgfusepath{}%
\end{pgfscope}%
\begin{pgfscope}%
\pgfpathrectangle{\pgfqpoint{0.590551in}{0.590551in}}{\pgfqpoint{2.165354in}{2.165354in}}%
\pgfusepath{clip}%
\pgfsetbuttcap%
\pgfsetroundjoin%
\definecolor{currentfill}{rgb}{0.862745,0.352941,0.156863}%
\pgfsetfillcolor{currentfill}%
\pgfsetfillopacity{0.220000}%
\pgfsetlinewidth{1.003750pt}%
\definecolor{currentstroke}{rgb}{0.862745,0.352941,0.156863}%
\pgfsetstrokecolor{currentstroke}%
\pgfsetstrokeopacity{0.220000}%
\pgfsetdash{}{0pt}%
\pgfsys@defobject{currentmarker}{\pgfqpoint{0.688976in}{0.688976in}}{\pgfqpoint{2.657480in}{2.657480in}}{%
\pgfpathmoveto{\pgfqpoint{0.688976in}{2.153245in}}%
\pgfpathlineto{\pgfqpoint{0.688976in}{1.419813in}}%
\pgfpathlineto{\pgfqpoint{0.885827in}{0.961417in}}%
\pgfpathlineto{\pgfqpoint{1.082677in}{1.603171in}}%
\pgfpathlineto{\pgfqpoint{1.279528in}{1.144775in}}%
\pgfpathlineto{\pgfqpoint{1.476378in}{0.838223in}}%
\pgfpathlineto{\pgfqpoint{1.673228in}{0.720760in}}%
\pgfpathlineto{\pgfqpoint{1.870079in}{0.694975in}}%
\pgfpathlineto{\pgfqpoint{2.066929in}{0.692110in}}%
\pgfpathlineto{\pgfqpoint{2.263780in}{0.701421in}}%
\pgfpathlineto{\pgfqpoint{2.460630in}{0.688976in}}%
\pgfpathlineto{\pgfqpoint{2.657480in}{0.691752in}}%
\pgfpathlineto{\pgfqpoint{2.657480in}{1.372093in}}%
\pgfpathlineto{\pgfqpoint{2.657480in}{1.372093in}}%
\pgfpathlineto{\pgfqpoint{2.460630in}{0.763555in}}%
\pgfpathlineto{\pgfqpoint{2.263780in}{1.241468in}}%
\pgfpathlineto{\pgfqpoint{2.066929in}{0.786654in}}%
\pgfpathlineto{\pgfqpoint{1.870079in}{0.898388in}}%
\pgfpathlineto{\pgfqpoint{1.673228in}{0.967147in}}%
\pgfpathlineto{\pgfqpoint{1.476378in}{1.285159in}}%
\pgfpathlineto{\pgfqpoint{1.279528in}{1.964157in}}%
\pgfpathlineto{\pgfqpoint{1.082677in}{2.657480in}}%
\pgfpathlineto{\pgfqpoint{0.885827in}{1.396893in}}%
\pgfpathlineto{\pgfqpoint{0.688976in}{2.153245in}}%
\pgfpathlineto{\pgfqpoint{0.688976in}{2.153245in}}%
\pgfpathclose%
\pgfusepath{stroke,fill}%
}%
\begin{pgfscope}%
\pgfsys@transformshift{0.000000in}{0.000000in}%
\pgfsys@useobject{currentmarker}{}%
\end{pgfscope}%
\end{pgfscope}%
\begin{pgfscope}%
\pgfpathrectangle{\pgfqpoint{0.590551in}{0.590551in}}{\pgfqpoint{2.165354in}{2.165354in}}%
\pgfusepath{clip}%
\pgfsetrectcap%
\pgfsetroundjoin%
\pgfsetlinewidth{0.803000pt}%
\definecolor{currentstroke}{rgb}{0.690196,0.690196,0.690196}%
\pgfsetstrokecolor{currentstroke}%
\pgfsetstrokeopacity{0.250000}%
\pgfsetdash{}{0pt}%
\pgfpathmoveto{\pgfqpoint{1.012570in}{0.590551in}}%
\pgfpathlineto{\pgfqpoint{1.012570in}{2.755906in}}%
\pgfusepath{stroke}%
\end{pgfscope}%
\begin{pgfscope}%
\pgfsetbuttcap%
\pgfsetroundjoin%
\definecolor{currentfill}{rgb}{0.000000,0.000000,0.000000}%
\pgfsetfillcolor{currentfill}%
\pgfsetlinewidth{0.803000pt}%
\definecolor{currentstroke}{rgb}{0.000000,0.000000,0.000000}%
\pgfsetstrokecolor{currentstroke}%
\pgfsetdash{}{0pt}%
\pgfsys@defobject{currentmarker}{\pgfqpoint{0.000000in}{-0.048611in}}{\pgfqpoint{0.000000in}{0.000000in}}{%
\pgfpathmoveto{\pgfqpoint{0.000000in}{0.000000in}}%
\pgfpathlineto{\pgfqpoint{0.000000in}{-0.048611in}}%
\pgfusepath{stroke,fill}%
}%
\begin{pgfscope}%
\pgfsys@transformshift{1.012570in}{0.590551in}%
\pgfsys@useobject{currentmarker}{}%
\end{pgfscope}%
\end{pgfscope}%
\begin{pgfscope}%
\definecolor{textcolor}{rgb}{0.000000,0.000000,0.000000}%
\pgfsetstrokecolor{textcolor}%
\pgfsetfillcolor{textcolor}%
\pgftext[x=1.012570in,y=0.493329in,,top]{\color{textcolor}{\rmfamily\fontsize{8.330000}{9.996000}\selectfont\catcode`\^=\active\def^{\ifmmode\sp\else\^{}\fi}\catcode`\%=\active\def%{\%}$\mathdefault{10^{2}}$}}%
\end{pgfscope}%
\begin{pgfscope}%
\pgfpathrectangle{\pgfqpoint{0.590551in}{0.590551in}}{\pgfqpoint{2.165354in}{2.165354in}}%
\pgfusepath{clip}%
\pgfsetrectcap%
\pgfsetroundjoin%
\pgfsetlinewidth{0.803000pt}%
\definecolor{currentstroke}{rgb}{0.690196,0.690196,0.690196}%
\pgfsetstrokecolor{currentstroke}%
\pgfsetstrokeopacity{0.250000}%
\pgfsetdash{}{0pt}%
\pgfpathmoveto{\pgfqpoint{1.666493in}{0.590551in}}%
\pgfpathlineto{\pgfqpoint{1.666493in}{2.755906in}}%
\pgfusepath{stroke}%
\end{pgfscope}%
\begin{pgfscope}%
\pgfsetbuttcap%
\pgfsetroundjoin%
\definecolor{currentfill}{rgb}{0.000000,0.000000,0.000000}%
\pgfsetfillcolor{currentfill}%
\pgfsetlinewidth{0.803000pt}%
\definecolor{currentstroke}{rgb}{0.000000,0.000000,0.000000}%
\pgfsetstrokecolor{currentstroke}%
\pgfsetdash{}{0pt}%
\pgfsys@defobject{currentmarker}{\pgfqpoint{0.000000in}{-0.048611in}}{\pgfqpoint{0.000000in}{0.000000in}}{%
\pgfpathmoveto{\pgfqpoint{0.000000in}{0.000000in}}%
\pgfpathlineto{\pgfqpoint{0.000000in}{-0.048611in}}%
\pgfusepath{stroke,fill}%
}%
\begin{pgfscope}%
\pgfsys@transformshift{1.666493in}{0.590551in}%
\pgfsys@useobject{currentmarker}{}%
\end{pgfscope}%
\end{pgfscope}%
\begin{pgfscope}%
\definecolor{textcolor}{rgb}{0.000000,0.000000,0.000000}%
\pgfsetstrokecolor{textcolor}%
\pgfsetfillcolor{textcolor}%
\pgftext[x=1.666493in,y=0.493329in,,top]{\color{textcolor}{\rmfamily\fontsize{8.330000}{9.996000}\selectfont\catcode`\^=\active\def^{\ifmmode\sp\else\^{}\fi}\catcode`\%=\active\def%{\%}$\mathdefault{10^{3}}$}}%
\end{pgfscope}%
\begin{pgfscope}%
\pgfpathrectangle{\pgfqpoint{0.590551in}{0.590551in}}{\pgfqpoint{2.165354in}{2.165354in}}%
\pgfusepath{clip}%
\pgfsetrectcap%
\pgfsetroundjoin%
\pgfsetlinewidth{0.803000pt}%
\definecolor{currentstroke}{rgb}{0.690196,0.690196,0.690196}%
\pgfsetstrokecolor{currentstroke}%
\pgfsetstrokeopacity{0.250000}%
\pgfsetdash{}{0pt}%
\pgfpathmoveto{\pgfqpoint{2.320416in}{0.590551in}}%
\pgfpathlineto{\pgfqpoint{2.320416in}{2.755906in}}%
\pgfusepath{stroke}%
\end{pgfscope}%
\begin{pgfscope}%
\pgfsetbuttcap%
\pgfsetroundjoin%
\definecolor{currentfill}{rgb}{0.000000,0.000000,0.000000}%
\pgfsetfillcolor{currentfill}%
\pgfsetlinewidth{0.803000pt}%
\definecolor{currentstroke}{rgb}{0.000000,0.000000,0.000000}%
\pgfsetstrokecolor{currentstroke}%
\pgfsetdash{}{0pt}%
\pgfsys@defobject{currentmarker}{\pgfqpoint{0.000000in}{-0.048611in}}{\pgfqpoint{0.000000in}{0.000000in}}{%
\pgfpathmoveto{\pgfqpoint{0.000000in}{0.000000in}}%
\pgfpathlineto{\pgfqpoint{0.000000in}{-0.048611in}}%
\pgfusepath{stroke,fill}%
}%
\begin{pgfscope}%
\pgfsys@transformshift{2.320416in}{0.590551in}%
\pgfsys@useobject{currentmarker}{}%
\end{pgfscope}%
\end{pgfscope}%
\begin{pgfscope}%
\definecolor{textcolor}{rgb}{0.000000,0.000000,0.000000}%
\pgfsetstrokecolor{textcolor}%
\pgfsetfillcolor{textcolor}%
\pgftext[x=2.320416in,y=0.493329in,,top]{\color{textcolor}{\rmfamily\fontsize{8.330000}{9.996000}\selectfont\catcode`\^=\active\def^{\ifmmode\sp\else\^{}\fi}\catcode`\%=\active\def%{\%}$\mathdefault{10^{4}}$}}%
\end{pgfscope}%
\begin{pgfscope}%
\pgfsetbuttcap%
\pgfsetroundjoin%
\definecolor{currentfill}{rgb}{0.000000,0.000000,0.000000}%
\pgfsetfillcolor{currentfill}%
\pgfsetlinewidth{0.602250pt}%
\definecolor{currentstroke}{rgb}{0.000000,0.000000,0.000000}%
\pgfsetstrokecolor{currentstroke}%
\pgfsetdash{}{0pt}%
\pgfsys@defobject{currentmarker}{\pgfqpoint{0.000000in}{-0.027778in}}{\pgfqpoint{0.000000in}{0.000000in}}{%
\pgfpathmoveto{\pgfqpoint{0.000000in}{0.000000in}}%
\pgfpathlineto{\pgfqpoint{0.000000in}{-0.027778in}}%
\pgfusepath{stroke,fill}%
}%
\begin{pgfscope}%
\pgfsys@transformshift{0.670648in}{0.590551in}%
\pgfsys@useobject{currentmarker}{}%
\end{pgfscope}%
\end{pgfscope}%
\begin{pgfscope}%
\pgfsetbuttcap%
\pgfsetroundjoin%
\definecolor{currentfill}{rgb}{0.000000,0.000000,0.000000}%
\pgfsetfillcolor{currentfill}%
\pgfsetlinewidth{0.602250pt}%
\definecolor{currentstroke}{rgb}{0.000000,0.000000,0.000000}%
\pgfsetstrokecolor{currentstroke}%
\pgfsetdash{}{0pt}%
\pgfsys@defobject{currentmarker}{\pgfqpoint{0.000000in}{-0.027778in}}{\pgfqpoint{0.000000in}{0.000000in}}{%
\pgfpathmoveto{\pgfqpoint{0.000000in}{0.000000in}}%
\pgfpathlineto{\pgfqpoint{0.000000in}{-0.027778in}}%
\pgfusepath{stroke,fill}%
}%
\begin{pgfscope}%
\pgfsys@transformshift{0.752348in}{0.590551in}%
\pgfsys@useobject{currentmarker}{}%
\end{pgfscope}%
\end{pgfscope}%
\begin{pgfscope}%
\pgfsetbuttcap%
\pgfsetroundjoin%
\definecolor{currentfill}{rgb}{0.000000,0.000000,0.000000}%
\pgfsetfillcolor{currentfill}%
\pgfsetlinewidth{0.602250pt}%
\definecolor{currentstroke}{rgb}{0.000000,0.000000,0.000000}%
\pgfsetstrokecolor{currentstroke}%
\pgfsetdash{}{0pt}%
\pgfsys@defobject{currentmarker}{\pgfqpoint{0.000000in}{-0.027778in}}{\pgfqpoint{0.000000in}{0.000000in}}{%
\pgfpathmoveto{\pgfqpoint{0.000000in}{0.000000in}}%
\pgfpathlineto{\pgfqpoint{0.000000in}{-0.027778in}}%
\pgfusepath{stroke,fill}%
}%
\begin{pgfscope}%
\pgfsys@transformshift{0.815720in}{0.590551in}%
\pgfsys@useobject{currentmarker}{}%
\end{pgfscope}%
\end{pgfscope}%
\begin{pgfscope}%
\pgfsetbuttcap%
\pgfsetroundjoin%
\definecolor{currentfill}{rgb}{0.000000,0.000000,0.000000}%
\pgfsetfillcolor{currentfill}%
\pgfsetlinewidth{0.602250pt}%
\definecolor{currentstroke}{rgb}{0.000000,0.000000,0.000000}%
\pgfsetstrokecolor{currentstroke}%
\pgfsetdash{}{0pt}%
\pgfsys@defobject{currentmarker}{\pgfqpoint{0.000000in}{-0.027778in}}{\pgfqpoint{0.000000in}{0.000000in}}{%
\pgfpathmoveto{\pgfqpoint{0.000000in}{0.000000in}}%
\pgfpathlineto{\pgfqpoint{0.000000in}{-0.027778in}}%
\pgfusepath{stroke,fill}%
}%
\begin{pgfscope}%
\pgfsys@transformshift{0.867498in}{0.590551in}%
\pgfsys@useobject{currentmarker}{}%
\end{pgfscope}%
\end{pgfscope}%
\begin{pgfscope}%
\pgfsetbuttcap%
\pgfsetroundjoin%
\definecolor{currentfill}{rgb}{0.000000,0.000000,0.000000}%
\pgfsetfillcolor{currentfill}%
\pgfsetlinewidth{0.602250pt}%
\definecolor{currentstroke}{rgb}{0.000000,0.000000,0.000000}%
\pgfsetstrokecolor{currentstroke}%
\pgfsetdash{}{0pt}%
\pgfsys@defobject{currentmarker}{\pgfqpoint{0.000000in}{-0.027778in}}{\pgfqpoint{0.000000in}{0.000000in}}{%
\pgfpathmoveto{\pgfqpoint{0.000000in}{0.000000in}}%
\pgfpathlineto{\pgfqpoint{0.000000in}{-0.027778in}}%
\pgfusepath{stroke,fill}%
}%
\begin{pgfscope}%
\pgfsys@transformshift{0.911276in}{0.590551in}%
\pgfsys@useobject{currentmarker}{}%
\end{pgfscope}%
\end{pgfscope}%
\begin{pgfscope}%
\pgfsetbuttcap%
\pgfsetroundjoin%
\definecolor{currentfill}{rgb}{0.000000,0.000000,0.000000}%
\pgfsetfillcolor{currentfill}%
\pgfsetlinewidth{0.602250pt}%
\definecolor{currentstroke}{rgb}{0.000000,0.000000,0.000000}%
\pgfsetstrokecolor{currentstroke}%
\pgfsetdash{}{0pt}%
\pgfsys@defobject{currentmarker}{\pgfqpoint{0.000000in}{-0.027778in}}{\pgfqpoint{0.000000in}{0.000000in}}{%
\pgfpathmoveto{\pgfqpoint{0.000000in}{0.000000in}}%
\pgfpathlineto{\pgfqpoint{0.000000in}{-0.027778in}}%
\pgfusepath{stroke,fill}%
}%
\begin{pgfscope}%
\pgfsys@transformshift{0.949198in}{0.590551in}%
\pgfsys@useobject{currentmarker}{}%
\end{pgfscope}%
\end{pgfscope}%
\begin{pgfscope}%
\pgfsetbuttcap%
\pgfsetroundjoin%
\definecolor{currentfill}{rgb}{0.000000,0.000000,0.000000}%
\pgfsetfillcolor{currentfill}%
\pgfsetlinewidth{0.602250pt}%
\definecolor{currentstroke}{rgb}{0.000000,0.000000,0.000000}%
\pgfsetstrokecolor{currentstroke}%
\pgfsetdash{}{0pt}%
\pgfsys@defobject{currentmarker}{\pgfqpoint{0.000000in}{-0.027778in}}{\pgfqpoint{0.000000in}{0.000000in}}{%
\pgfpathmoveto{\pgfqpoint{0.000000in}{0.000000in}}%
\pgfpathlineto{\pgfqpoint{0.000000in}{-0.027778in}}%
\pgfusepath{stroke,fill}%
}%
\begin{pgfscope}%
\pgfsys@transformshift{0.982648in}{0.590551in}%
\pgfsys@useobject{currentmarker}{}%
\end{pgfscope}%
\end{pgfscope}%
\begin{pgfscope}%
\pgfsetbuttcap%
\pgfsetroundjoin%
\definecolor{currentfill}{rgb}{0.000000,0.000000,0.000000}%
\pgfsetfillcolor{currentfill}%
\pgfsetlinewidth{0.602250pt}%
\definecolor{currentstroke}{rgb}{0.000000,0.000000,0.000000}%
\pgfsetstrokecolor{currentstroke}%
\pgfsetdash{}{0pt}%
\pgfsys@defobject{currentmarker}{\pgfqpoint{0.000000in}{-0.027778in}}{\pgfqpoint{0.000000in}{0.000000in}}{%
\pgfpathmoveto{\pgfqpoint{0.000000in}{0.000000in}}%
\pgfpathlineto{\pgfqpoint{0.000000in}{-0.027778in}}%
\pgfusepath{stroke,fill}%
}%
\begin{pgfscope}%
\pgfsys@transformshift{1.209421in}{0.590551in}%
\pgfsys@useobject{currentmarker}{}%
\end{pgfscope}%
\end{pgfscope}%
\begin{pgfscope}%
\pgfsetbuttcap%
\pgfsetroundjoin%
\definecolor{currentfill}{rgb}{0.000000,0.000000,0.000000}%
\pgfsetfillcolor{currentfill}%
\pgfsetlinewidth{0.602250pt}%
\definecolor{currentstroke}{rgb}{0.000000,0.000000,0.000000}%
\pgfsetstrokecolor{currentstroke}%
\pgfsetdash{}{0pt}%
\pgfsys@defobject{currentmarker}{\pgfqpoint{0.000000in}{-0.027778in}}{\pgfqpoint{0.000000in}{0.000000in}}{%
\pgfpathmoveto{\pgfqpoint{0.000000in}{0.000000in}}%
\pgfpathlineto{\pgfqpoint{0.000000in}{-0.027778in}}%
\pgfusepath{stroke,fill}%
}%
\begin{pgfscope}%
\pgfsys@transformshift{1.324571in}{0.590551in}%
\pgfsys@useobject{currentmarker}{}%
\end{pgfscope}%
\end{pgfscope}%
\begin{pgfscope}%
\pgfsetbuttcap%
\pgfsetroundjoin%
\definecolor{currentfill}{rgb}{0.000000,0.000000,0.000000}%
\pgfsetfillcolor{currentfill}%
\pgfsetlinewidth{0.602250pt}%
\definecolor{currentstroke}{rgb}{0.000000,0.000000,0.000000}%
\pgfsetstrokecolor{currentstroke}%
\pgfsetdash{}{0pt}%
\pgfsys@defobject{currentmarker}{\pgfqpoint{0.000000in}{-0.027778in}}{\pgfqpoint{0.000000in}{0.000000in}}{%
\pgfpathmoveto{\pgfqpoint{0.000000in}{0.000000in}}%
\pgfpathlineto{\pgfqpoint{0.000000in}{-0.027778in}}%
\pgfusepath{stroke,fill}%
}%
\begin{pgfscope}%
\pgfsys@transformshift{1.406271in}{0.590551in}%
\pgfsys@useobject{currentmarker}{}%
\end{pgfscope}%
\end{pgfscope}%
\begin{pgfscope}%
\pgfsetbuttcap%
\pgfsetroundjoin%
\definecolor{currentfill}{rgb}{0.000000,0.000000,0.000000}%
\pgfsetfillcolor{currentfill}%
\pgfsetlinewidth{0.602250pt}%
\definecolor{currentstroke}{rgb}{0.000000,0.000000,0.000000}%
\pgfsetstrokecolor{currentstroke}%
\pgfsetdash{}{0pt}%
\pgfsys@defobject{currentmarker}{\pgfqpoint{0.000000in}{-0.027778in}}{\pgfqpoint{0.000000in}{0.000000in}}{%
\pgfpathmoveto{\pgfqpoint{0.000000in}{0.000000in}}%
\pgfpathlineto{\pgfqpoint{0.000000in}{-0.027778in}}%
\pgfusepath{stroke,fill}%
}%
\begin{pgfscope}%
\pgfsys@transformshift{1.469643in}{0.590551in}%
\pgfsys@useobject{currentmarker}{}%
\end{pgfscope}%
\end{pgfscope}%
\begin{pgfscope}%
\pgfsetbuttcap%
\pgfsetroundjoin%
\definecolor{currentfill}{rgb}{0.000000,0.000000,0.000000}%
\pgfsetfillcolor{currentfill}%
\pgfsetlinewidth{0.602250pt}%
\definecolor{currentstroke}{rgb}{0.000000,0.000000,0.000000}%
\pgfsetstrokecolor{currentstroke}%
\pgfsetdash{}{0pt}%
\pgfsys@defobject{currentmarker}{\pgfqpoint{0.000000in}{-0.027778in}}{\pgfqpoint{0.000000in}{0.000000in}}{%
\pgfpathmoveto{\pgfqpoint{0.000000in}{0.000000in}}%
\pgfpathlineto{\pgfqpoint{0.000000in}{-0.027778in}}%
\pgfusepath{stroke,fill}%
}%
\begin{pgfscope}%
\pgfsys@transformshift{1.521421in}{0.590551in}%
\pgfsys@useobject{currentmarker}{}%
\end{pgfscope}%
\end{pgfscope}%
\begin{pgfscope}%
\pgfsetbuttcap%
\pgfsetroundjoin%
\definecolor{currentfill}{rgb}{0.000000,0.000000,0.000000}%
\pgfsetfillcolor{currentfill}%
\pgfsetlinewidth{0.602250pt}%
\definecolor{currentstroke}{rgb}{0.000000,0.000000,0.000000}%
\pgfsetstrokecolor{currentstroke}%
\pgfsetdash{}{0pt}%
\pgfsys@defobject{currentmarker}{\pgfqpoint{0.000000in}{-0.027778in}}{\pgfqpoint{0.000000in}{0.000000in}}{%
\pgfpathmoveto{\pgfqpoint{0.000000in}{0.000000in}}%
\pgfpathlineto{\pgfqpoint{0.000000in}{-0.027778in}}%
\pgfusepath{stroke,fill}%
}%
\begin{pgfscope}%
\pgfsys@transformshift{1.565199in}{0.590551in}%
\pgfsys@useobject{currentmarker}{}%
\end{pgfscope}%
\end{pgfscope}%
\begin{pgfscope}%
\pgfsetbuttcap%
\pgfsetroundjoin%
\definecolor{currentfill}{rgb}{0.000000,0.000000,0.000000}%
\pgfsetfillcolor{currentfill}%
\pgfsetlinewidth{0.602250pt}%
\definecolor{currentstroke}{rgb}{0.000000,0.000000,0.000000}%
\pgfsetstrokecolor{currentstroke}%
\pgfsetdash{}{0pt}%
\pgfsys@defobject{currentmarker}{\pgfqpoint{0.000000in}{-0.027778in}}{\pgfqpoint{0.000000in}{0.000000in}}{%
\pgfpathmoveto{\pgfqpoint{0.000000in}{0.000000in}}%
\pgfpathlineto{\pgfqpoint{0.000000in}{-0.027778in}}%
\pgfusepath{stroke,fill}%
}%
\begin{pgfscope}%
\pgfsys@transformshift{1.603121in}{0.590551in}%
\pgfsys@useobject{currentmarker}{}%
\end{pgfscope}%
\end{pgfscope}%
\begin{pgfscope}%
\pgfsetbuttcap%
\pgfsetroundjoin%
\definecolor{currentfill}{rgb}{0.000000,0.000000,0.000000}%
\pgfsetfillcolor{currentfill}%
\pgfsetlinewidth{0.602250pt}%
\definecolor{currentstroke}{rgb}{0.000000,0.000000,0.000000}%
\pgfsetstrokecolor{currentstroke}%
\pgfsetdash{}{0pt}%
\pgfsys@defobject{currentmarker}{\pgfqpoint{0.000000in}{-0.027778in}}{\pgfqpoint{0.000000in}{0.000000in}}{%
\pgfpathmoveto{\pgfqpoint{0.000000in}{0.000000in}}%
\pgfpathlineto{\pgfqpoint{0.000000in}{-0.027778in}}%
\pgfusepath{stroke,fill}%
}%
\begin{pgfscope}%
\pgfsys@transformshift{1.636571in}{0.590551in}%
\pgfsys@useobject{currentmarker}{}%
\end{pgfscope}%
\end{pgfscope}%
\begin{pgfscope}%
\pgfsetbuttcap%
\pgfsetroundjoin%
\definecolor{currentfill}{rgb}{0.000000,0.000000,0.000000}%
\pgfsetfillcolor{currentfill}%
\pgfsetlinewidth{0.602250pt}%
\definecolor{currentstroke}{rgb}{0.000000,0.000000,0.000000}%
\pgfsetstrokecolor{currentstroke}%
\pgfsetdash{}{0pt}%
\pgfsys@defobject{currentmarker}{\pgfqpoint{0.000000in}{-0.027778in}}{\pgfqpoint{0.000000in}{0.000000in}}{%
\pgfpathmoveto{\pgfqpoint{0.000000in}{0.000000in}}%
\pgfpathlineto{\pgfqpoint{0.000000in}{-0.027778in}}%
\pgfusepath{stroke,fill}%
}%
\begin{pgfscope}%
\pgfsys@transformshift{1.863343in}{0.590551in}%
\pgfsys@useobject{currentmarker}{}%
\end{pgfscope}%
\end{pgfscope}%
\begin{pgfscope}%
\pgfsetbuttcap%
\pgfsetroundjoin%
\definecolor{currentfill}{rgb}{0.000000,0.000000,0.000000}%
\pgfsetfillcolor{currentfill}%
\pgfsetlinewidth{0.602250pt}%
\definecolor{currentstroke}{rgb}{0.000000,0.000000,0.000000}%
\pgfsetstrokecolor{currentstroke}%
\pgfsetdash{}{0pt}%
\pgfsys@defobject{currentmarker}{\pgfqpoint{0.000000in}{-0.027778in}}{\pgfqpoint{0.000000in}{0.000000in}}{%
\pgfpathmoveto{\pgfqpoint{0.000000in}{0.000000in}}%
\pgfpathlineto{\pgfqpoint{0.000000in}{-0.027778in}}%
\pgfusepath{stroke,fill}%
}%
\begin{pgfscope}%
\pgfsys@transformshift{1.978493in}{0.590551in}%
\pgfsys@useobject{currentmarker}{}%
\end{pgfscope}%
\end{pgfscope}%
\begin{pgfscope}%
\pgfsetbuttcap%
\pgfsetroundjoin%
\definecolor{currentfill}{rgb}{0.000000,0.000000,0.000000}%
\pgfsetfillcolor{currentfill}%
\pgfsetlinewidth{0.602250pt}%
\definecolor{currentstroke}{rgb}{0.000000,0.000000,0.000000}%
\pgfsetstrokecolor{currentstroke}%
\pgfsetdash{}{0pt}%
\pgfsys@defobject{currentmarker}{\pgfqpoint{0.000000in}{-0.027778in}}{\pgfqpoint{0.000000in}{0.000000in}}{%
\pgfpathmoveto{\pgfqpoint{0.000000in}{0.000000in}}%
\pgfpathlineto{\pgfqpoint{0.000000in}{-0.027778in}}%
\pgfusepath{stroke,fill}%
}%
\begin{pgfscope}%
\pgfsys@transformshift{2.060194in}{0.590551in}%
\pgfsys@useobject{currentmarker}{}%
\end{pgfscope}%
\end{pgfscope}%
\begin{pgfscope}%
\pgfsetbuttcap%
\pgfsetroundjoin%
\definecolor{currentfill}{rgb}{0.000000,0.000000,0.000000}%
\pgfsetfillcolor{currentfill}%
\pgfsetlinewidth{0.602250pt}%
\definecolor{currentstroke}{rgb}{0.000000,0.000000,0.000000}%
\pgfsetstrokecolor{currentstroke}%
\pgfsetdash{}{0pt}%
\pgfsys@defobject{currentmarker}{\pgfqpoint{0.000000in}{-0.027778in}}{\pgfqpoint{0.000000in}{0.000000in}}{%
\pgfpathmoveto{\pgfqpoint{0.000000in}{0.000000in}}%
\pgfpathlineto{\pgfqpoint{0.000000in}{-0.027778in}}%
\pgfusepath{stroke,fill}%
}%
\begin{pgfscope}%
\pgfsys@transformshift{2.123565in}{0.590551in}%
\pgfsys@useobject{currentmarker}{}%
\end{pgfscope}%
\end{pgfscope}%
\begin{pgfscope}%
\pgfsetbuttcap%
\pgfsetroundjoin%
\definecolor{currentfill}{rgb}{0.000000,0.000000,0.000000}%
\pgfsetfillcolor{currentfill}%
\pgfsetlinewidth{0.602250pt}%
\definecolor{currentstroke}{rgb}{0.000000,0.000000,0.000000}%
\pgfsetstrokecolor{currentstroke}%
\pgfsetdash{}{0pt}%
\pgfsys@defobject{currentmarker}{\pgfqpoint{0.000000in}{-0.027778in}}{\pgfqpoint{0.000000in}{0.000000in}}{%
\pgfpathmoveto{\pgfqpoint{0.000000in}{0.000000in}}%
\pgfpathlineto{\pgfqpoint{0.000000in}{-0.027778in}}%
\pgfusepath{stroke,fill}%
}%
\begin{pgfscope}%
\pgfsys@transformshift{2.175344in}{0.590551in}%
\pgfsys@useobject{currentmarker}{}%
\end{pgfscope}%
\end{pgfscope}%
\begin{pgfscope}%
\pgfsetbuttcap%
\pgfsetroundjoin%
\definecolor{currentfill}{rgb}{0.000000,0.000000,0.000000}%
\pgfsetfillcolor{currentfill}%
\pgfsetlinewidth{0.602250pt}%
\definecolor{currentstroke}{rgb}{0.000000,0.000000,0.000000}%
\pgfsetstrokecolor{currentstroke}%
\pgfsetdash{}{0pt}%
\pgfsys@defobject{currentmarker}{\pgfqpoint{0.000000in}{-0.027778in}}{\pgfqpoint{0.000000in}{0.000000in}}{%
\pgfpathmoveto{\pgfqpoint{0.000000in}{0.000000in}}%
\pgfpathlineto{\pgfqpoint{0.000000in}{-0.027778in}}%
\pgfusepath{stroke,fill}%
}%
\begin{pgfscope}%
\pgfsys@transformshift{2.219122in}{0.590551in}%
\pgfsys@useobject{currentmarker}{}%
\end{pgfscope}%
\end{pgfscope}%
\begin{pgfscope}%
\pgfsetbuttcap%
\pgfsetroundjoin%
\definecolor{currentfill}{rgb}{0.000000,0.000000,0.000000}%
\pgfsetfillcolor{currentfill}%
\pgfsetlinewidth{0.602250pt}%
\definecolor{currentstroke}{rgb}{0.000000,0.000000,0.000000}%
\pgfsetstrokecolor{currentstroke}%
\pgfsetdash{}{0pt}%
\pgfsys@defobject{currentmarker}{\pgfqpoint{0.000000in}{-0.027778in}}{\pgfqpoint{0.000000in}{0.000000in}}{%
\pgfpathmoveto{\pgfqpoint{0.000000in}{0.000000in}}%
\pgfpathlineto{\pgfqpoint{0.000000in}{-0.027778in}}%
\pgfusepath{stroke,fill}%
}%
\begin{pgfscope}%
\pgfsys@transformshift{2.257044in}{0.590551in}%
\pgfsys@useobject{currentmarker}{}%
\end{pgfscope}%
\end{pgfscope}%
\begin{pgfscope}%
\pgfsetbuttcap%
\pgfsetroundjoin%
\definecolor{currentfill}{rgb}{0.000000,0.000000,0.000000}%
\pgfsetfillcolor{currentfill}%
\pgfsetlinewidth{0.602250pt}%
\definecolor{currentstroke}{rgb}{0.000000,0.000000,0.000000}%
\pgfsetstrokecolor{currentstroke}%
\pgfsetdash{}{0pt}%
\pgfsys@defobject{currentmarker}{\pgfqpoint{0.000000in}{-0.027778in}}{\pgfqpoint{0.000000in}{0.000000in}}{%
\pgfpathmoveto{\pgfqpoint{0.000000in}{0.000000in}}%
\pgfpathlineto{\pgfqpoint{0.000000in}{-0.027778in}}%
\pgfusepath{stroke,fill}%
}%
\begin{pgfscope}%
\pgfsys@transformshift{2.290494in}{0.590551in}%
\pgfsys@useobject{currentmarker}{}%
\end{pgfscope}%
\end{pgfscope}%
\begin{pgfscope}%
\pgfsetbuttcap%
\pgfsetroundjoin%
\definecolor{currentfill}{rgb}{0.000000,0.000000,0.000000}%
\pgfsetfillcolor{currentfill}%
\pgfsetlinewidth{0.602250pt}%
\definecolor{currentstroke}{rgb}{0.000000,0.000000,0.000000}%
\pgfsetstrokecolor{currentstroke}%
\pgfsetdash{}{0pt}%
\pgfsys@defobject{currentmarker}{\pgfqpoint{0.000000in}{-0.027778in}}{\pgfqpoint{0.000000in}{0.000000in}}{%
\pgfpathmoveto{\pgfqpoint{0.000000in}{0.000000in}}%
\pgfpathlineto{\pgfqpoint{0.000000in}{-0.027778in}}%
\pgfusepath{stroke,fill}%
}%
\begin{pgfscope}%
\pgfsys@transformshift{2.517266in}{0.590551in}%
\pgfsys@useobject{currentmarker}{}%
\end{pgfscope}%
\end{pgfscope}%
\begin{pgfscope}%
\pgfsetbuttcap%
\pgfsetroundjoin%
\definecolor{currentfill}{rgb}{0.000000,0.000000,0.000000}%
\pgfsetfillcolor{currentfill}%
\pgfsetlinewidth{0.602250pt}%
\definecolor{currentstroke}{rgb}{0.000000,0.000000,0.000000}%
\pgfsetstrokecolor{currentstroke}%
\pgfsetdash{}{0pt}%
\pgfsys@defobject{currentmarker}{\pgfqpoint{0.000000in}{-0.027778in}}{\pgfqpoint{0.000000in}{0.000000in}}{%
\pgfpathmoveto{\pgfqpoint{0.000000in}{0.000000in}}%
\pgfpathlineto{\pgfqpoint{0.000000in}{-0.027778in}}%
\pgfusepath{stroke,fill}%
}%
\begin{pgfscope}%
\pgfsys@transformshift{2.632416in}{0.590551in}%
\pgfsys@useobject{currentmarker}{}%
\end{pgfscope}%
\end{pgfscope}%
\begin{pgfscope}%
\pgfsetbuttcap%
\pgfsetroundjoin%
\definecolor{currentfill}{rgb}{0.000000,0.000000,0.000000}%
\pgfsetfillcolor{currentfill}%
\pgfsetlinewidth{0.602250pt}%
\definecolor{currentstroke}{rgb}{0.000000,0.000000,0.000000}%
\pgfsetstrokecolor{currentstroke}%
\pgfsetdash{}{0pt}%
\pgfsys@defobject{currentmarker}{\pgfqpoint{0.000000in}{-0.027778in}}{\pgfqpoint{0.000000in}{0.000000in}}{%
\pgfpathmoveto{\pgfqpoint{0.000000in}{0.000000in}}%
\pgfpathlineto{\pgfqpoint{0.000000in}{-0.027778in}}%
\pgfusepath{stroke,fill}%
}%
\begin{pgfscope}%
\pgfsys@transformshift{2.714117in}{0.590551in}%
\pgfsys@useobject{currentmarker}{}%
\end{pgfscope}%
\end{pgfscope}%
\begin{pgfscope}%
\definecolor{textcolor}{rgb}{0.000000,0.000000,0.000000}%
\pgfsetstrokecolor{textcolor}%
\pgfsetfillcolor{textcolor}%
\pgftext[x=1.673228in,y=0.337912in,,top]{\color{textcolor}{\rmfamily\fontsize{10.000000}{12.000000}\selectfont\catcode`\^=\active\def^{\ifmmode\sp\else\^{}\fi}\catcode`\%=\active\def%{\%}network size $n$}}%
\end{pgfscope}%
\begin{pgfscope}%
\pgfpathrectangle{\pgfqpoint{0.590551in}{0.590551in}}{\pgfqpoint{2.165354in}{2.165354in}}%
\pgfusepath{clip}%
\pgfsetrectcap%
\pgfsetroundjoin%
\pgfsetlinewidth{0.803000pt}%
\definecolor{currentstroke}{rgb}{0.690196,0.690196,0.690196}%
\pgfsetstrokecolor{currentstroke}%
\pgfsetstrokeopacity{0.250000}%
\pgfsetdash{}{0pt}%
\pgfpathmoveto{\pgfqpoint{0.590551in}{0.686380in}}%
\pgfpathlineto{\pgfqpoint{2.755906in}{0.686380in}}%
\pgfusepath{stroke}%
\end{pgfscope}%
\begin{pgfscope}%
\pgfsetbuttcap%
\pgfsetroundjoin%
\definecolor{currentfill}{rgb}{0.000000,0.000000,0.000000}%
\pgfsetfillcolor{currentfill}%
\pgfsetlinewidth{0.803000pt}%
\definecolor{currentstroke}{rgb}{0.000000,0.000000,0.000000}%
\pgfsetstrokecolor{currentstroke}%
\pgfsetdash{}{0pt}%
\pgfsys@defobject{currentmarker}{\pgfqpoint{-0.048611in}{0.000000in}}{\pgfqpoint{-0.000000in}{0.000000in}}{%
\pgfpathmoveto{\pgfqpoint{-0.000000in}{0.000000in}}%
\pgfpathlineto{\pgfqpoint{-0.048611in}{0.000000in}}%
\pgfusepath{stroke,fill}%
}%
\begin{pgfscope}%
\pgfsys@transformshift{0.590551in}{0.686380in}%
\pgfsys@useobject{currentmarker}{}%
\end{pgfscope}%
\end{pgfscope}%
\begin{pgfscope}%
\definecolor{textcolor}{rgb}{0.000000,0.000000,0.000000}%
\pgfsetstrokecolor{textcolor}%
\pgfsetfillcolor{textcolor}%
\pgftext[x=0.283449in, y=0.647800in, left, base]{\color{textcolor}{\rmfamily\fontsize{8.330000}{9.996000}\selectfont\catcode`\^=\active\def^{\ifmmode\sp\else\^{}\fi}\catcode`\%=\active\def%{\%}$\mathdefault{0.00}$}}%
\end{pgfscope}%
\begin{pgfscope}%
\pgfpathrectangle{\pgfqpoint{0.590551in}{0.590551in}}{\pgfqpoint{2.165354in}{2.165354in}}%
\pgfusepath{clip}%
\pgfsetrectcap%
\pgfsetroundjoin%
\pgfsetlinewidth{0.803000pt}%
\definecolor{currentstroke}{rgb}{0.690196,0.690196,0.690196}%
\pgfsetstrokecolor{currentstroke}%
\pgfsetstrokeopacity{0.250000}%
\pgfsetdash{}{0pt}%
\pgfpathmoveto{\pgfqpoint{0.590551in}{0.979753in}}%
\pgfpathlineto{\pgfqpoint{2.755906in}{0.979753in}}%
\pgfusepath{stroke}%
\end{pgfscope}%
\begin{pgfscope}%
\pgfsetbuttcap%
\pgfsetroundjoin%
\definecolor{currentfill}{rgb}{0.000000,0.000000,0.000000}%
\pgfsetfillcolor{currentfill}%
\pgfsetlinewidth{0.803000pt}%
\definecolor{currentstroke}{rgb}{0.000000,0.000000,0.000000}%
\pgfsetstrokecolor{currentstroke}%
\pgfsetdash{}{0pt}%
\pgfsys@defobject{currentmarker}{\pgfqpoint{-0.048611in}{0.000000in}}{\pgfqpoint{-0.000000in}{0.000000in}}{%
\pgfpathmoveto{\pgfqpoint{-0.000000in}{0.000000in}}%
\pgfpathlineto{\pgfqpoint{-0.048611in}{0.000000in}}%
\pgfusepath{stroke,fill}%
}%
\begin{pgfscope}%
\pgfsys@transformshift{0.590551in}{0.979753in}%
\pgfsys@useobject{currentmarker}{}%
\end{pgfscope}%
\end{pgfscope}%
\begin{pgfscope}%
\definecolor{textcolor}{rgb}{0.000000,0.000000,0.000000}%
\pgfsetstrokecolor{textcolor}%
\pgfsetfillcolor{textcolor}%
\pgftext[x=0.283449in, y=0.941173in, left, base]{\color{textcolor}{\rmfamily\fontsize{8.330000}{9.996000}\selectfont\catcode`\^=\active\def^{\ifmmode\sp\else\^{}\fi}\catcode`\%=\active\def%{\%}$\mathdefault{0.05}$}}%
\end{pgfscope}%
\begin{pgfscope}%
\pgfpathrectangle{\pgfqpoint{0.590551in}{0.590551in}}{\pgfqpoint{2.165354in}{2.165354in}}%
\pgfusepath{clip}%
\pgfsetrectcap%
\pgfsetroundjoin%
\pgfsetlinewidth{0.803000pt}%
\definecolor{currentstroke}{rgb}{0.690196,0.690196,0.690196}%
\pgfsetstrokecolor{currentstroke}%
\pgfsetstrokeopacity{0.250000}%
\pgfsetdash{}{0pt}%
\pgfpathmoveto{\pgfqpoint{0.590551in}{1.273126in}}%
\pgfpathlineto{\pgfqpoint{2.755906in}{1.273126in}}%
\pgfusepath{stroke}%
\end{pgfscope}%
\begin{pgfscope}%
\pgfsetbuttcap%
\pgfsetroundjoin%
\definecolor{currentfill}{rgb}{0.000000,0.000000,0.000000}%
\pgfsetfillcolor{currentfill}%
\pgfsetlinewidth{0.803000pt}%
\definecolor{currentstroke}{rgb}{0.000000,0.000000,0.000000}%
\pgfsetstrokecolor{currentstroke}%
\pgfsetdash{}{0pt}%
\pgfsys@defobject{currentmarker}{\pgfqpoint{-0.048611in}{0.000000in}}{\pgfqpoint{-0.000000in}{0.000000in}}{%
\pgfpathmoveto{\pgfqpoint{-0.000000in}{0.000000in}}%
\pgfpathlineto{\pgfqpoint{-0.048611in}{0.000000in}}%
\pgfusepath{stroke,fill}%
}%
\begin{pgfscope}%
\pgfsys@transformshift{0.590551in}{1.273126in}%
\pgfsys@useobject{currentmarker}{}%
\end{pgfscope}%
\end{pgfscope}%
\begin{pgfscope}%
\definecolor{textcolor}{rgb}{0.000000,0.000000,0.000000}%
\pgfsetstrokecolor{textcolor}%
\pgfsetfillcolor{textcolor}%
\pgftext[x=0.283449in, y=1.234546in, left, base]{\color{textcolor}{\rmfamily\fontsize{8.330000}{9.996000}\selectfont\catcode`\^=\active\def^{\ifmmode\sp\else\^{}\fi}\catcode`\%=\active\def%{\%}$\mathdefault{0.10}$}}%
\end{pgfscope}%
\begin{pgfscope}%
\pgfpathrectangle{\pgfqpoint{0.590551in}{0.590551in}}{\pgfqpoint{2.165354in}{2.165354in}}%
\pgfusepath{clip}%
\pgfsetrectcap%
\pgfsetroundjoin%
\pgfsetlinewidth{0.803000pt}%
\definecolor{currentstroke}{rgb}{0.690196,0.690196,0.690196}%
\pgfsetstrokecolor{currentstroke}%
\pgfsetstrokeopacity{0.250000}%
\pgfsetdash{}{0pt}%
\pgfpathmoveto{\pgfqpoint{0.590551in}{1.566499in}}%
\pgfpathlineto{\pgfqpoint{2.755906in}{1.566499in}}%
\pgfusepath{stroke}%
\end{pgfscope}%
\begin{pgfscope}%
\pgfsetbuttcap%
\pgfsetroundjoin%
\definecolor{currentfill}{rgb}{0.000000,0.000000,0.000000}%
\pgfsetfillcolor{currentfill}%
\pgfsetlinewidth{0.803000pt}%
\definecolor{currentstroke}{rgb}{0.000000,0.000000,0.000000}%
\pgfsetstrokecolor{currentstroke}%
\pgfsetdash{}{0pt}%
\pgfsys@defobject{currentmarker}{\pgfqpoint{-0.048611in}{0.000000in}}{\pgfqpoint{-0.000000in}{0.000000in}}{%
\pgfpathmoveto{\pgfqpoint{-0.000000in}{0.000000in}}%
\pgfpathlineto{\pgfqpoint{-0.048611in}{0.000000in}}%
\pgfusepath{stroke,fill}%
}%
\begin{pgfscope}%
\pgfsys@transformshift{0.590551in}{1.566499in}%
\pgfsys@useobject{currentmarker}{}%
\end{pgfscope}%
\end{pgfscope}%
\begin{pgfscope}%
\definecolor{textcolor}{rgb}{0.000000,0.000000,0.000000}%
\pgfsetstrokecolor{textcolor}%
\pgfsetfillcolor{textcolor}%
\pgftext[x=0.283449in, y=1.527919in, left, base]{\color{textcolor}{\rmfamily\fontsize{8.330000}{9.996000}\selectfont\catcode`\^=\active\def^{\ifmmode\sp\else\^{}\fi}\catcode`\%=\active\def%{\%}$\mathdefault{0.15}$}}%
\end{pgfscope}%
\begin{pgfscope}%
\pgfpathrectangle{\pgfqpoint{0.590551in}{0.590551in}}{\pgfqpoint{2.165354in}{2.165354in}}%
\pgfusepath{clip}%
\pgfsetrectcap%
\pgfsetroundjoin%
\pgfsetlinewidth{0.803000pt}%
\definecolor{currentstroke}{rgb}{0.690196,0.690196,0.690196}%
\pgfsetstrokecolor{currentstroke}%
\pgfsetstrokeopacity{0.250000}%
\pgfsetdash{}{0pt}%
\pgfpathmoveto{\pgfqpoint{0.590551in}{1.859872in}}%
\pgfpathlineto{\pgfqpoint{2.755906in}{1.859872in}}%
\pgfusepath{stroke}%
\end{pgfscope}%
\begin{pgfscope}%
\pgfsetbuttcap%
\pgfsetroundjoin%
\definecolor{currentfill}{rgb}{0.000000,0.000000,0.000000}%
\pgfsetfillcolor{currentfill}%
\pgfsetlinewidth{0.803000pt}%
\definecolor{currentstroke}{rgb}{0.000000,0.000000,0.000000}%
\pgfsetstrokecolor{currentstroke}%
\pgfsetdash{}{0pt}%
\pgfsys@defobject{currentmarker}{\pgfqpoint{-0.048611in}{0.000000in}}{\pgfqpoint{-0.000000in}{0.000000in}}{%
\pgfpathmoveto{\pgfqpoint{-0.000000in}{0.000000in}}%
\pgfpathlineto{\pgfqpoint{-0.048611in}{0.000000in}}%
\pgfusepath{stroke,fill}%
}%
\begin{pgfscope}%
\pgfsys@transformshift{0.590551in}{1.859872in}%
\pgfsys@useobject{currentmarker}{}%
\end{pgfscope}%
\end{pgfscope}%
\begin{pgfscope}%
\definecolor{textcolor}{rgb}{0.000000,0.000000,0.000000}%
\pgfsetstrokecolor{textcolor}%
\pgfsetfillcolor{textcolor}%
\pgftext[x=0.283449in, y=1.821292in, left, base]{\color{textcolor}{\rmfamily\fontsize{8.330000}{9.996000}\selectfont\catcode`\^=\active\def^{\ifmmode\sp\else\^{}\fi}\catcode`\%=\active\def%{\%}$\mathdefault{0.20}$}}%
\end{pgfscope}%
\begin{pgfscope}%
\pgfpathrectangle{\pgfqpoint{0.590551in}{0.590551in}}{\pgfqpoint{2.165354in}{2.165354in}}%
\pgfusepath{clip}%
\pgfsetrectcap%
\pgfsetroundjoin%
\pgfsetlinewidth{0.803000pt}%
\definecolor{currentstroke}{rgb}{0.690196,0.690196,0.690196}%
\pgfsetstrokecolor{currentstroke}%
\pgfsetstrokeopacity{0.250000}%
\pgfsetdash{}{0pt}%
\pgfpathmoveto{\pgfqpoint{0.590551in}{2.153245in}}%
\pgfpathlineto{\pgfqpoint{2.755906in}{2.153245in}}%
\pgfusepath{stroke}%
\end{pgfscope}%
\begin{pgfscope}%
\pgfsetbuttcap%
\pgfsetroundjoin%
\definecolor{currentfill}{rgb}{0.000000,0.000000,0.000000}%
\pgfsetfillcolor{currentfill}%
\pgfsetlinewidth{0.803000pt}%
\definecolor{currentstroke}{rgb}{0.000000,0.000000,0.000000}%
\pgfsetstrokecolor{currentstroke}%
\pgfsetdash{}{0pt}%
\pgfsys@defobject{currentmarker}{\pgfqpoint{-0.048611in}{0.000000in}}{\pgfqpoint{-0.000000in}{0.000000in}}{%
\pgfpathmoveto{\pgfqpoint{-0.000000in}{0.000000in}}%
\pgfpathlineto{\pgfqpoint{-0.048611in}{0.000000in}}%
\pgfusepath{stroke,fill}%
}%
\begin{pgfscope}%
\pgfsys@transformshift{0.590551in}{2.153245in}%
\pgfsys@useobject{currentmarker}{}%
\end{pgfscope}%
\end{pgfscope}%
\begin{pgfscope}%
\definecolor{textcolor}{rgb}{0.000000,0.000000,0.000000}%
\pgfsetstrokecolor{textcolor}%
\pgfsetfillcolor{textcolor}%
\pgftext[x=0.283449in, y=2.114665in, left, base]{\color{textcolor}{\rmfamily\fontsize{8.330000}{9.996000}\selectfont\catcode`\^=\active\def^{\ifmmode\sp\else\^{}\fi}\catcode`\%=\active\def%{\%}$\mathdefault{0.25}$}}%
\end{pgfscope}%
\begin{pgfscope}%
\pgfpathrectangle{\pgfqpoint{0.590551in}{0.590551in}}{\pgfqpoint{2.165354in}{2.165354in}}%
\pgfusepath{clip}%
\pgfsetrectcap%
\pgfsetroundjoin%
\pgfsetlinewidth{0.803000pt}%
\definecolor{currentstroke}{rgb}{0.690196,0.690196,0.690196}%
\pgfsetstrokecolor{currentstroke}%
\pgfsetstrokeopacity{0.250000}%
\pgfsetdash{}{0pt}%
\pgfpathmoveto{\pgfqpoint{0.590551in}{2.446618in}}%
\pgfpathlineto{\pgfqpoint{2.755906in}{2.446618in}}%
\pgfusepath{stroke}%
\end{pgfscope}%
\begin{pgfscope}%
\pgfsetbuttcap%
\pgfsetroundjoin%
\definecolor{currentfill}{rgb}{0.000000,0.000000,0.000000}%
\pgfsetfillcolor{currentfill}%
\pgfsetlinewidth{0.803000pt}%
\definecolor{currentstroke}{rgb}{0.000000,0.000000,0.000000}%
\pgfsetstrokecolor{currentstroke}%
\pgfsetdash{}{0pt}%
\pgfsys@defobject{currentmarker}{\pgfqpoint{-0.048611in}{0.000000in}}{\pgfqpoint{-0.000000in}{0.000000in}}{%
\pgfpathmoveto{\pgfqpoint{-0.000000in}{0.000000in}}%
\pgfpathlineto{\pgfqpoint{-0.048611in}{0.000000in}}%
\pgfusepath{stroke,fill}%
}%
\begin{pgfscope}%
\pgfsys@transformshift{0.590551in}{2.446618in}%
\pgfsys@useobject{currentmarker}{}%
\end{pgfscope}%
\end{pgfscope}%
\begin{pgfscope}%
\definecolor{textcolor}{rgb}{0.000000,0.000000,0.000000}%
\pgfsetstrokecolor{textcolor}%
\pgfsetfillcolor{textcolor}%
\pgftext[x=0.283449in, y=2.408038in, left, base]{\color{textcolor}{\rmfamily\fontsize{8.330000}{9.996000}\selectfont\catcode`\^=\active\def^{\ifmmode\sp\else\^{}\fi}\catcode`\%=\active\def%{\%}$\mathdefault{0.30}$}}%
\end{pgfscope}%
\begin{pgfscope}%
\pgfpathrectangle{\pgfqpoint{0.590551in}{0.590551in}}{\pgfqpoint{2.165354in}{2.165354in}}%
\pgfusepath{clip}%
\pgfsetrectcap%
\pgfsetroundjoin%
\pgfsetlinewidth{0.803000pt}%
\definecolor{currentstroke}{rgb}{0.690196,0.690196,0.690196}%
\pgfsetstrokecolor{currentstroke}%
\pgfsetstrokeopacity{0.250000}%
\pgfsetdash{}{0pt}%
\pgfpathmoveto{\pgfqpoint{0.590551in}{2.739991in}}%
\pgfpathlineto{\pgfqpoint{2.755906in}{2.739991in}}%
\pgfusepath{stroke}%
\end{pgfscope}%
\begin{pgfscope}%
\pgfsetbuttcap%
\pgfsetroundjoin%
\definecolor{currentfill}{rgb}{0.000000,0.000000,0.000000}%
\pgfsetfillcolor{currentfill}%
\pgfsetlinewidth{0.803000pt}%
\definecolor{currentstroke}{rgb}{0.000000,0.000000,0.000000}%
\pgfsetstrokecolor{currentstroke}%
\pgfsetdash{}{0pt}%
\pgfsys@defobject{currentmarker}{\pgfqpoint{-0.048611in}{0.000000in}}{\pgfqpoint{-0.000000in}{0.000000in}}{%
\pgfpathmoveto{\pgfqpoint{-0.000000in}{0.000000in}}%
\pgfpathlineto{\pgfqpoint{-0.048611in}{0.000000in}}%
\pgfusepath{stroke,fill}%
}%
\begin{pgfscope}%
\pgfsys@transformshift{0.590551in}{2.739991in}%
\pgfsys@useobject{currentmarker}{}%
\end{pgfscope}%
\end{pgfscope}%
\begin{pgfscope}%
\definecolor{textcolor}{rgb}{0.000000,0.000000,0.000000}%
\pgfsetstrokecolor{textcolor}%
\pgfsetfillcolor{textcolor}%
\pgftext[x=0.283449in, y=2.701411in, left, base]{\color{textcolor}{\rmfamily\fontsize{8.330000}{9.996000}\selectfont\catcode`\^=\active\def^{\ifmmode\sp\else\^{}\fi}\catcode`\%=\active\def%{\%}$\mathdefault{0.35}$}}%
\end{pgfscope}%
\begin{pgfscope}%
\definecolor{textcolor}{rgb}{0.000000,0.000000,0.000000}%
\pgfsetstrokecolor{textcolor}%
\pgfsetfillcolor{textcolor}%
\pgftext[x=0.227894in,y=1.673228in,,bottom,rotate=90.000000]{\color{textcolor}{\rmfamily\fontsize{10.000000}{12.000000}\selectfont\catcode`\^=\active\def^{\ifmmode\sp\else\^{}\fi}\catcode`\%=\active\def%{\%}rank (normalized)}}%
\end{pgfscope}%
\begin{pgfscope}%
\pgfpathrectangle{\pgfqpoint{0.590551in}{0.590551in}}{\pgfqpoint{2.165354in}{2.165354in}}%
\pgfusepath{clip}%
\pgfsetrectcap%
\pgfsetroundjoin%
\pgfsetlinewidth{1.405250pt}%
\definecolor{currentstroke}{rgb}{0.862745,0.352941,0.156863}%
\pgfsetstrokecolor{currentstroke}%
\pgfsetdash{}{0pt}%
\pgfpathmoveto{\pgfqpoint{0.688976in}{1.786529in}}%
\pgfpathlineto{\pgfqpoint{0.885827in}{1.144775in}}%
\pgfpathlineto{\pgfqpoint{1.082677in}{2.222005in}}%
\pgfpathlineto{\pgfqpoint{1.279528in}{1.408353in}}%
\pgfpathlineto{\pgfqpoint{1.476378in}{1.070286in}}%
\pgfpathlineto{\pgfqpoint{1.673228in}{0.846818in}}%
\pgfpathlineto{\pgfqpoint{1.870079in}{0.763734in}}%
\pgfpathlineto{\pgfqpoint{2.066929in}{0.723625in}}%
\pgfpathlineto{\pgfqpoint{2.263780in}{0.799546in}}%
\pgfpathlineto{\pgfqpoint{2.460630in}{0.705360in}}%
\pgfpathlineto{\pgfqpoint{2.657480in}{0.744843in}}%
\pgfusepath{stroke}%
\end{pgfscope}%
\begin{pgfscope}%
\pgfsetrectcap%
\pgfsetmiterjoin%
\pgfsetlinewidth{0.803000pt}%
\definecolor{currentstroke}{rgb}{0.000000,0.000000,0.000000}%
\pgfsetstrokecolor{currentstroke}%
\pgfsetdash{}{0pt}%
\pgfpathmoveto{\pgfqpoint{0.590551in}{0.590551in}}%
\pgfpathlineto{\pgfqpoint{0.590551in}{2.755906in}}%
\pgfusepath{stroke}%
\end{pgfscope}%
\begin{pgfscope}%
\pgfsetrectcap%
\pgfsetmiterjoin%
\pgfsetlinewidth{0.803000pt}%
\definecolor{currentstroke}{rgb}{0.000000,0.000000,0.000000}%
\pgfsetstrokecolor{currentstroke}%
\pgfsetdash{}{0pt}%
\pgfpathmoveto{\pgfqpoint{2.755906in}{0.590551in}}%
\pgfpathlineto{\pgfqpoint{2.755906in}{2.755906in}}%
\pgfusepath{stroke}%
\end{pgfscope}%
\begin{pgfscope}%
\pgfsetrectcap%
\pgfsetmiterjoin%
\pgfsetlinewidth{0.803000pt}%
\definecolor{currentstroke}{rgb}{0.000000,0.000000,0.000000}%
\pgfsetstrokecolor{currentstroke}%
\pgfsetdash{}{0pt}%
\pgfpathmoveto{\pgfqpoint{0.590551in}{0.590551in}}%
\pgfpathlineto{\pgfqpoint{2.755906in}{0.590551in}}%
\pgfusepath{stroke}%
\end{pgfscope}%
\begin{pgfscope}%
\pgfsetrectcap%
\pgfsetmiterjoin%
\pgfsetlinewidth{0.803000pt}%
\definecolor{currentstroke}{rgb}{0.000000,0.000000,0.000000}%
\pgfsetstrokecolor{currentstroke}%
\pgfsetdash{}{0pt}%
\pgfpathmoveto{\pgfqpoint{0.590551in}{2.755906in}}%
\pgfpathlineto{\pgfqpoint{2.755906in}{2.755906in}}%
\pgfusepath{stroke}%
\end{pgfscope}%
\end{pgfpicture}%
\makeatother%
\endgroup%

%% file: figures/p.pgf
%% Creator: Matplotlib, PGF backend
%%
%% To include the figure in your LaTeX document, write
%%   \input{<filename>.pgf}
%%
%% Make sure the required packages are loaded in your preamble
%%   \usepackage{pgf}
%%
%% Also ensure that all the required font packages are loaded; for instance,
%% the lmodern package is sometimes necessary when using math font.
%%   \usepackage{lmodern}
%%
%% Figures using additional raster images can only be included by \input if
%% they are in the same directory as the main LaTeX file. For loading figures
%% from other directories you can use the `import` package
%%   \usepackage{import}
%%
%% and then include the figures with
%%   \import{<path to file>}{<filename>.pgf}
%%
%% Matplotlib used the following preamble
%%   \def\mathdefault#1{#1}
%%   \everymath=\expandafter{\the\everymath\displaystyle}
%%   \IfFileExists{scrextend.sty}{
%%     \usepackage[fontsize=10.000000pt]{scrextend}
%%   }{
%%     \renewcommand{\normalsize}{\fontsize{10.000000}{12.000000}\selectfont}
%%     \normalsize
%%   }
%%   
%%   \makeatletter\@ifpackageloaded{underscore}{}{\usepackage[strings]{underscore}}\makeatother
%%
\begingroup%
\makeatletter%
\begin{pgfpicture}%
\pgfpathrectangle{\pgfpointorigin}{\pgfqpoint{2.952756in}{2.952756in}}%
\pgfusepath{use as bounding box, clip}%
\begin{pgfscope}%
\pgfsetbuttcap%
\pgfsetmiterjoin%
\pgfsetlinewidth{0.000000pt}%
\definecolor{currentstroke}{rgb}{1.000000,1.000000,1.000000}%
\pgfsetstrokecolor{currentstroke}%
\pgfsetdash{}{0pt}%
\pgfpathmoveto{\pgfqpoint{0.000000in}{0.000000in}}%
\pgfpathlineto{\pgfqpoint{2.952756in}{0.000000in}}%
\pgfpathlineto{\pgfqpoint{2.952756in}{2.952756in}}%
\pgfpathlineto{\pgfqpoint{0.000000in}{2.952756in}}%
\pgfpathlineto{\pgfqpoint{0.000000in}{0.000000in}}%
\pgfpathclose%
\pgfusepath{}%
\end{pgfscope}%
\begin{pgfscope}%
\pgfsetbuttcap%
\pgfsetmiterjoin%
\pgfsetlinewidth{0.000000pt}%
\definecolor{currentstroke}{rgb}{0.000000,0.000000,0.000000}%
\pgfsetstrokecolor{currentstroke}%
\pgfsetstrokeopacity{0.000000}%
\pgfsetdash{}{0pt}%
\pgfpathmoveto{\pgfqpoint{0.590551in}{0.590551in}}%
\pgfpathlineto{\pgfqpoint{2.755906in}{0.590551in}}%
\pgfpathlineto{\pgfqpoint{2.755906in}{2.755906in}}%
\pgfpathlineto{\pgfqpoint{0.590551in}{2.755906in}}%
\pgfpathlineto{\pgfqpoint{0.590551in}{0.590551in}}%
\pgfpathclose%
\pgfusepath{}%
\end{pgfscope}%
\begin{pgfscope}%
\pgfpathrectangle{\pgfqpoint{0.590551in}{0.590551in}}{\pgfqpoint{2.165354in}{2.165354in}}%
\pgfusepath{clip}%
\pgfsetbuttcap%
\pgfsetroundjoin%
\definecolor{currentfill}{rgb}{0.470588,0.274510,0.686275}%
\pgfsetfillcolor{currentfill}%
\pgfsetfillopacity{0.220000}%
\pgfsetlinewidth{1.003750pt}%
\definecolor{currentstroke}{rgb}{0.470588,0.274510,0.686275}%
\pgfsetstrokecolor{currentstroke}%
\pgfsetstrokeopacity{0.220000}%
\pgfsetdash{}{0pt}%
\pgfsys@defobject{currentmarker}{\pgfqpoint{0.688976in}{0.688976in}}{\pgfqpoint{2.657480in}{2.657480in}}{%
\pgfpathmoveto{\pgfqpoint{0.688976in}{2.201038in}}%
\pgfpathlineto{\pgfqpoint{0.688976in}{2.155065in}}%
\pgfpathlineto{\pgfqpoint{0.707732in}{1.694682in}}%
\pgfpathlineto{\pgfqpoint{0.728315in}{1.597994in}}%
\pgfpathlineto{\pgfqpoint{0.750906in}{1.435922in}}%
\pgfpathlineto{\pgfqpoint{0.775699in}{1.280929in}}%
\pgfpathlineto{\pgfqpoint{0.802910in}{1.130022in}}%
\pgfpathlineto{\pgfqpoint{0.832774in}{1.045228in}}%
\pgfpathlineto{\pgfqpoint{0.865549in}{0.698536in}}%
\pgfpathlineto{\pgfqpoint{0.901520in}{0.708971in}}%
\pgfpathlineto{\pgfqpoint{0.940998in}{0.696711in}}%
\pgfpathlineto{\pgfqpoint{0.984325in}{0.692625in}}%
\pgfpathlineto{\pgfqpoint{1.031876in}{0.702549in}}%
\pgfpathlineto{\pgfqpoint{1.084064in}{0.696274in}}%
\pgfpathlineto{\pgfqpoint{1.141339in}{0.691457in}}%
\pgfpathlineto{\pgfqpoint{1.204199in}{0.699484in}}%
\pgfpathlineto{\pgfqpoint{1.273188in}{0.694814in}}%
\pgfpathlineto{\pgfqpoint{1.348903in}{0.690728in}}%
\pgfpathlineto{\pgfqpoint{1.432001in}{0.689049in}}%
\pgfpathlineto{\pgfqpoint{1.523200in}{0.689122in}}%
\pgfpathlineto{\pgfqpoint{1.623291in}{0.688976in}}%
\pgfpathlineto{\pgfqpoint{1.733141in}{0.689268in}}%
\pgfpathlineto{\pgfqpoint{1.853701in}{0.689560in}}%
\pgfpathlineto{\pgfqpoint{1.986015in}{0.689706in}}%
\pgfpathlineto{\pgfqpoint{2.131230in}{0.693939in}}%
\pgfpathlineto{\pgfqpoint{2.290603in}{0.772749in}}%
\pgfpathlineto{\pgfqpoint{2.465515in}{0.921685in}}%
\pgfpathlineto{\pgfqpoint{2.657480in}{0.992542in}}%
\pgfpathlineto{\pgfqpoint{2.657480in}{2.657480in}}%
\pgfpathlineto{\pgfqpoint{2.657480in}{2.657480in}}%
\pgfpathlineto{\pgfqpoint{2.465515in}{2.115952in}}%
\pgfpathlineto{\pgfqpoint{2.290603in}{1.611056in}}%
\pgfpathlineto{\pgfqpoint{2.131230in}{1.042017in}}%
\pgfpathlineto{\pgfqpoint{1.986015in}{0.808213in}}%
\pgfpathlineto{\pgfqpoint{1.853701in}{0.734438in}}%
\pgfpathlineto{\pgfqpoint{1.733141in}{0.737138in}}%
\pgfpathlineto{\pgfqpoint{1.623291in}{0.731446in}}%
\pgfpathlineto{\pgfqpoint{1.523200in}{0.743414in}}%
\pgfpathlineto{\pgfqpoint{1.432001in}{0.731154in}}%
\pgfpathlineto{\pgfqpoint{1.348903in}{0.748376in}}%
\pgfpathlineto{\pgfqpoint{1.273188in}{0.755527in}}%
\pgfpathlineto{\pgfqpoint{1.204199in}{0.752827in}}%
\pgfpathlineto{\pgfqpoint{1.141339in}{0.751003in}}%
\pgfpathlineto{\pgfqpoint{1.084064in}{0.762606in}}%
\pgfpathlineto{\pgfqpoint{1.031876in}{0.782089in}}%
\pgfpathlineto{\pgfqpoint{0.984325in}{0.806827in}}%
\pgfpathlineto{\pgfqpoint{0.940998in}{0.842510in}}%
\pgfpathlineto{\pgfqpoint{0.901520in}{0.894321in}}%
\pgfpathlineto{\pgfqpoint{0.865549in}{0.951969in}}%
\pgfpathlineto{\pgfqpoint{0.832774in}{1.056028in}}%
\pgfpathlineto{\pgfqpoint{0.802910in}{1.156292in}}%
\pgfpathlineto{\pgfqpoint{0.775699in}{1.296764in}}%
\pgfpathlineto{\pgfqpoint{0.750906in}{1.435922in}}%
\pgfpathlineto{\pgfqpoint{0.728315in}{1.600475in}}%
\pgfpathlineto{\pgfqpoint{0.707732in}{1.838584in}}%
\pgfpathlineto{\pgfqpoint{0.688976in}{2.201038in}}%
\pgfpathlineto{\pgfqpoint{0.688976in}{2.201038in}}%
\pgfpathclose%
\pgfusepath{stroke,fill}%
}%
\begin{pgfscope}%
\pgfsys@transformshift{0.000000in}{0.000000in}%
\pgfsys@useobject{currentmarker}{}%
\end{pgfscope}%
\end{pgfscope}%
\begin{pgfscope}%
\pgfpathrectangle{\pgfqpoint{0.590551in}{0.590551in}}{\pgfqpoint{2.165354in}{2.165354in}}%
\pgfusepath{clip}%
\pgfsetrectcap%
\pgfsetroundjoin%
\pgfsetlinewidth{0.803000pt}%
\definecolor{currentstroke}{rgb}{0.690196,0.690196,0.690196}%
\pgfsetstrokecolor{currentstroke}%
\pgfsetstrokeopacity{0.250000}%
\pgfsetdash{}{0pt}%
\pgfpathmoveto{\pgfqpoint{1.191123in}{0.590551in}}%
\pgfpathlineto{\pgfqpoint{1.191123in}{2.755906in}}%
\pgfusepath{stroke}%
\end{pgfscope}%
\begin{pgfscope}%
\pgfsetbuttcap%
\pgfsetroundjoin%
\definecolor{currentfill}{rgb}{0.000000,0.000000,0.000000}%
\pgfsetfillcolor{currentfill}%
\pgfsetlinewidth{0.803000pt}%
\definecolor{currentstroke}{rgb}{0.000000,0.000000,0.000000}%
\pgfsetstrokecolor{currentstroke}%
\pgfsetdash{}{0pt}%
\pgfsys@defobject{currentmarker}{\pgfqpoint{0.000000in}{-0.048611in}}{\pgfqpoint{0.000000in}{0.000000in}}{%
\pgfpathmoveto{\pgfqpoint{0.000000in}{0.000000in}}%
\pgfpathlineto{\pgfqpoint{0.000000in}{-0.048611in}}%
\pgfusepath{stroke,fill}%
}%
\begin{pgfscope}%
\pgfsys@transformshift{1.191123in}{0.590551in}%
\pgfsys@useobject{currentmarker}{}%
\end{pgfscope}%
\end{pgfscope}%
\begin{pgfscope}%
\definecolor{textcolor}{rgb}{0.000000,0.000000,0.000000}%
\pgfsetstrokecolor{textcolor}%
\pgfsetfillcolor{textcolor}%
\pgftext[x=1.191123in,y=0.493329in,,top]{\color{textcolor}{\rmfamily\fontsize{8.330000}{9.996000}\selectfont\catcode`\^=\active\def^{\ifmmode\sp\else\^{}\fi}\catcode`\%=\active\def%{\%}$\mathdefault{0.0005}$}}%
\end{pgfscope}%
\begin{pgfscope}%
\pgfpathrectangle{\pgfqpoint{0.590551in}{0.590551in}}{\pgfqpoint{2.165354in}{2.165354in}}%
\pgfusepath{clip}%
\pgfsetrectcap%
\pgfsetroundjoin%
\pgfsetlinewidth{0.803000pt}%
\definecolor{currentstroke}{rgb}{0.690196,0.690196,0.690196}%
\pgfsetstrokecolor{currentstroke}%
\pgfsetstrokeopacity{0.250000}%
\pgfsetdash{}{0pt}%
\pgfpathmoveto{\pgfqpoint{1.885634in}{0.590551in}}%
\pgfpathlineto{\pgfqpoint{1.885634in}{2.755906in}}%
\pgfusepath{stroke}%
\end{pgfscope}%
\begin{pgfscope}%
\pgfsetbuttcap%
\pgfsetroundjoin%
\definecolor{currentfill}{rgb}{0.000000,0.000000,0.000000}%
\pgfsetfillcolor{currentfill}%
\pgfsetlinewidth{0.803000pt}%
\definecolor{currentstroke}{rgb}{0.000000,0.000000,0.000000}%
\pgfsetstrokecolor{currentstroke}%
\pgfsetdash{}{0pt}%
\pgfsys@defobject{currentmarker}{\pgfqpoint{0.000000in}{-0.048611in}}{\pgfqpoint{0.000000in}{0.000000in}}{%
\pgfpathmoveto{\pgfqpoint{0.000000in}{0.000000in}}%
\pgfpathlineto{\pgfqpoint{0.000000in}{-0.048611in}}%
\pgfusepath{stroke,fill}%
}%
\begin{pgfscope}%
\pgfsys@transformshift{1.885634in}{0.590551in}%
\pgfsys@useobject{currentmarker}{}%
\end{pgfscope}%
\end{pgfscope}%
\begin{pgfscope}%
\definecolor{textcolor}{rgb}{0.000000,0.000000,0.000000}%
\pgfsetstrokecolor{textcolor}%
\pgfsetfillcolor{textcolor}%
\pgftext[x=1.885634in,y=0.493329in,,top]{\color{textcolor}{\rmfamily\fontsize{8.330000}{9.996000}\selectfont\catcode`\^=\active\def^{\ifmmode\sp\else\^{}\fi}\catcode`\%=\active\def%{\%}$\mathdefault{0.0010}$}}%
\end{pgfscope}%
\begin{pgfscope}%
\pgfpathrectangle{\pgfqpoint{0.590551in}{0.590551in}}{\pgfqpoint{2.165354in}{2.165354in}}%
\pgfusepath{clip}%
\pgfsetrectcap%
\pgfsetroundjoin%
\pgfsetlinewidth{0.803000pt}%
\definecolor{currentstroke}{rgb}{0.690196,0.690196,0.690196}%
\pgfsetstrokecolor{currentstroke}%
\pgfsetstrokeopacity{0.250000}%
\pgfsetdash{}{0pt}%
\pgfpathmoveto{\pgfqpoint{2.580145in}{0.590551in}}%
\pgfpathlineto{\pgfqpoint{2.580145in}{2.755906in}}%
\pgfusepath{stroke}%
\end{pgfscope}%
\begin{pgfscope}%
\pgfsetbuttcap%
\pgfsetroundjoin%
\definecolor{currentfill}{rgb}{0.000000,0.000000,0.000000}%
\pgfsetfillcolor{currentfill}%
\pgfsetlinewidth{0.803000pt}%
\definecolor{currentstroke}{rgb}{0.000000,0.000000,0.000000}%
\pgfsetstrokecolor{currentstroke}%
\pgfsetdash{}{0pt}%
\pgfsys@defobject{currentmarker}{\pgfqpoint{0.000000in}{-0.048611in}}{\pgfqpoint{0.000000in}{0.000000in}}{%
\pgfpathmoveto{\pgfqpoint{0.000000in}{0.000000in}}%
\pgfpathlineto{\pgfqpoint{0.000000in}{-0.048611in}}%
\pgfusepath{stroke,fill}%
}%
\begin{pgfscope}%
\pgfsys@transformshift{2.580145in}{0.590551in}%
\pgfsys@useobject{currentmarker}{}%
\end{pgfscope}%
\end{pgfscope}%
\begin{pgfscope}%
\definecolor{textcolor}{rgb}{0.000000,0.000000,0.000000}%
\pgfsetstrokecolor{textcolor}%
\pgfsetfillcolor{textcolor}%
\pgftext[x=2.580145in,y=0.493329in,,top]{\color{textcolor}{\rmfamily\fontsize{8.330000}{9.996000}\selectfont\catcode`\^=\active\def^{\ifmmode\sp\else\^{}\fi}\catcode`\%=\active\def%{\%}$\mathdefault{0.0015}$}}%
\end{pgfscope}%
\begin{pgfscope}%
\definecolor{textcolor}{rgb}{0.000000,0.000000,0.000000}%
\pgfsetstrokecolor{textcolor}%
\pgfsetfillcolor{textcolor}%
\pgftext[x=1.673228in,y=0.339008in,,top]{\color{textcolor}{\rmfamily\fontsize{10.000000}{12.000000}\selectfont\catcode`\^=\active\def^{\ifmmode\sp\else\^{}\fi}\catcode`\%=\active\def%{\%}edge probability $p$}}%
\end{pgfscope}%
\begin{pgfscope}%
\pgfpathrectangle{\pgfqpoint{0.590551in}{0.590551in}}{\pgfqpoint{2.165354in}{2.165354in}}%
\pgfusepath{clip}%
\pgfsetrectcap%
\pgfsetroundjoin%
\pgfsetlinewidth{0.803000pt}%
\definecolor{currentstroke}{rgb}{0.690196,0.690196,0.690196}%
\pgfsetstrokecolor{currentstroke}%
\pgfsetstrokeopacity{0.250000}%
\pgfsetdash{}{0pt}%
\pgfpathmoveto{\pgfqpoint{0.590551in}{0.688393in}}%
\pgfpathlineto{\pgfqpoint{2.755906in}{0.688393in}}%
\pgfusepath{stroke}%
\end{pgfscope}%
\begin{pgfscope}%
\pgfsetbuttcap%
\pgfsetroundjoin%
\definecolor{currentfill}{rgb}{0.000000,0.000000,0.000000}%
\pgfsetfillcolor{currentfill}%
\pgfsetlinewidth{0.803000pt}%
\definecolor{currentstroke}{rgb}{0.000000,0.000000,0.000000}%
\pgfsetstrokecolor{currentstroke}%
\pgfsetdash{}{0pt}%
\pgfsys@defobject{currentmarker}{\pgfqpoint{-0.048611in}{0.000000in}}{\pgfqpoint{-0.000000in}{0.000000in}}{%
\pgfpathmoveto{\pgfqpoint{-0.000000in}{0.000000in}}%
\pgfpathlineto{\pgfqpoint{-0.048611in}{0.000000in}}%
\pgfusepath{stroke,fill}%
}%
\begin{pgfscope}%
\pgfsys@transformshift{0.590551in}{0.688393in}%
\pgfsys@useobject{currentmarker}{}%
\end{pgfscope}%
\end{pgfscope}%
\begin{pgfscope}%
\definecolor{textcolor}{rgb}{0.000000,0.000000,0.000000}%
\pgfsetstrokecolor{textcolor}%
\pgfsetfillcolor{textcolor}%
\pgftext[x=0.342478in, y=0.649812in, left, base]{\color{textcolor}{\rmfamily\fontsize{8.330000}{9.996000}\selectfont\catcode`\^=\active\def^{\ifmmode\sp\else\^{}\fi}\catcode`\%=\active\def%{\%}$\mathdefault{0.0}$}}%
\end{pgfscope}%
\begin{pgfscope}%
\pgfpathrectangle{\pgfqpoint{0.590551in}{0.590551in}}{\pgfqpoint{2.165354in}{2.165354in}}%
\pgfusepath{clip}%
\pgfsetrectcap%
\pgfsetroundjoin%
\pgfsetlinewidth{0.803000pt}%
\definecolor{currentstroke}{rgb}{0.690196,0.690196,0.690196}%
\pgfsetstrokecolor{currentstroke}%
\pgfsetstrokeopacity{0.250000}%
\pgfsetdash{}{0pt}%
\pgfpathmoveto{\pgfqpoint{0.590551in}{0.980282in}}%
\pgfpathlineto{\pgfqpoint{2.755906in}{0.980282in}}%
\pgfusepath{stroke}%
\end{pgfscope}%
\begin{pgfscope}%
\pgfsetbuttcap%
\pgfsetroundjoin%
\definecolor{currentfill}{rgb}{0.000000,0.000000,0.000000}%
\pgfsetfillcolor{currentfill}%
\pgfsetlinewidth{0.803000pt}%
\definecolor{currentstroke}{rgb}{0.000000,0.000000,0.000000}%
\pgfsetstrokecolor{currentstroke}%
\pgfsetdash{}{0pt}%
\pgfsys@defobject{currentmarker}{\pgfqpoint{-0.048611in}{0.000000in}}{\pgfqpoint{-0.000000in}{0.000000in}}{%
\pgfpathmoveto{\pgfqpoint{-0.000000in}{0.000000in}}%
\pgfpathlineto{\pgfqpoint{-0.048611in}{0.000000in}}%
\pgfusepath{stroke,fill}%
}%
\begin{pgfscope}%
\pgfsys@transformshift{0.590551in}{0.980282in}%
\pgfsys@useobject{currentmarker}{}%
\end{pgfscope}%
\end{pgfscope}%
\begin{pgfscope}%
\definecolor{textcolor}{rgb}{0.000000,0.000000,0.000000}%
\pgfsetstrokecolor{textcolor}%
\pgfsetfillcolor{textcolor}%
\pgftext[x=0.342478in, y=0.941702in, left, base]{\color{textcolor}{\rmfamily\fontsize{8.330000}{9.996000}\selectfont\catcode`\^=\active\def^{\ifmmode\sp\else\^{}\fi}\catcode`\%=\active\def%{\%}$\mathdefault{0.1}$}}%
\end{pgfscope}%
\begin{pgfscope}%
\pgfpathrectangle{\pgfqpoint{0.590551in}{0.590551in}}{\pgfqpoint{2.165354in}{2.165354in}}%
\pgfusepath{clip}%
\pgfsetrectcap%
\pgfsetroundjoin%
\pgfsetlinewidth{0.803000pt}%
\definecolor{currentstroke}{rgb}{0.690196,0.690196,0.690196}%
\pgfsetstrokecolor{currentstroke}%
\pgfsetstrokeopacity{0.250000}%
\pgfsetdash{}{0pt}%
\pgfpathmoveto{\pgfqpoint{0.590551in}{1.272172in}}%
\pgfpathlineto{\pgfqpoint{2.755906in}{1.272172in}}%
\pgfusepath{stroke}%
\end{pgfscope}%
\begin{pgfscope}%
\pgfsetbuttcap%
\pgfsetroundjoin%
\definecolor{currentfill}{rgb}{0.000000,0.000000,0.000000}%
\pgfsetfillcolor{currentfill}%
\pgfsetlinewidth{0.803000pt}%
\definecolor{currentstroke}{rgb}{0.000000,0.000000,0.000000}%
\pgfsetstrokecolor{currentstroke}%
\pgfsetdash{}{0pt}%
\pgfsys@defobject{currentmarker}{\pgfqpoint{-0.048611in}{0.000000in}}{\pgfqpoint{-0.000000in}{0.000000in}}{%
\pgfpathmoveto{\pgfqpoint{-0.000000in}{0.000000in}}%
\pgfpathlineto{\pgfqpoint{-0.048611in}{0.000000in}}%
\pgfusepath{stroke,fill}%
}%
\begin{pgfscope}%
\pgfsys@transformshift{0.590551in}{1.272172in}%
\pgfsys@useobject{currentmarker}{}%
\end{pgfscope}%
\end{pgfscope}%
\begin{pgfscope}%
\definecolor{textcolor}{rgb}{0.000000,0.000000,0.000000}%
\pgfsetstrokecolor{textcolor}%
\pgfsetfillcolor{textcolor}%
\pgftext[x=0.342478in, y=1.233592in, left, base]{\color{textcolor}{\rmfamily\fontsize{8.330000}{9.996000}\selectfont\catcode`\^=\active\def^{\ifmmode\sp\else\^{}\fi}\catcode`\%=\active\def%{\%}$\mathdefault{0.2}$}}%
\end{pgfscope}%
\begin{pgfscope}%
\pgfpathrectangle{\pgfqpoint{0.590551in}{0.590551in}}{\pgfqpoint{2.165354in}{2.165354in}}%
\pgfusepath{clip}%
\pgfsetrectcap%
\pgfsetroundjoin%
\pgfsetlinewidth{0.803000pt}%
\definecolor{currentstroke}{rgb}{0.690196,0.690196,0.690196}%
\pgfsetstrokecolor{currentstroke}%
\pgfsetstrokeopacity{0.250000}%
\pgfsetdash{}{0pt}%
\pgfpathmoveto{\pgfqpoint{0.590551in}{1.564062in}}%
\pgfpathlineto{\pgfqpoint{2.755906in}{1.564062in}}%
\pgfusepath{stroke}%
\end{pgfscope}%
\begin{pgfscope}%
\pgfsetbuttcap%
\pgfsetroundjoin%
\definecolor{currentfill}{rgb}{0.000000,0.000000,0.000000}%
\pgfsetfillcolor{currentfill}%
\pgfsetlinewidth{0.803000pt}%
\definecolor{currentstroke}{rgb}{0.000000,0.000000,0.000000}%
\pgfsetstrokecolor{currentstroke}%
\pgfsetdash{}{0pt}%
\pgfsys@defobject{currentmarker}{\pgfqpoint{-0.048611in}{0.000000in}}{\pgfqpoint{-0.000000in}{0.000000in}}{%
\pgfpathmoveto{\pgfqpoint{-0.000000in}{0.000000in}}%
\pgfpathlineto{\pgfqpoint{-0.048611in}{0.000000in}}%
\pgfusepath{stroke,fill}%
}%
\begin{pgfscope}%
\pgfsys@transformshift{0.590551in}{1.564062in}%
\pgfsys@useobject{currentmarker}{}%
\end{pgfscope}%
\end{pgfscope}%
\begin{pgfscope}%
\definecolor{textcolor}{rgb}{0.000000,0.000000,0.000000}%
\pgfsetstrokecolor{textcolor}%
\pgfsetfillcolor{textcolor}%
\pgftext[x=0.342478in, y=1.525481in, left, base]{\color{textcolor}{\rmfamily\fontsize{8.330000}{9.996000}\selectfont\catcode`\^=\active\def^{\ifmmode\sp\else\^{}\fi}\catcode`\%=\active\def%{\%}$\mathdefault{0.3}$}}%
\end{pgfscope}%
\begin{pgfscope}%
\pgfpathrectangle{\pgfqpoint{0.590551in}{0.590551in}}{\pgfqpoint{2.165354in}{2.165354in}}%
\pgfusepath{clip}%
\pgfsetrectcap%
\pgfsetroundjoin%
\pgfsetlinewidth{0.803000pt}%
\definecolor{currentstroke}{rgb}{0.690196,0.690196,0.690196}%
\pgfsetstrokecolor{currentstroke}%
\pgfsetstrokeopacity{0.250000}%
\pgfsetdash{}{0pt}%
\pgfpathmoveto{\pgfqpoint{0.590551in}{1.855951in}}%
\pgfpathlineto{\pgfqpoint{2.755906in}{1.855951in}}%
\pgfusepath{stroke}%
\end{pgfscope}%
\begin{pgfscope}%
\pgfsetbuttcap%
\pgfsetroundjoin%
\definecolor{currentfill}{rgb}{0.000000,0.000000,0.000000}%
\pgfsetfillcolor{currentfill}%
\pgfsetlinewidth{0.803000pt}%
\definecolor{currentstroke}{rgb}{0.000000,0.000000,0.000000}%
\pgfsetstrokecolor{currentstroke}%
\pgfsetdash{}{0pt}%
\pgfsys@defobject{currentmarker}{\pgfqpoint{-0.048611in}{0.000000in}}{\pgfqpoint{-0.000000in}{0.000000in}}{%
\pgfpathmoveto{\pgfqpoint{-0.000000in}{0.000000in}}%
\pgfpathlineto{\pgfqpoint{-0.048611in}{0.000000in}}%
\pgfusepath{stroke,fill}%
}%
\begin{pgfscope}%
\pgfsys@transformshift{0.590551in}{1.855951in}%
\pgfsys@useobject{currentmarker}{}%
\end{pgfscope}%
\end{pgfscope}%
\begin{pgfscope}%
\definecolor{textcolor}{rgb}{0.000000,0.000000,0.000000}%
\pgfsetstrokecolor{textcolor}%
\pgfsetfillcolor{textcolor}%
\pgftext[x=0.342478in, y=1.817371in, left, base]{\color{textcolor}{\rmfamily\fontsize{8.330000}{9.996000}\selectfont\catcode`\^=\active\def^{\ifmmode\sp\else\^{}\fi}\catcode`\%=\active\def%{\%}$\mathdefault{0.4}$}}%
\end{pgfscope}%
\begin{pgfscope}%
\pgfpathrectangle{\pgfqpoint{0.590551in}{0.590551in}}{\pgfqpoint{2.165354in}{2.165354in}}%
\pgfusepath{clip}%
\pgfsetrectcap%
\pgfsetroundjoin%
\pgfsetlinewidth{0.803000pt}%
\definecolor{currentstroke}{rgb}{0.690196,0.690196,0.690196}%
\pgfsetstrokecolor{currentstroke}%
\pgfsetstrokeopacity{0.250000}%
\pgfsetdash{}{0pt}%
\pgfpathmoveto{\pgfqpoint{0.590551in}{2.147841in}}%
\pgfpathlineto{\pgfqpoint{2.755906in}{2.147841in}}%
\pgfusepath{stroke}%
\end{pgfscope}%
\begin{pgfscope}%
\pgfsetbuttcap%
\pgfsetroundjoin%
\definecolor{currentfill}{rgb}{0.000000,0.000000,0.000000}%
\pgfsetfillcolor{currentfill}%
\pgfsetlinewidth{0.803000pt}%
\definecolor{currentstroke}{rgb}{0.000000,0.000000,0.000000}%
\pgfsetstrokecolor{currentstroke}%
\pgfsetdash{}{0pt}%
\pgfsys@defobject{currentmarker}{\pgfqpoint{-0.048611in}{0.000000in}}{\pgfqpoint{-0.000000in}{0.000000in}}{%
\pgfpathmoveto{\pgfqpoint{-0.000000in}{0.000000in}}%
\pgfpathlineto{\pgfqpoint{-0.048611in}{0.000000in}}%
\pgfusepath{stroke,fill}%
}%
\begin{pgfscope}%
\pgfsys@transformshift{0.590551in}{2.147841in}%
\pgfsys@useobject{currentmarker}{}%
\end{pgfscope}%
\end{pgfscope}%
\begin{pgfscope}%
\definecolor{textcolor}{rgb}{0.000000,0.000000,0.000000}%
\pgfsetstrokecolor{textcolor}%
\pgfsetfillcolor{textcolor}%
\pgftext[x=0.342478in, y=2.109261in, left, base]{\color{textcolor}{\rmfamily\fontsize{8.330000}{9.996000}\selectfont\catcode`\^=\active\def^{\ifmmode\sp\else\^{}\fi}\catcode`\%=\active\def%{\%}$\mathdefault{0.5}$}}%
\end{pgfscope}%
\begin{pgfscope}%
\pgfpathrectangle{\pgfqpoint{0.590551in}{0.590551in}}{\pgfqpoint{2.165354in}{2.165354in}}%
\pgfusepath{clip}%
\pgfsetrectcap%
\pgfsetroundjoin%
\pgfsetlinewidth{0.803000pt}%
\definecolor{currentstroke}{rgb}{0.690196,0.690196,0.690196}%
\pgfsetstrokecolor{currentstroke}%
\pgfsetstrokeopacity{0.250000}%
\pgfsetdash{}{0pt}%
\pgfpathmoveto{\pgfqpoint{0.590551in}{2.439731in}}%
\pgfpathlineto{\pgfqpoint{2.755906in}{2.439731in}}%
\pgfusepath{stroke}%
\end{pgfscope}%
\begin{pgfscope}%
\pgfsetbuttcap%
\pgfsetroundjoin%
\definecolor{currentfill}{rgb}{0.000000,0.000000,0.000000}%
\pgfsetfillcolor{currentfill}%
\pgfsetlinewidth{0.803000pt}%
\definecolor{currentstroke}{rgb}{0.000000,0.000000,0.000000}%
\pgfsetstrokecolor{currentstroke}%
\pgfsetdash{}{0pt}%
\pgfsys@defobject{currentmarker}{\pgfqpoint{-0.048611in}{0.000000in}}{\pgfqpoint{-0.000000in}{0.000000in}}{%
\pgfpathmoveto{\pgfqpoint{-0.000000in}{0.000000in}}%
\pgfpathlineto{\pgfqpoint{-0.048611in}{0.000000in}}%
\pgfusepath{stroke,fill}%
}%
\begin{pgfscope}%
\pgfsys@transformshift{0.590551in}{2.439731in}%
\pgfsys@useobject{currentmarker}{}%
\end{pgfscope}%
\end{pgfscope}%
\begin{pgfscope}%
\definecolor{textcolor}{rgb}{0.000000,0.000000,0.000000}%
\pgfsetstrokecolor{textcolor}%
\pgfsetfillcolor{textcolor}%
\pgftext[x=0.342478in, y=2.401150in, left, base]{\color{textcolor}{\rmfamily\fontsize{8.330000}{9.996000}\selectfont\catcode`\^=\active\def^{\ifmmode\sp\else\^{}\fi}\catcode`\%=\active\def%{\%}$\mathdefault{0.6}$}}%
\end{pgfscope}%
\begin{pgfscope}%
\pgfpathrectangle{\pgfqpoint{0.590551in}{0.590551in}}{\pgfqpoint{2.165354in}{2.165354in}}%
\pgfusepath{clip}%
\pgfsetrectcap%
\pgfsetroundjoin%
\pgfsetlinewidth{0.803000pt}%
\definecolor{currentstroke}{rgb}{0.690196,0.690196,0.690196}%
\pgfsetstrokecolor{currentstroke}%
\pgfsetstrokeopacity{0.250000}%
\pgfsetdash{}{0pt}%
\pgfpathmoveto{\pgfqpoint{0.590551in}{2.731620in}}%
\pgfpathlineto{\pgfqpoint{2.755906in}{2.731620in}}%
\pgfusepath{stroke}%
\end{pgfscope}%
\begin{pgfscope}%
\pgfsetbuttcap%
\pgfsetroundjoin%
\definecolor{currentfill}{rgb}{0.000000,0.000000,0.000000}%
\pgfsetfillcolor{currentfill}%
\pgfsetlinewidth{0.803000pt}%
\definecolor{currentstroke}{rgb}{0.000000,0.000000,0.000000}%
\pgfsetstrokecolor{currentstroke}%
\pgfsetdash{}{0pt}%
\pgfsys@defobject{currentmarker}{\pgfqpoint{-0.048611in}{0.000000in}}{\pgfqpoint{-0.000000in}{0.000000in}}{%
\pgfpathmoveto{\pgfqpoint{-0.000000in}{0.000000in}}%
\pgfpathlineto{\pgfqpoint{-0.048611in}{0.000000in}}%
\pgfusepath{stroke,fill}%
}%
\begin{pgfscope}%
\pgfsys@transformshift{0.590551in}{2.731620in}%
\pgfsys@useobject{currentmarker}{}%
\end{pgfscope}%
\end{pgfscope}%
\begin{pgfscope}%
\definecolor{textcolor}{rgb}{0.000000,0.000000,0.000000}%
\pgfsetstrokecolor{textcolor}%
\pgfsetfillcolor{textcolor}%
\pgftext[x=0.342478in, y=2.693040in, left, base]{\color{textcolor}{\rmfamily\fontsize{8.330000}{9.996000}\selectfont\catcode`\^=\active\def^{\ifmmode\sp\else\^{}\fi}\catcode`\%=\active\def%{\%}$\mathdefault{0.7}$}}%
\end{pgfscope}%
\begin{pgfscope}%
\definecolor{textcolor}{rgb}{0.000000,0.000000,0.000000}%
\pgfsetstrokecolor{textcolor}%
\pgfsetfillcolor{textcolor}%
\pgftext[x=0.286922in,y=1.673228in,,bottom,rotate=90.000000]{\color{textcolor}{\rmfamily\fontsize{10.000000}{12.000000}\selectfont\catcode`\^=\active\def^{\ifmmode\sp\else\^{}\fi}\catcode`\%=\active\def%{\%}rank (normalized)}}%
\end{pgfscope}%
\begin{pgfscope}%
\pgfpathrectangle{\pgfqpoint{0.590551in}{0.590551in}}{\pgfqpoint{2.165354in}{2.165354in}}%
\pgfusepath{clip}%
\pgfsetrectcap%
\pgfsetroundjoin%
\pgfsetlinewidth{1.405250pt}%
\definecolor{currentstroke}{rgb}{0.470588,0.274510,0.686275}%
\pgfsetstrokecolor{currentstroke}%
\pgfsetdash{}{0pt}%
\pgfpathmoveto{\pgfqpoint{0.688976in}{2.178052in}}%
\pgfpathlineto{\pgfqpoint{0.707732in}{1.766633in}}%
\pgfpathlineto{\pgfqpoint{0.728315in}{1.599234in}}%
\pgfpathlineto{\pgfqpoint{0.750906in}{1.435922in}}%
\pgfpathlineto{\pgfqpoint{0.775699in}{1.289539in}}%
\pgfpathlineto{\pgfqpoint{0.802910in}{1.151330in}}%
\pgfpathlineto{\pgfqpoint{0.832774in}{1.050044in}}%
\pgfpathlineto{\pgfqpoint{0.865549in}{0.838716in}}%
\pgfpathlineto{\pgfqpoint{0.901520in}{0.886294in}}%
\pgfpathlineto{\pgfqpoint{0.940998in}{0.832878in}}%
\pgfpathlineto{\pgfqpoint{0.984325in}{0.725463in}}%
\pgfpathlineto{\pgfqpoint{1.031876in}{0.758446in}}%
\pgfpathlineto{\pgfqpoint{1.084064in}{0.720209in}}%
\pgfpathlineto{\pgfqpoint{1.141339in}{0.705906in}}%
\pgfpathlineto{\pgfqpoint{1.204199in}{0.712619in}}%
\pgfpathlineto{\pgfqpoint{1.273188in}{0.713203in}}%
\pgfpathlineto{\pgfqpoint{1.348903in}{0.704738in}}%
\pgfpathlineto{\pgfqpoint{1.432001in}{0.697441in}}%
\pgfpathlineto{\pgfqpoint{1.523200in}{0.698025in}}%
\pgfpathlineto{\pgfqpoint{1.623291in}{0.691311in}}%
\pgfpathlineto{\pgfqpoint{1.733141in}{0.693939in}}%
\pgfpathlineto{\pgfqpoint{1.853701in}{0.698025in}}%
\pgfpathlineto{\pgfqpoint{1.986015in}{0.699776in}}%
\pgfpathlineto{\pgfqpoint{2.131230in}{0.714079in}}%
\pgfpathlineto{\pgfqpoint{2.290603in}{1.095871in}}%
\pgfpathlineto{\pgfqpoint{2.465515in}{1.443365in}}%
\pgfpathlineto{\pgfqpoint{2.657480in}{1.756125in}}%
\pgfusepath{stroke}%
\end{pgfscope}%
\begin{pgfscope}%
\pgfsetrectcap%
\pgfsetmiterjoin%
\pgfsetlinewidth{0.803000pt}%
\definecolor{currentstroke}{rgb}{0.000000,0.000000,0.000000}%
\pgfsetstrokecolor{currentstroke}%
\pgfsetdash{}{0pt}%
\pgfpathmoveto{\pgfqpoint{0.590551in}{0.590551in}}%
\pgfpathlineto{\pgfqpoint{0.590551in}{2.755906in}}%
\pgfusepath{stroke}%
\end{pgfscope}%
\begin{pgfscope}%
\pgfsetrectcap%
\pgfsetmiterjoin%
\pgfsetlinewidth{0.803000pt}%
\definecolor{currentstroke}{rgb}{0.000000,0.000000,0.000000}%
\pgfsetstrokecolor{currentstroke}%
\pgfsetdash{}{0pt}%
\pgfpathmoveto{\pgfqpoint{2.755906in}{0.590551in}}%
\pgfpathlineto{\pgfqpoint{2.755906in}{2.755906in}}%
\pgfusepath{stroke}%
\end{pgfscope}%
\begin{pgfscope}%
\pgfsetrectcap%
\pgfsetmiterjoin%
\pgfsetlinewidth{0.803000pt}%
\definecolor{currentstroke}{rgb}{0.000000,0.000000,0.000000}%
\pgfsetstrokecolor{currentstroke}%
\pgfsetdash{}{0pt}%
\pgfpathmoveto{\pgfqpoint{0.590551in}{0.590551in}}%
\pgfpathlineto{\pgfqpoint{2.755906in}{0.590551in}}%
\pgfusepath{stroke}%
\end{pgfscope}%
\begin{pgfscope}%
\pgfsetrectcap%
\pgfsetmiterjoin%
\pgfsetlinewidth{0.803000pt}%
\definecolor{currentstroke}{rgb}{0.000000,0.000000,0.000000}%
\pgfsetstrokecolor{currentstroke}%
\pgfsetdash{}{0pt}%
\pgfpathmoveto{\pgfqpoint{0.590551in}{2.755906in}}%
\pgfpathlineto{\pgfqpoint{2.755906in}{2.755906in}}%
\pgfusepath{stroke}%
\end{pgfscope}%
\end{pgfpicture}%
\makeatother%
\endgroup%

%% file: figures/beta.pgf
%% Creator: Matplotlib, PGF backend
%%
%% To include the figure in your LaTeX document, write
%%   \input{<filename>.pgf}
%%
%% Make sure the required packages are loaded in your preamble
%%   \usepackage{pgf}
%%
%% Also ensure that all the required font packages are loaded; for instance,
%% the lmodern package is sometimes necessary when using math font.
%%   \usepackage{lmodern}
%%
%% Figures using additional raster images can only be included by \input if
%% they are in the same directory as the main LaTeX file. For loading figures
%% from other directories you can use the `import` package
%%   \usepackage{import}
%%
%% and then include the figures with
%%   \import{<path to file>}{<filename>.pgf}
%%
%% Matplotlib used the following preamble
%%   \def\mathdefault#1{#1}
%%   \everymath=\expandafter{\the\everymath\displaystyle}
%%   \IfFileExists{scrextend.sty}{
%%     \usepackage[fontsize=10.000000pt]{scrextend}
%%   }{
%%     \renewcommand{\normalsize}{\fontsize{10.000000}{12.000000}\selectfont}
%%     \normalsize
%%   }
%%   
%%   \makeatletter\@ifpackageloaded{underscore}{}{\usepackage[strings]{underscore}}\makeatother
%%
\begingroup%
\makeatletter%
\begin{pgfpicture}%
\pgfpathrectangle{\pgfpointorigin}{\pgfqpoint{2.952756in}{2.952756in}}%
\pgfusepath{use as bounding box, clip}%
\begin{pgfscope}%
\pgfsetbuttcap%
\pgfsetmiterjoin%
\pgfsetlinewidth{0.000000pt}%
\definecolor{currentstroke}{rgb}{1.000000,1.000000,1.000000}%
\pgfsetstrokecolor{currentstroke}%
\pgfsetdash{}{0pt}%
\pgfpathmoveto{\pgfqpoint{0.000000in}{0.000000in}}%
\pgfpathlineto{\pgfqpoint{2.952756in}{0.000000in}}%
\pgfpathlineto{\pgfqpoint{2.952756in}{2.952756in}}%
\pgfpathlineto{\pgfqpoint{0.000000in}{2.952756in}}%
\pgfpathlineto{\pgfqpoint{0.000000in}{0.000000in}}%
\pgfpathclose%
\pgfusepath{}%
\end{pgfscope}%
\begin{pgfscope}%
\pgfsetbuttcap%
\pgfsetmiterjoin%
\pgfsetlinewidth{0.000000pt}%
\definecolor{currentstroke}{rgb}{0.000000,0.000000,0.000000}%
\pgfsetstrokecolor{currentstroke}%
\pgfsetstrokeopacity{0.000000}%
\pgfsetdash{}{0pt}%
\pgfpathmoveto{\pgfqpoint{0.590551in}{0.590551in}}%
\pgfpathlineto{\pgfqpoint{2.755906in}{0.590551in}}%
\pgfpathlineto{\pgfqpoint{2.755906in}{2.755906in}}%
\pgfpathlineto{\pgfqpoint{0.590551in}{2.755906in}}%
\pgfpathlineto{\pgfqpoint{0.590551in}{0.590551in}}%
\pgfpathclose%
\pgfusepath{}%
\end{pgfscope}%
\begin{pgfscope}%
\pgfpathrectangle{\pgfqpoint{0.590551in}{0.590551in}}{\pgfqpoint{2.165354in}{2.165354in}}%
\pgfusepath{clip}%
\pgfsetbuttcap%
\pgfsetroundjoin%
\definecolor{currentfill}{rgb}{0.196078,0.431373,0.745098}%
\pgfsetfillcolor{currentfill}%
\pgfsetfillopacity{0.220000}%
\pgfsetlinewidth{1.003750pt}%
\definecolor{currentstroke}{rgb}{0.196078,0.431373,0.745098}%
\pgfsetstrokecolor{currentstroke}%
\pgfsetstrokeopacity{0.220000}%
\pgfsetdash{}{0pt}%
\pgfsys@defobject{currentmarker}{\pgfqpoint{0.688976in}{0.688976in}}{\pgfqpoint{2.657480in}{2.657480in}}{%
\pgfpathmoveto{\pgfqpoint{0.688976in}{1.599531in}}%
\pgfpathlineto{\pgfqpoint{0.688976in}{1.599531in}}%
\pgfpathlineto{\pgfqpoint{0.709270in}{1.534075in}}%
\pgfpathlineto{\pgfqpoint{0.729564in}{0.822796in}}%
\pgfpathlineto{\pgfqpoint{0.749858in}{1.811169in}}%
\pgfpathlineto{\pgfqpoint{0.770152in}{1.489469in}}%
\pgfpathlineto{\pgfqpoint{0.790446in}{1.476863in}}%
\pgfpathlineto{\pgfqpoint{0.810740in}{1.137950in}}%
\pgfpathlineto{\pgfqpoint{0.831033in}{1.176739in}}%
\pgfpathlineto{\pgfqpoint{0.851327in}{1.602197in}}%
\pgfpathlineto{\pgfqpoint{0.871621in}{1.174072in}}%
\pgfpathlineto{\pgfqpoint{0.891915in}{1.520015in}}%
\pgfpathlineto{\pgfqpoint{0.912209in}{1.219406in}}%
\pgfpathlineto{\pgfqpoint{0.932503in}{1.025949in}}%
\pgfpathlineto{\pgfqpoint{0.952796in}{0.953464in}}%
\pgfpathlineto{\pgfqpoint{0.973090in}{0.814069in}}%
\pgfpathlineto{\pgfqpoint{0.993384in}{0.995888in}}%
\pgfpathlineto{\pgfqpoint{1.013678in}{0.871281in}}%
\pgfpathlineto{\pgfqpoint{1.033972in}{0.736250in}}%
\pgfpathlineto{\pgfqpoint{1.054266in}{0.724856in}}%
\pgfpathlineto{\pgfqpoint{1.074560in}{0.718552in}}%
\pgfpathlineto{\pgfqpoint{1.094853in}{0.713219in}}%
\pgfpathlineto{\pgfqpoint{1.115147in}{0.707886in}}%
\pgfpathlineto{\pgfqpoint{1.135441in}{0.712249in}}%
\pgfpathlineto{\pgfqpoint{1.155735in}{0.701340in}}%
\pgfpathlineto{\pgfqpoint{1.176029in}{0.709340in}}%
\pgfpathlineto{\pgfqpoint{1.196323in}{0.702795in}}%
\pgfpathlineto{\pgfqpoint{1.216617in}{0.698673in}}%
\pgfpathlineto{\pgfqpoint{1.236910in}{0.693825in}}%
\pgfpathlineto{\pgfqpoint{1.257204in}{0.712249in}}%
\pgfpathlineto{\pgfqpoint{1.277498in}{0.701583in}}%
\pgfpathlineto{\pgfqpoint{1.297792in}{0.695764in}}%
\pgfpathlineto{\pgfqpoint{1.318086in}{0.692855in}}%
\pgfpathlineto{\pgfqpoint{1.338380in}{0.696007in}}%
\pgfpathlineto{\pgfqpoint{1.358674in}{0.704007in}}%
\pgfpathlineto{\pgfqpoint{1.378967in}{0.695037in}}%
\pgfpathlineto{\pgfqpoint{1.399261in}{0.691158in}}%
\pgfpathlineto{\pgfqpoint{1.419555in}{0.696007in}}%
\pgfpathlineto{\pgfqpoint{1.439849in}{0.710310in}}%
\pgfpathlineto{\pgfqpoint{1.460143in}{0.700855in}}%
\pgfpathlineto{\pgfqpoint{1.480437in}{0.705704in}}%
\pgfpathlineto{\pgfqpoint{1.500731in}{0.696249in}}%
\pgfpathlineto{\pgfqpoint{1.521024in}{0.705461in}}%
\pgfpathlineto{\pgfqpoint{1.541318in}{0.702310in}}%
\pgfpathlineto{\pgfqpoint{1.561612in}{0.695764in}}%
\pgfpathlineto{\pgfqpoint{1.581906in}{0.703037in}}%
\pgfpathlineto{\pgfqpoint{1.602200in}{0.696734in}}%
\pgfpathlineto{\pgfqpoint{1.622494in}{0.692855in}}%
\pgfpathlineto{\pgfqpoint{1.642788in}{0.695522in}}%
\pgfpathlineto{\pgfqpoint{1.663081in}{0.706674in}}%
\pgfpathlineto{\pgfqpoint{1.683375in}{0.702795in}}%
\pgfpathlineto{\pgfqpoint{1.703669in}{0.696492in}}%
\pgfpathlineto{\pgfqpoint{1.723963in}{0.699886in}}%
\pgfpathlineto{\pgfqpoint{1.744257in}{0.691885in}}%
\pgfpathlineto{\pgfqpoint{1.764551in}{0.691885in}}%
\pgfpathlineto{\pgfqpoint{1.784845in}{0.694310in}}%
\pgfpathlineto{\pgfqpoint{1.805138in}{0.700370in}}%
\pgfpathlineto{\pgfqpoint{1.825432in}{0.693825in}}%
\pgfpathlineto{\pgfqpoint{1.845726in}{0.696734in}}%
\pgfpathlineto{\pgfqpoint{1.866020in}{0.694795in}}%
\pgfpathlineto{\pgfqpoint{1.886314in}{0.698673in}}%
\pgfpathlineto{\pgfqpoint{1.906608in}{0.693825in}}%
\pgfpathlineto{\pgfqpoint{1.926902in}{0.696249in}}%
\pgfpathlineto{\pgfqpoint{1.947195in}{0.697704in}}%
\pgfpathlineto{\pgfqpoint{1.967489in}{0.692855in}}%
\pgfpathlineto{\pgfqpoint{1.987783in}{0.692855in}}%
\pgfpathlineto{\pgfqpoint{2.008077in}{0.691885in}}%
\pgfpathlineto{\pgfqpoint{2.028371in}{0.689946in}}%
\pgfpathlineto{\pgfqpoint{2.048665in}{0.693825in}}%
\pgfpathlineto{\pgfqpoint{2.068959in}{0.692855in}}%
\pgfpathlineto{\pgfqpoint{2.089252in}{0.691643in}}%
\pgfpathlineto{\pgfqpoint{2.109546in}{0.690916in}}%
\pgfpathlineto{\pgfqpoint{2.129840in}{0.690916in}}%
\pgfpathlineto{\pgfqpoint{2.150134in}{0.689946in}}%
\pgfpathlineto{\pgfqpoint{2.170428in}{0.690916in}}%
\pgfpathlineto{\pgfqpoint{2.190722in}{0.689946in}}%
\pgfpathlineto{\pgfqpoint{2.211016in}{0.691885in}}%
\pgfpathlineto{\pgfqpoint{2.231309in}{0.689946in}}%
\pgfpathlineto{\pgfqpoint{2.251603in}{0.690916in}}%
\pgfpathlineto{\pgfqpoint{2.271897in}{0.689946in}}%
\pgfpathlineto{\pgfqpoint{2.292191in}{0.689946in}}%
\pgfpathlineto{\pgfqpoint{2.312485in}{0.689946in}}%
\pgfpathlineto{\pgfqpoint{2.332779in}{0.688976in}}%
\pgfpathlineto{\pgfqpoint{2.353072in}{0.689946in}}%
\pgfpathlineto{\pgfqpoint{2.373366in}{0.688976in}}%
\pgfpathlineto{\pgfqpoint{2.393660in}{0.688976in}}%
\pgfpathlineto{\pgfqpoint{2.413954in}{0.688976in}}%
\pgfpathlineto{\pgfqpoint{2.434248in}{0.688976in}}%
\pgfpathlineto{\pgfqpoint{2.454542in}{0.688976in}}%
\pgfpathlineto{\pgfqpoint{2.474836in}{0.688976in}}%
\pgfpathlineto{\pgfqpoint{2.495129in}{0.688976in}}%
\pgfpathlineto{\pgfqpoint{2.515423in}{0.688976in}}%
\pgfpathlineto{\pgfqpoint{2.535717in}{0.688976in}}%
\pgfpathlineto{\pgfqpoint{2.556011in}{0.688976in}}%
\pgfpathlineto{\pgfqpoint{2.576305in}{0.688976in}}%
\pgfpathlineto{\pgfqpoint{2.596599in}{0.688976in}}%
\pgfpathlineto{\pgfqpoint{2.616893in}{0.688976in}}%
\pgfpathlineto{\pgfqpoint{2.637186in}{0.688976in}}%
\pgfpathlineto{\pgfqpoint{2.657480in}{0.688976in}}%
\pgfpathlineto{\pgfqpoint{2.657480in}{0.689219in}}%
\pgfpathlineto{\pgfqpoint{2.657480in}{0.689219in}}%
\pgfpathlineto{\pgfqpoint{2.637186in}{0.689946in}}%
\pgfpathlineto{\pgfqpoint{2.616893in}{0.689946in}}%
\pgfpathlineto{\pgfqpoint{2.596599in}{0.689946in}}%
\pgfpathlineto{\pgfqpoint{2.576305in}{0.689946in}}%
\pgfpathlineto{\pgfqpoint{2.556011in}{0.691885in}}%
\pgfpathlineto{\pgfqpoint{2.535717in}{0.694067in}}%
\pgfpathlineto{\pgfqpoint{2.515423in}{0.694067in}}%
\pgfpathlineto{\pgfqpoint{2.495129in}{0.696734in}}%
\pgfpathlineto{\pgfqpoint{2.474836in}{0.693825in}}%
\pgfpathlineto{\pgfqpoint{2.454542in}{0.698916in}}%
\pgfpathlineto{\pgfqpoint{2.434248in}{0.698431in}}%
\pgfpathlineto{\pgfqpoint{2.413954in}{0.717825in}}%
\pgfpathlineto{\pgfqpoint{2.393660in}{0.708613in}}%
\pgfpathlineto{\pgfqpoint{2.373366in}{0.714916in}}%
\pgfpathlineto{\pgfqpoint{2.353072in}{0.745462in}}%
\pgfpathlineto{\pgfqpoint{2.332779in}{0.733583in}}%
\pgfpathlineto{\pgfqpoint{2.312485in}{0.736007in}}%
\pgfpathlineto{\pgfqpoint{2.292191in}{0.722189in}}%
\pgfpathlineto{\pgfqpoint{2.271897in}{0.791523in}}%
\pgfpathlineto{\pgfqpoint{2.251603in}{0.760977in}}%
\pgfpathlineto{\pgfqpoint{2.231309in}{0.747159in}}%
\pgfpathlineto{\pgfqpoint{2.211016in}{0.799765in}}%
\pgfpathlineto{\pgfqpoint{2.190722in}{0.784492in}}%
\pgfpathlineto{\pgfqpoint{2.170428in}{0.816493in}}%
\pgfpathlineto{\pgfqpoint{2.150134in}{0.815281in}}%
\pgfpathlineto{\pgfqpoint{2.129840in}{0.745219in}}%
\pgfpathlineto{\pgfqpoint{2.109546in}{0.877827in}}%
\pgfpathlineto{\pgfqpoint{2.089252in}{0.812614in}}%
\pgfpathlineto{\pgfqpoint{2.068959in}{0.911039in}}%
\pgfpathlineto{\pgfqpoint{2.048665in}{1.010676in}}%
\pgfpathlineto{\pgfqpoint{2.028371in}{0.880493in}}%
\pgfpathlineto{\pgfqpoint{2.008077in}{0.888251in}}%
\pgfpathlineto{\pgfqpoint{1.987783in}{0.854311in}}%
\pgfpathlineto{\pgfqpoint{1.967489in}{0.886069in}}%
\pgfpathlineto{\pgfqpoint{1.947195in}{0.986919in}}%
\pgfpathlineto{\pgfqpoint{1.926902in}{0.989585in}}%
\pgfpathlineto{\pgfqpoint{1.906608in}{1.123647in}}%
\pgfpathlineto{\pgfqpoint{1.886314in}{1.050677in}}%
\pgfpathlineto{\pgfqpoint{1.866020in}{1.041707in}}%
\pgfpathlineto{\pgfqpoint{1.845726in}{1.111768in}}%
\pgfpathlineto{\pgfqpoint{1.825432in}{1.178921in}}%
\pgfpathlineto{\pgfqpoint{1.805138in}{0.998555in}}%
\pgfpathlineto{\pgfqpoint{1.784845in}{1.017707in}}%
\pgfpathlineto{\pgfqpoint{1.764551in}{1.127284in}}%
\pgfpathlineto{\pgfqpoint{1.744257in}{0.876615in}}%
\pgfpathlineto{\pgfqpoint{1.723963in}{1.165345in}}%
\pgfpathlineto{\pgfqpoint{1.703669in}{1.031768in}}%
\pgfpathlineto{\pgfqpoint{1.683375in}{1.242436in}}%
\pgfpathlineto{\pgfqpoint{1.663081in}{1.208497in}}%
\pgfpathlineto{\pgfqpoint{1.642788in}{1.108132in}}%
\pgfpathlineto{\pgfqpoint{1.622494in}{1.141587in}}%
\pgfpathlineto{\pgfqpoint{1.602200in}{1.006070in}}%
\pgfpathlineto{\pgfqpoint{1.581906in}{1.556864in}}%
\pgfpathlineto{\pgfqpoint{1.561612in}{1.091889in}}%
\pgfpathlineto{\pgfqpoint{1.541318in}{1.315892in}}%
\pgfpathlineto{\pgfqpoint{1.521024in}{1.632501in}}%
\pgfpathlineto{\pgfqpoint{1.500731in}{1.106435in}}%
\pgfpathlineto{\pgfqpoint{1.480437in}{1.462075in}}%
\pgfpathlineto{\pgfqpoint{1.460143in}{1.231285in}}%
\pgfpathlineto{\pgfqpoint{1.439849in}{1.167769in}}%
\pgfpathlineto{\pgfqpoint{1.419555in}{1.008979in}}%
\pgfpathlineto{\pgfqpoint{1.399261in}{1.465954in}}%
\pgfpathlineto{\pgfqpoint{1.378967in}{1.006313in}}%
\pgfpathlineto{\pgfqpoint{1.358674in}{1.353225in}}%
\pgfpathlineto{\pgfqpoint{1.338380in}{1.030313in}}%
\pgfpathlineto{\pgfqpoint{1.318086in}{1.168011in}}%
\pgfpathlineto{\pgfqpoint{1.297792in}{1.106920in}}%
\pgfpathlineto{\pgfqpoint{1.277498in}{1.187405in}}%
\pgfpathlineto{\pgfqpoint{1.257204in}{1.457226in}}%
\pgfpathlineto{\pgfqpoint{1.236910in}{1.113223in}}%
\pgfpathlineto{\pgfqpoint{1.216617in}{1.228618in}}%
\pgfpathlineto{\pgfqpoint{1.196323in}{1.271043in}}%
\pgfpathlineto{\pgfqpoint{1.176029in}{1.235891in}}%
\pgfpathlineto{\pgfqpoint{1.155735in}{1.865230in}}%
\pgfpathlineto{\pgfqpoint{1.135441in}{1.094314in}}%
\pgfpathlineto{\pgfqpoint{1.115147in}{1.168496in}}%
\pgfpathlineto{\pgfqpoint{1.094853in}{1.802199in}}%
\pgfpathlineto{\pgfqpoint{1.074560in}{1.440984in}}%
\pgfpathlineto{\pgfqpoint{1.054266in}{1.779896in}}%
\pgfpathlineto{\pgfqpoint{1.033972in}{1.706198in}}%
\pgfpathlineto{\pgfqpoint{1.013678in}{2.017474in}}%
\pgfpathlineto{\pgfqpoint{0.993384in}{2.123172in}}%
\pgfpathlineto{\pgfqpoint{0.973090in}{2.213355in}}%
\pgfpathlineto{\pgfqpoint{0.952796in}{2.036626in}}%
\pgfpathlineto{\pgfqpoint{0.932503in}{2.316629in}}%
\pgfpathlineto{\pgfqpoint{0.912209in}{2.278083in}}%
\pgfpathlineto{\pgfqpoint{0.891915in}{2.657480in}}%
\pgfpathlineto{\pgfqpoint{0.871621in}{2.455054in}}%
\pgfpathlineto{\pgfqpoint{0.851327in}{2.209234in}}%
\pgfpathlineto{\pgfqpoint{0.831033in}{2.070808in}}%
\pgfpathlineto{\pgfqpoint{0.810740in}{2.108384in}}%
\pgfpathlineto{\pgfqpoint{0.790446in}{2.077111in}}%
\pgfpathlineto{\pgfqpoint{0.770152in}{2.456509in}}%
\pgfpathlineto{\pgfqpoint{0.749858in}{2.264264in}}%
\pgfpathlineto{\pgfqpoint{0.729564in}{2.067899in}}%
\pgfpathlineto{\pgfqpoint{0.709270in}{1.629592in}}%
\pgfpathlineto{\pgfqpoint{0.688976in}{1.599531in}}%
\pgfpathlineto{\pgfqpoint{0.688976in}{1.599531in}}%
\pgfpathclose%
\pgfusepath{stroke,fill}%
}%
\begin{pgfscope}%
\pgfsys@transformshift{0.000000in}{0.000000in}%
\pgfsys@useobject{currentmarker}{}%
\end{pgfscope}%
\end{pgfscope}%
\begin{pgfscope}%
\pgfpathrectangle{\pgfqpoint{0.590551in}{0.590551in}}{\pgfqpoint{2.165354in}{2.165354in}}%
\pgfusepath{clip}%
\pgfsetrectcap%
\pgfsetroundjoin%
\pgfsetlinewidth{0.803000pt}%
\definecolor{currentstroke}{rgb}{0.690196,0.690196,0.690196}%
\pgfsetstrokecolor{currentstroke}%
\pgfsetstrokeopacity{0.250000}%
\pgfsetdash{}{0pt}%
\pgfpathmoveto{\pgfqpoint{0.628095in}{0.590551in}}%
\pgfpathlineto{\pgfqpoint{0.628095in}{2.755906in}}%
\pgfusepath{stroke}%
\end{pgfscope}%
\begin{pgfscope}%
\pgfsetbuttcap%
\pgfsetroundjoin%
\definecolor{currentfill}{rgb}{0.000000,0.000000,0.000000}%
\pgfsetfillcolor{currentfill}%
\pgfsetlinewidth{0.803000pt}%
\definecolor{currentstroke}{rgb}{0.000000,0.000000,0.000000}%
\pgfsetstrokecolor{currentstroke}%
\pgfsetdash{}{0pt}%
\pgfsys@defobject{currentmarker}{\pgfqpoint{0.000000in}{-0.048611in}}{\pgfqpoint{0.000000in}{0.000000in}}{%
\pgfpathmoveto{\pgfqpoint{0.000000in}{0.000000in}}%
\pgfpathlineto{\pgfqpoint{0.000000in}{-0.048611in}}%
\pgfusepath{stroke,fill}%
}%
\begin{pgfscope}%
\pgfsys@transformshift{0.628095in}{0.590551in}%
\pgfsys@useobject{currentmarker}{}%
\end{pgfscope}%
\end{pgfscope}%
\begin{pgfscope}%
\definecolor{textcolor}{rgb}{0.000000,0.000000,0.000000}%
\pgfsetstrokecolor{textcolor}%
\pgfsetfillcolor{textcolor}%
\pgftext[x=0.628095in,y=0.493329in,,top]{\color{textcolor}{\rmfamily\fontsize{8.330000}{9.996000}\selectfont\catcode`\^=\active\def^{\ifmmode\sp\else\^{}\fi}\catcode`\%=\active\def%{\%}$\mathdefault{0.00}$}}%
\end{pgfscope}%
\begin{pgfscope}%
\pgfpathrectangle{\pgfqpoint{0.590551in}{0.590551in}}{\pgfqpoint{2.165354in}{2.165354in}}%
\pgfusepath{clip}%
\pgfsetrectcap%
\pgfsetroundjoin%
\pgfsetlinewidth{0.803000pt}%
\definecolor{currentstroke}{rgb}{0.690196,0.690196,0.690196}%
\pgfsetstrokecolor{currentstroke}%
\pgfsetstrokeopacity{0.250000}%
\pgfsetdash{}{0pt}%
\pgfpathmoveto{\pgfqpoint{1.135441in}{0.590551in}}%
\pgfpathlineto{\pgfqpoint{1.135441in}{2.755906in}}%
\pgfusepath{stroke}%
\end{pgfscope}%
\begin{pgfscope}%
\pgfsetbuttcap%
\pgfsetroundjoin%
\definecolor{currentfill}{rgb}{0.000000,0.000000,0.000000}%
\pgfsetfillcolor{currentfill}%
\pgfsetlinewidth{0.803000pt}%
\definecolor{currentstroke}{rgb}{0.000000,0.000000,0.000000}%
\pgfsetstrokecolor{currentstroke}%
\pgfsetdash{}{0pt}%
\pgfsys@defobject{currentmarker}{\pgfqpoint{0.000000in}{-0.048611in}}{\pgfqpoint{0.000000in}{0.000000in}}{%
\pgfpathmoveto{\pgfqpoint{0.000000in}{0.000000in}}%
\pgfpathlineto{\pgfqpoint{0.000000in}{-0.048611in}}%
\pgfusepath{stroke,fill}%
}%
\begin{pgfscope}%
\pgfsys@transformshift{1.135441in}{0.590551in}%
\pgfsys@useobject{currentmarker}{}%
\end{pgfscope}%
\end{pgfscope}%
\begin{pgfscope}%
\definecolor{textcolor}{rgb}{0.000000,0.000000,0.000000}%
\pgfsetstrokecolor{textcolor}%
\pgfsetfillcolor{textcolor}%
\pgftext[x=1.135441in,y=0.493329in,,top]{\color{textcolor}{\rmfamily\fontsize{8.330000}{9.996000}\selectfont\catcode`\^=\active\def^{\ifmmode\sp\else\^{}\fi}\catcode`\%=\active\def%{\%}$\mathdefault{0.25}$}}%
\end{pgfscope}%
\begin{pgfscope}%
\pgfpathrectangle{\pgfqpoint{0.590551in}{0.590551in}}{\pgfqpoint{2.165354in}{2.165354in}}%
\pgfusepath{clip}%
\pgfsetrectcap%
\pgfsetroundjoin%
\pgfsetlinewidth{0.803000pt}%
\definecolor{currentstroke}{rgb}{0.690196,0.690196,0.690196}%
\pgfsetstrokecolor{currentstroke}%
\pgfsetstrokeopacity{0.250000}%
\pgfsetdash{}{0pt}%
\pgfpathmoveto{\pgfqpoint{1.642788in}{0.590551in}}%
\pgfpathlineto{\pgfqpoint{1.642788in}{2.755906in}}%
\pgfusepath{stroke}%
\end{pgfscope}%
\begin{pgfscope}%
\pgfsetbuttcap%
\pgfsetroundjoin%
\definecolor{currentfill}{rgb}{0.000000,0.000000,0.000000}%
\pgfsetfillcolor{currentfill}%
\pgfsetlinewidth{0.803000pt}%
\definecolor{currentstroke}{rgb}{0.000000,0.000000,0.000000}%
\pgfsetstrokecolor{currentstroke}%
\pgfsetdash{}{0pt}%
\pgfsys@defobject{currentmarker}{\pgfqpoint{0.000000in}{-0.048611in}}{\pgfqpoint{0.000000in}{0.000000in}}{%
\pgfpathmoveto{\pgfqpoint{0.000000in}{0.000000in}}%
\pgfpathlineto{\pgfqpoint{0.000000in}{-0.048611in}}%
\pgfusepath{stroke,fill}%
}%
\begin{pgfscope}%
\pgfsys@transformshift{1.642788in}{0.590551in}%
\pgfsys@useobject{currentmarker}{}%
\end{pgfscope}%
\end{pgfscope}%
\begin{pgfscope}%
\definecolor{textcolor}{rgb}{0.000000,0.000000,0.000000}%
\pgfsetstrokecolor{textcolor}%
\pgfsetfillcolor{textcolor}%
\pgftext[x=1.642788in,y=0.493329in,,top]{\color{textcolor}{\rmfamily\fontsize{8.330000}{9.996000}\selectfont\catcode`\^=\active\def^{\ifmmode\sp\else\^{}\fi}\catcode`\%=\active\def%{\%}$\mathdefault{0.50}$}}%
\end{pgfscope}%
\begin{pgfscope}%
\pgfpathrectangle{\pgfqpoint{0.590551in}{0.590551in}}{\pgfqpoint{2.165354in}{2.165354in}}%
\pgfusepath{clip}%
\pgfsetrectcap%
\pgfsetroundjoin%
\pgfsetlinewidth{0.803000pt}%
\definecolor{currentstroke}{rgb}{0.690196,0.690196,0.690196}%
\pgfsetstrokecolor{currentstroke}%
\pgfsetstrokeopacity{0.250000}%
\pgfsetdash{}{0pt}%
\pgfpathmoveto{\pgfqpoint{2.150134in}{0.590551in}}%
\pgfpathlineto{\pgfqpoint{2.150134in}{2.755906in}}%
\pgfusepath{stroke}%
\end{pgfscope}%
\begin{pgfscope}%
\pgfsetbuttcap%
\pgfsetroundjoin%
\definecolor{currentfill}{rgb}{0.000000,0.000000,0.000000}%
\pgfsetfillcolor{currentfill}%
\pgfsetlinewidth{0.803000pt}%
\definecolor{currentstroke}{rgb}{0.000000,0.000000,0.000000}%
\pgfsetstrokecolor{currentstroke}%
\pgfsetdash{}{0pt}%
\pgfsys@defobject{currentmarker}{\pgfqpoint{0.000000in}{-0.048611in}}{\pgfqpoint{0.000000in}{0.000000in}}{%
\pgfpathmoveto{\pgfqpoint{0.000000in}{0.000000in}}%
\pgfpathlineto{\pgfqpoint{0.000000in}{-0.048611in}}%
\pgfusepath{stroke,fill}%
}%
\begin{pgfscope}%
\pgfsys@transformshift{2.150134in}{0.590551in}%
\pgfsys@useobject{currentmarker}{}%
\end{pgfscope}%
\end{pgfscope}%
\begin{pgfscope}%
\definecolor{textcolor}{rgb}{0.000000,0.000000,0.000000}%
\pgfsetstrokecolor{textcolor}%
\pgfsetfillcolor{textcolor}%
\pgftext[x=2.150134in,y=0.493329in,,top]{\color{textcolor}{\rmfamily\fontsize{8.330000}{9.996000}\selectfont\catcode`\^=\active\def^{\ifmmode\sp\else\^{}\fi}\catcode`\%=\active\def%{\%}$\mathdefault{0.75}$}}%
\end{pgfscope}%
\begin{pgfscope}%
\pgfpathrectangle{\pgfqpoint{0.590551in}{0.590551in}}{\pgfqpoint{2.165354in}{2.165354in}}%
\pgfusepath{clip}%
\pgfsetrectcap%
\pgfsetroundjoin%
\pgfsetlinewidth{0.803000pt}%
\definecolor{currentstroke}{rgb}{0.690196,0.690196,0.690196}%
\pgfsetstrokecolor{currentstroke}%
\pgfsetstrokeopacity{0.250000}%
\pgfsetdash{}{0pt}%
\pgfpathmoveto{\pgfqpoint{2.657480in}{0.590551in}}%
\pgfpathlineto{\pgfqpoint{2.657480in}{2.755906in}}%
\pgfusepath{stroke}%
\end{pgfscope}%
\begin{pgfscope}%
\pgfsetbuttcap%
\pgfsetroundjoin%
\definecolor{currentfill}{rgb}{0.000000,0.000000,0.000000}%
\pgfsetfillcolor{currentfill}%
\pgfsetlinewidth{0.803000pt}%
\definecolor{currentstroke}{rgb}{0.000000,0.000000,0.000000}%
\pgfsetstrokecolor{currentstroke}%
\pgfsetdash{}{0pt}%
\pgfsys@defobject{currentmarker}{\pgfqpoint{0.000000in}{-0.048611in}}{\pgfqpoint{0.000000in}{0.000000in}}{%
\pgfpathmoveto{\pgfqpoint{0.000000in}{0.000000in}}%
\pgfpathlineto{\pgfqpoint{0.000000in}{-0.048611in}}%
\pgfusepath{stroke,fill}%
}%
\begin{pgfscope}%
\pgfsys@transformshift{2.657480in}{0.590551in}%
\pgfsys@useobject{currentmarker}{}%
\end{pgfscope}%
\end{pgfscope}%
\begin{pgfscope}%
\definecolor{textcolor}{rgb}{0.000000,0.000000,0.000000}%
\pgfsetstrokecolor{textcolor}%
\pgfsetfillcolor{textcolor}%
\pgftext[x=2.657480in,y=0.493329in,,top]{\color{textcolor}{\rmfamily\fontsize{8.330000}{9.996000}\selectfont\catcode`\^=\active\def^{\ifmmode\sp\else\^{}\fi}\catcode`\%=\active\def%{\%}$\mathdefault{1.00}$}}%
\end{pgfscope}%
\begin{pgfscope}%
\definecolor{textcolor}{rgb}{0.000000,0.000000,0.000000}%
\pgfsetstrokecolor{textcolor}%
\pgfsetfillcolor{textcolor}%
\pgftext[x=1.673228in,y=0.314317in,,top]{\color{textcolor}{\rmfamily\fontsize{10.000000}{12.000000}\selectfont\catcode`\^=\active\def^{\ifmmode\sp\else\^{}\fi}\catcode`\%=\active\def%{\%}infection probability $\beta$}}%
\end{pgfscope}%
\begin{pgfscope}%
\pgfpathrectangle{\pgfqpoint{0.590551in}{0.590551in}}{\pgfqpoint{2.165354in}{2.165354in}}%
\pgfusepath{clip}%
\pgfsetrectcap%
\pgfsetroundjoin%
\pgfsetlinewidth{0.803000pt}%
\definecolor{currentstroke}{rgb}{0.690196,0.690196,0.690196}%
\pgfsetstrokecolor{currentstroke}%
\pgfsetstrokeopacity{0.250000}%
\pgfsetdash{}{0pt}%
\pgfpathmoveto{\pgfqpoint{0.590551in}{0.688007in}}%
\pgfpathlineto{\pgfqpoint{2.755906in}{0.688007in}}%
\pgfusepath{stroke}%
\end{pgfscope}%
\begin{pgfscope}%
\pgfsetbuttcap%
\pgfsetroundjoin%
\definecolor{currentfill}{rgb}{0.000000,0.000000,0.000000}%
\pgfsetfillcolor{currentfill}%
\pgfsetlinewidth{0.803000pt}%
\definecolor{currentstroke}{rgb}{0.000000,0.000000,0.000000}%
\pgfsetstrokecolor{currentstroke}%
\pgfsetdash{}{0pt}%
\pgfsys@defobject{currentmarker}{\pgfqpoint{-0.048611in}{0.000000in}}{\pgfqpoint{-0.000000in}{0.000000in}}{%
\pgfpathmoveto{\pgfqpoint{-0.000000in}{0.000000in}}%
\pgfpathlineto{\pgfqpoint{-0.048611in}{0.000000in}}%
\pgfusepath{stroke,fill}%
}%
\begin{pgfscope}%
\pgfsys@transformshift{0.590551in}{0.688007in}%
\pgfsys@useobject{currentmarker}{}%
\end{pgfscope}%
\end{pgfscope}%
\begin{pgfscope}%
\definecolor{textcolor}{rgb}{0.000000,0.000000,0.000000}%
\pgfsetstrokecolor{textcolor}%
\pgfsetfillcolor{textcolor}%
\pgftext[x=0.246415in, y=0.639781in, left, base]{\color{textcolor}{\rmfamily\fontsize{10.000000}{12.000000}\selectfont\catcode`\^=\active\def^{\ifmmode\sp\else\^{}\fi}\catcode`\%=\active\def%{\%}$\mathdefault{0.00}$}}%
\end{pgfscope}%
\begin{pgfscope}%
\pgfpathrectangle{\pgfqpoint{0.590551in}{0.590551in}}{\pgfqpoint{2.165354in}{2.165354in}}%
\pgfusepath{clip}%
\pgfsetrectcap%
\pgfsetroundjoin%
\pgfsetlinewidth{0.803000pt}%
\definecolor{currentstroke}{rgb}{0.690196,0.690196,0.690196}%
\pgfsetstrokecolor{currentstroke}%
\pgfsetstrokeopacity{0.250000}%
\pgfsetdash{}{0pt}%
\pgfpathmoveto{\pgfqpoint{0.590551in}{1.172860in}}%
\pgfpathlineto{\pgfqpoint{2.755906in}{1.172860in}}%
\pgfusepath{stroke}%
\end{pgfscope}%
\begin{pgfscope}%
\pgfsetbuttcap%
\pgfsetroundjoin%
\definecolor{currentfill}{rgb}{0.000000,0.000000,0.000000}%
\pgfsetfillcolor{currentfill}%
\pgfsetlinewidth{0.803000pt}%
\definecolor{currentstroke}{rgb}{0.000000,0.000000,0.000000}%
\pgfsetstrokecolor{currentstroke}%
\pgfsetdash{}{0pt}%
\pgfsys@defobject{currentmarker}{\pgfqpoint{-0.048611in}{0.000000in}}{\pgfqpoint{-0.000000in}{0.000000in}}{%
\pgfpathmoveto{\pgfqpoint{-0.000000in}{0.000000in}}%
\pgfpathlineto{\pgfqpoint{-0.048611in}{0.000000in}}%
\pgfusepath{stroke,fill}%
}%
\begin{pgfscope}%
\pgfsys@transformshift{0.590551in}{1.172860in}%
\pgfsys@useobject{currentmarker}{}%
\end{pgfscope}%
\end{pgfscope}%
\begin{pgfscope}%
\definecolor{textcolor}{rgb}{0.000000,0.000000,0.000000}%
\pgfsetstrokecolor{textcolor}%
\pgfsetfillcolor{textcolor}%
\pgftext[x=0.246415in, y=1.124635in, left, base]{\color{textcolor}{\rmfamily\fontsize{8.330000}{9.996000}\selectfont\catcode`\^=\active\def^{\ifmmode\sp\else\^{}\fi}\catcode`\%=\active\def%{\%}$\mathdefault{0.05}$}}%
\end{pgfscope}%
\begin{pgfscope}%
\pgfpathrectangle{\pgfqpoint{0.590551in}{0.590551in}}{\pgfqpoint{2.165354in}{2.165354in}}%
\pgfusepath{clip}%
\pgfsetrectcap%
\pgfsetroundjoin%
\pgfsetlinewidth{0.803000pt}%
\definecolor{currentstroke}{rgb}{0.690196,0.690196,0.690196}%
\pgfsetstrokecolor{currentstroke}%
\pgfsetstrokeopacity{0.250000}%
\pgfsetdash{}{0pt}%
\pgfpathmoveto{\pgfqpoint{0.590551in}{1.657713in}}%
\pgfpathlineto{\pgfqpoint{2.755906in}{1.657713in}}%
\pgfusepath{stroke}%
\end{pgfscope}%
\begin{pgfscope}%
\pgfsetbuttcap%
\pgfsetroundjoin%
\definecolor{currentfill}{rgb}{0.000000,0.000000,0.000000}%
\pgfsetfillcolor{currentfill}%
\pgfsetlinewidth{0.803000pt}%
\definecolor{currentstroke}{rgb}{0.000000,0.000000,0.000000}%
\pgfsetstrokecolor{currentstroke}%
\pgfsetdash{}{0pt}%
\pgfsys@defobject{currentmarker}{\pgfqpoint{-0.048611in}{0.000000in}}{\pgfqpoint{-0.000000in}{0.000000in}}{%
\pgfpathmoveto{\pgfqpoint{-0.000000in}{0.000000in}}%
\pgfpathlineto{\pgfqpoint{-0.048611in}{0.000000in}}%
\pgfusepath{stroke,fill}%
}%
\begin{pgfscope}%
\pgfsys@transformshift{0.590551in}{1.657713in}%
\pgfsys@useobject{currentmarker}{}%
\end{pgfscope}%
\end{pgfscope}%
\begin{pgfscope}%
\definecolor{textcolor}{rgb}{0.000000,0.000000,0.000000}%
\pgfsetstrokecolor{textcolor}%
\pgfsetfillcolor{textcolor}%
\pgftext[x=0.246415in, y=1.609488in, left, base]{\color{textcolor}{\rmfamily\fontsize{8.330000}{9.996000}\selectfont\catcode`\^=\active\def^{\ifmmode\sp\else\^{}\fi}\catcode`\%=\active\def%{\%}$\mathdefault{0.10}$}}%
\end{pgfscope}%
\begin{pgfscope}%
\pgfpathrectangle{\pgfqpoint{0.590551in}{0.590551in}}{\pgfqpoint{2.165354in}{2.165354in}}%
\pgfusepath{clip}%
\pgfsetrectcap%
\pgfsetroundjoin%
\pgfsetlinewidth{0.803000pt}%
\definecolor{currentstroke}{rgb}{0.690196,0.690196,0.690196}%
\pgfsetstrokecolor{currentstroke}%
\pgfsetstrokeopacity{0.250000}%
\pgfsetdash{}{0pt}%
\pgfpathmoveto{\pgfqpoint{0.590551in}{2.142566in}}%
\pgfpathlineto{\pgfqpoint{2.755906in}{2.142566in}}%
\pgfusepath{stroke}%
\end{pgfscope}%
\begin{pgfscope}%
\pgfsetbuttcap%
\pgfsetroundjoin%
\definecolor{currentfill}{rgb}{0.000000,0.000000,0.000000}%
\pgfsetfillcolor{currentfill}%
\pgfsetlinewidth{0.803000pt}%
\definecolor{currentstroke}{rgb}{0.000000,0.000000,0.000000}%
\pgfsetstrokecolor{currentstroke}%
\pgfsetdash{}{0pt}%
\pgfsys@defobject{currentmarker}{\pgfqpoint{-0.048611in}{0.000000in}}{\pgfqpoint{-0.000000in}{0.000000in}}{%
\pgfpathmoveto{\pgfqpoint{-0.000000in}{0.000000in}}%
\pgfpathlineto{\pgfqpoint{-0.048611in}{0.000000in}}%
\pgfusepath{stroke,fill}%
}%
\begin{pgfscope}%
\pgfsys@transformshift{0.590551in}{2.142566in}%
\pgfsys@useobject{currentmarker}{}%
\end{pgfscope}%
\end{pgfscope}%
\begin{pgfscope}%
\definecolor{textcolor}{rgb}{0.000000,0.000000,0.000000}%
\pgfsetstrokecolor{textcolor}%
\pgfsetfillcolor{textcolor}%
\pgftext[x=0.246415in, y=2.094341in, left, base]{\color{textcolor}{\rmfamily\fontsize{8.330000}{9.996000}\selectfont\catcode`\^=\active\def^{\ifmmode\sp\else\^{}\fi}\catcode`\%=\active\def%{\%}$\mathdefault{0.15}$}}%
\end{pgfscope}%
\begin{pgfscope}%
\pgfpathrectangle{\pgfqpoint{0.590551in}{0.590551in}}{\pgfqpoint{2.165354in}{2.165354in}}%
\pgfusepath{clip}%
\pgfsetrectcap%
\pgfsetroundjoin%
\pgfsetlinewidth{0.803000pt}%
\definecolor{currentstroke}{rgb}{0.690196,0.690196,0.690196}%
\pgfsetstrokecolor{currentstroke}%
\pgfsetstrokeopacity{0.250000}%
\pgfsetdash{}{0pt}%
\pgfpathmoveto{\pgfqpoint{0.590551in}{2.627419in}}%
\pgfpathlineto{\pgfqpoint{2.755906in}{2.627419in}}%
\pgfusepath{stroke}%
\end{pgfscope}%
\begin{pgfscope}%
\pgfsetbuttcap%
\pgfsetroundjoin%
\definecolor{currentfill}{rgb}{0.000000,0.000000,0.000000}%
\pgfsetfillcolor{currentfill}%
\pgfsetlinewidth{0.803000pt}%
\definecolor{currentstroke}{rgb}{0.000000,0.000000,0.000000}%
\pgfsetstrokecolor{currentstroke}%
\pgfsetdash{}{0pt}%
\pgfsys@defobject{currentmarker}{\pgfqpoint{-0.048611in}{0.000000in}}{\pgfqpoint{-0.000000in}{0.000000in}}{%
\pgfpathmoveto{\pgfqpoint{-0.000000in}{0.000000in}}%
\pgfpathlineto{\pgfqpoint{-0.048611in}{0.000000in}}%
\pgfusepath{stroke,fill}%
}%
\begin{pgfscope}%
\pgfsys@transformshift{0.590551in}{2.627419in}%
\pgfsys@useobject{currentmarker}{}%
\end{pgfscope}%
\end{pgfscope}%
\begin{pgfscope}%
\definecolor{textcolor}{rgb}{0.000000,0.000000,0.000000}%
\pgfsetstrokecolor{textcolor}%
\pgfsetfillcolor{textcolor}%
\pgftext[x=0.246415in, y=2.579194in, left, base]{\color{textcolor}{\rmfamily\fontsize{8.330000}{9.996000}\selectfont\catcode`\^=\active\def^{\ifmmode\sp\else\^{}\fi}\catcode`\%=\active\def%{\%}$\mathdefault{0.20}$}}%
\end{pgfscope}%
\begin{pgfscope}%
\definecolor{textcolor}{rgb}{0.000000,0.000000,0.000000}%
\pgfsetstrokecolor{textcolor}%
\pgfsetfillcolor{textcolor}%
\pgftext[x=0.190859in,y=1.673228in,,bottom,rotate=90.000000]{\color{textcolor}{\rmfamily\fontsize{10.000000}{12.000000}\selectfont\catcode`\^=\active\def^{\ifmmode\sp\else\^{}\fi}\catcode`\%=\active\def%{\%}rank (normalized)}}%
\end{pgfscope}%
\begin{pgfscope}%
\pgfpathrectangle{\pgfqpoint{0.590551in}{0.590551in}}{\pgfqpoint{2.165354in}{2.165354in}}%
\pgfusepath{clip}%
\pgfsetrectcap%
\pgfsetroundjoin%
\pgfsetlinewidth{1.405250pt}%
\definecolor{currentstroke}{rgb}{0.196078,0.431373,0.745098}%
\pgfsetstrokecolor{currentstroke}%
\pgfsetdash{}{0pt}%
\pgfpathmoveto{\pgfqpoint{0.688976in}{1.599531in}}%
\pgfpathlineto{\pgfqpoint{0.709270in}{1.556864in}}%
\pgfpathlineto{\pgfqpoint{0.729564in}{1.562682in}}%
\pgfpathlineto{\pgfqpoint{0.749858in}{2.149354in}}%
\pgfpathlineto{\pgfqpoint{0.770152in}{1.810442in}}%
\pgfpathlineto{\pgfqpoint{0.790446in}{1.847291in}}%
\pgfpathlineto{\pgfqpoint{0.810740in}{1.675168in}}%
\pgfpathlineto{\pgfqpoint{0.831033in}{1.749835in}}%
\pgfpathlineto{\pgfqpoint{0.851327in}{1.890927in}}%
\pgfpathlineto{\pgfqpoint{0.871621in}{1.999535in}}%
\pgfpathlineto{\pgfqpoint{0.891915in}{2.178445in}}%
\pgfpathlineto{\pgfqpoint{0.912209in}{1.681471in}}%
\pgfpathlineto{\pgfqpoint{0.932503in}{1.801230in}}%
\pgfpathlineto{\pgfqpoint{0.952796in}{1.407529in}}%
\pgfpathlineto{\pgfqpoint{0.973090in}{1.488984in}}%
\pgfpathlineto{\pgfqpoint{0.993384in}{1.444378in}}%
\pgfpathlineto{\pgfqpoint{1.013678in}{1.276618in}}%
\pgfpathlineto{\pgfqpoint{1.033972in}{0.959524in}}%
\pgfpathlineto{\pgfqpoint{1.054266in}{0.859645in}}%
\pgfpathlineto{\pgfqpoint{1.074560in}{0.914918in}}%
\pgfpathlineto{\pgfqpoint{1.094853in}{0.912979in}}%
\pgfpathlineto{\pgfqpoint{1.115147in}{0.833463in}}%
\pgfpathlineto{\pgfqpoint{1.135441in}{0.791765in}}%
\pgfpathlineto{\pgfqpoint{1.155735in}{0.829099in}}%
\pgfpathlineto{\pgfqpoint{1.176029in}{0.791765in}}%
\pgfpathlineto{\pgfqpoint{1.196323in}{0.772371in}}%
\pgfpathlineto{\pgfqpoint{1.216617in}{0.827644in}}%
\pgfpathlineto{\pgfqpoint{1.236910in}{0.760250in}}%
\pgfpathlineto{\pgfqpoint{1.257204in}{0.830069in}}%
\pgfpathlineto{\pgfqpoint{1.277498in}{0.755886in}}%
\pgfpathlineto{\pgfqpoint{1.297792in}{0.761704in}}%
\pgfpathlineto{\pgfqpoint{1.318086in}{0.760735in}}%
\pgfpathlineto{\pgfqpoint{1.338380in}{0.789341in}}%
\pgfpathlineto{\pgfqpoint{1.358674in}{0.796129in}}%
\pgfpathlineto{\pgfqpoint{1.378967in}{0.728249in}}%
\pgfpathlineto{\pgfqpoint{1.399261in}{0.742795in}}%
\pgfpathlineto{\pgfqpoint{1.419555in}{0.755886in}}%
\pgfpathlineto{\pgfqpoint{1.439849in}{0.797583in}}%
\pgfpathlineto{\pgfqpoint{1.460143in}{0.804856in}}%
\pgfpathlineto{\pgfqpoint{1.480437in}{0.856251in}}%
\pgfpathlineto{\pgfqpoint{1.500731in}{0.744250in}}%
\pgfpathlineto{\pgfqpoint{1.521024in}{0.809220in}}%
\pgfpathlineto{\pgfqpoint{1.541318in}{0.793220in}}%
\pgfpathlineto{\pgfqpoint{1.561612in}{0.755886in}}%
\pgfpathlineto{\pgfqpoint{1.581906in}{0.782068in}}%
\pgfpathlineto{\pgfqpoint{1.602200in}{0.739401in}}%
\pgfpathlineto{\pgfqpoint{1.622494in}{0.763159in}}%
\pgfpathlineto{\pgfqpoint{1.642788in}{0.802432in}}%
\pgfpathlineto{\pgfqpoint{1.663081in}{0.777705in}}%
\pgfpathlineto{\pgfqpoint{1.683375in}{0.787886in}}%
\pgfpathlineto{\pgfqpoint{1.703669in}{0.745219in}}%
\pgfpathlineto{\pgfqpoint{1.723963in}{0.754432in}}%
\pgfpathlineto{\pgfqpoint{1.744257in}{0.725340in}}%
\pgfpathlineto{\pgfqpoint{1.764551in}{0.738431in}}%
\pgfpathlineto{\pgfqpoint{1.784845in}{0.736977in}}%
\pgfpathlineto{\pgfqpoint{1.805138in}{0.761220in}}%
\pgfpathlineto{\pgfqpoint{1.825432in}{0.760250in}}%
\pgfpathlineto{\pgfqpoint{1.845726in}{0.758310in}}%
\pgfpathlineto{\pgfqpoint{1.866020in}{0.781583in}}%
\pgfpathlineto{\pgfqpoint{1.886314in}{0.765583in}}%
\pgfpathlineto{\pgfqpoint{1.906608in}{0.772856in}}%
\pgfpathlineto{\pgfqpoint{1.926902in}{0.729704in}}%
\pgfpathlineto{\pgfqpoint{1.947195in}{0.761220in}}%
\pgfpathlineto{\pgfqpoint{1.967489in}{0.717583in}}%
\pgfpathlineto{\pgfqpoint{1.987783in}{0.719037in}}%
\pgfpathlineto{\pgfqpoint{2.008077in}{0.718068in}}%
\pgfpathlineto{\pgfqpoint{2.028371in}{0.715158in}}%
\pgfpathlineto{\pgfqpoint{2.048665in}{0.757341in}}%
\pgfpathlineto{\pgfqpoint{2.068959in}{0.723401in}}%
\pgfpathlineto{\pgfqpoint{2.089252in}{0.703037in}}%
\pgfpathlineto{\pgfqpoint{2.109546in}{0.710310in}}%
\pgfpathlineto{\pgfqpoint{2.129840in}{0.700613in}}%
\pgfpathlineto{\pgfqpoint{2.150134in}{0.707886in}}%
\pgfpathlineto{\pgfqpoint{2.170428in}{0.711764in}}%
\pgfpathlineto{\pgfqpoint{2.190722in}{0.698189in}}%
\pgfpathlineto{\pgfqpoint{2.211016in}{0.711764in}}%
\pgfpathlineto{\pgfqpoint{2.231309in}{0.696249in}}%
\pgfpathlineto{\pgfqpoint{2.251603in}{0.701583in}}%
\pgfpathlineto{\pgfqpoint{2.271897in}{0.701098in}}%
\pgfpathlineto{\pgfqpoint{2.292191in}{0.698189in}}%
\pgfpathlineto{\pgfqpoint{2.312485in}{0.696734in}}%
\pgfpathlineto{\pgfqpoint{2.332779in}{0.695279in}}%
\pgfpathlineto{\pgfqpoint{2.353072in}{0.692855in}}%
\pgfpathlineto{\pgfqpoint{2.373366in}{0.691885in}}%
\pgfpathlineto{\pgfqpoint{2.393660in}{0.690916in}}%
\pgfpathlineto{\pgfqpoint{2.413954in}{0.691885in}}%
\pgfpathlineto{\pgfqpoint{2.434248in}{0.691885in}}%
\pgfpathlineto{\pgfqpoint{2.454542in}{0.690916in}}%
\pgfpathlineto{\pgfqpoint{2.474836in}{0.689946in}}%
\pgfpathlineto{\pgfqpoint{2.495129in}{0.689946in}}%
\pgfpathlineto{\pgfqpoint{2.515423in}{0.689946in}}%
\pgfpathlineto{\pgfqpoint{2.535717in}{0.689946in}}%
\pgfpathlineto{\pgfqpoint{2.556011in}{0.688976in}}%
\pgfpathlineto{\pgfqpoint{2.576305in}{0.688976in}}%
\pgfpathlineto{\pgfqpoint{2.596599in}{0.688976in}}%
\pgfpathlineto{\pgfqpoint{2.616893in}{0.688976in}}%
\pgfpathlineto{\pgfqpoint{2.637186in}{0.688976in}}%
\pgfpathlineto{\pgfqpoint{2.657480in}{0.688976in}}%
\pgfusepath{stroke}%
\end{pgfscope}%
\begin{pgfscope}%
\pgfsetrectcap%
\pgfsetmiterjoin%
\pgfsetlinewidth{0.803000pt}%
\definecolor{currentstroke}{rgb}{0.000000,0.000000,0.000000}%
\pgfsetstrokecolor{currentstroke}%
\pgfsetdash{}{0pt}%
\pgfpathmoveto{\pgfqpoint{0.590551in}{0.590551in}}%
\pgfpathlineto{\pgfqpoint{0.590551in}{2.755906in}}%
\pgfusepath{stroke}%
\end{pgfscope}%
\begin{pgfscope}%
\pgfsetrectcap%
\pgfsetmiterjoin%
\pgfsetlinewidth{0.803000pt}%
\definecolor{currentstroke}{rgb}{0.000000,0.000000,0.000000}%
\pgfsetstrokecolor{currentstroke}%
\pgfsetdash{}{0pt}%
\pgfpathmoveto{\pgfqpoint{2.755906in}{0.590551in}}%
\pgfpathlineto{\pgfqpoint{2.755906in}{2.755906in}}%
\pgfusepath{stroke}%
\end{pgfscope}%
\begin{pgfscope}%
\pgfsetrectcap%
\pgfsetmiterjoin%
\pgfsetlinewidth{0.803000pt}%
\definecolor{currentstroke}{rgb}{0.000000,0.000000,0.000000}%
\pgfsetstrokecolor{currentstroke}%
\pgfsetdash{}{0pt}%
\pgfpathmoveto{\pgfqpoint{0.590551in}{0.590551in}}%
\pgfpathlineto{\pgfqpoint{2.755906in}{0.590551in}}%
\pgfusepath{stroke}%
\end{pgfscope}%
\begin{pgfscope}%
\pgfsetrectcap%
\pgfsetmiterjoin%
\pgfsetlinewidth{0.803000pt}%
\definecolor{currentstroke}{rgb}{0.000000,0.000000,0.000000}%
\pgfsetstrokecolor{currentstroke}%
\pgfsetdash{}{0pt}%
\pgfpathmoveto{\pgfqpoint{0.590551in}{2.755906in}}%
\pgfpathlineto{\pgfqpoint{2.755906in}{2.755906in}}%
\pgfusepath{stroke}%
\end{pgfscope}%
\end{pgfpicture}%
\makeatother%
\endgroup%

%% file: figures/k.pgf
%% Creator: Matplotlib, PGF backend
%%
%% To include the figure in your LaTeX document, write
%%   \input{<filename>.pgf}
%%
%% Make sure the required packages are loaded in your preamble
%%   \usepackage{pgf}
%%
%% Also ensure that all the required font packages are loaded; for instance,
%% the lmodern package is sometimes necessary when using math font.
%%   \usepackage{lmodern}
%%
%% Figures using additional raster images can only be included by \input if
%% they are in the same directory as the main LaTeX file. For loading figures
%% from other directories you can use the `import` package
%%   \usepackage{import}
%%
%% and then include the figures with
%%   \import{<path to file>}{<filename>.pgf}
%%
%% Matplotlib used the following preamble
%%   \def\mathdefault#1{#1}
%%   \everymath=\expandafter{\the\everymath\displaystyle}
%%   \IfFileExists{scrextend.sty}{
%%     \usepackage[fontsize=10.000000pt]{scrextend}
%%   }{
%%     \renewcommand{\normalsize}{\fontsize{10.000000}{12.000000}\selectfont}
%%     \normalsize
%%   }
%%   
%%   \makeatletter\@ifpackageloaded{underscore}{}{\usepackage[strings]{underscore}}\makeatother
%%
\begingroup%
\makeatletter%
\begin{pgfpicture}%
\pgfpathrectangle{\pgfpointorigin}{\pgfqpoint{2.952756in}{2.952756in}}%
\pgfusepath{use as bounding box, clip}%
\begin{pgfscope}%
\pgfsetbuttcap%
\pgfsetmiterjoin%
\pgfsetlinewidth{0.000000pt}%
\definecolor{currentstroke}{rgb}{1.000000,1.000000,1.000000}%
\pgfsetstrokecolor{currentstroke}%
\pgfsetdash{}{0pt}%
\pgfpathmoveto{\pgfqpoint{0.000000in}{0.000000in}}%
\pgfpathlineto{\pgfqpoint{2.952756in}{0.000000in}}%
\pgfpathlineto{\pgfqpoint{2.952756in}{2.952756in}}%
\pgfpathlineto{\pgfqpoint{0.000000in}{2.952756in}}%
\pgfpathlineto{\pgfqpoint{0.000000in}{0.000000in}}%
\pgfpathclose%
\pgfusepath{}%
\end{pgfscope}%
\begin{pgfscope}%
\pgfsetbuttcap%
\pgfsetmiterjoin%
\pgfsetlinewidth{0.000000pt}%
\definecolor{currentstroke}{rgb}{0.000000,0.000000,0.000000}%
\pgfsetstrokecolor{currentstroke}%
\pgfsetstrokeopacity{0.000000}%
\pgfsetdash{}{0pt}%
\pgfpathmoveto{\pgfqpoint{0.590551in}{0.590551in}}%
\pgfpathlineto{\pgfqpoint{2.755906in}{0.590551in}}%
\pgfpathlineto{\pgfqpoint{2.755906in}{2.755906in}}%
\pgfpathlineto{\pgfqpoint{0.590551in}{2.755906in}}%
\pgfpathlineto{\pgfqpoint{0.590551in}{0.590551in}}%
\pgfpathclose%
\pgfusepath{}%
\end{pgfscope}%
\begin{pgfscope}%
\pgfpathrectangle{\pgfqpoint{0.590551in}{0.590551in}}{\pgfqpoint{2.165354in}{2.165354in}}%
\pgfusepath{clip}%
\pgfsetbuttcap%
\pgfsetroundjoin%
\definecolor{currentfill}{rgb}{0.235294,0.607843,0.352941}%
\pgfsetfillcolor{currentfill}%
\pgfsetfillopacity{0.220000}%
\pgfsetlinewidth{1.003750pt}%
\definecolor{currentstroke}{rgb}{0.235294,0.607843,0.352941}%
\pgfsetstrokecolor{currentstroke}%
\pgfsetstrokeopacity{0.220000}%
\pgfsetdash{}{0pt}%
\pgfsys@defobject{currentmarker}{\pgfqpoint{0.688976in}{0.688976in}}{\pgfqpoint{2.657480in}{2.657480in}}{%
\pgfpathmoveto{\pgfqpoint{0.688976in}{2.657480in}}%
\pgfpathlineto{\pgfqpoint{0.688976in}{1.237655in}}%
\pgfpathlineto{\pgfqpoint{0.907699in}{0.961578in}}%
\pgfpathlineto{\pgfqpoint{0.978112in}{0.861976in}}%
\pgfpathlineto{\pgfqpoint{1.126422in}{0.742989in}}%
\pgfpathlineto{\pgfqpoint{1.226910in}{0.737106in}}%
\pgfpathlineto{\pgfqpoint{1.345144in}{0.707025in}}%
\pgfpathlineto{\pgfqpoint{1.445632in}{0.698469in}}%
\pgfpathlineto{\pgfqpoint{1.563867in}{0.692185in}}%
\pgfpathlineto{\pgfqpoint{1.671446in}{0.691115in}}%
\pgfpathlineto{\pgfqpoint{1.782590in}{0.690581in}}%
\pgfpathlineto{\pgfqpoint{1.890169in}{0.689511in}}%
\pgfpathlineto{\pgfqpoint{2.001312in}{0.689511in}}%
\pgfpathlineto{\pgfqpoint{2.110640in}{0.688976in}}%
\pgfpathlineto{\pgfqpoint{2.220035in}{0.689511in}}%
\pgfpathlineto{\pgfqpoint{2.329363in}{0.688976in}}%
\pgfpathlineto{\pgfqpoint{2.438758in}{0.688976in}}%
\pgfpathlineto{\pgfqpoint{2.548085in}{0.688976in}}%
\pgfpathlineto{\pgfqpoint{2.657480in}{0.689511in}}%
\pgfpathlineto{\pgfqpoint{2.657480in}{0.742587in}}%
\pgfpathlineto{\pgfqpoint{2.657480in}{0.742587in}}%
\pgfpathlineto{\pgfqpoint{2.548085in}{0.767321in}}%
\pgfpathlineto{\pgfqpoint{2.438758in}{0.771332in}}%
\pgfpathlineto{\pgfqpoint{2.329363in}{0.755823in}}%
\pgfpathlineto{\pgfqpoint{2.220035in}{0.804087in}}%
\pgfpathlineto{\pgfqpoint{2.110640in}{0.879891in}}%
\pgfpathlineto{\pgfqpoint{2.001312in}{0.865586in}}%
\pgfpathlineto{\pgfqpoint{1.890169in}{0.767588in}}%
\pgfpathlineto{\pgfqpoint{1.782590in}{0.843526in}}%
\pgfpathlineto{\pgfqpoint{1.671446in}{0.832296in}}%
\pgfpathlineto{\pgfqpoint{1.563867in}{0.872939in}}%
\pgfpathlineto{\pgfqpoint{1.445632in}{0.950748in}}%
\pgfpathlineto{\pgfqpoint{1.345144in}{1.018798in}}%
\pgfpathlineto{\pgfqpoint{1.226910in}{1.424826in}}%
\pgfpathlineto{\pgfqpoint{1.126422in}{1.424291in}}%
\pgfpathlineto{\pgfqpoint{0.978112in}{1.798767in}}%
\pgfpathlineto{\pgfqpoint{0.907699in}{2.511086in}}%
\pgfpathlineto{\pgfqpoint{0.688976in}{2.657480in}}%
\pgfpathlineto{\pgfqpoint{0.688976in}{2.657480in}}%
\pgfpathclose%
\pgfusepath{stroke,fill}%
}%
\begin{pgfscope}%
\pgfsys@transformshift{0.000000in}{0.000000in}%
\pgfsys@useobject{currentmarker}{}%
\end{pgfscope}%
\end{pgfscope}%
\begin{pgfscope}%
\pgfpathrectangle{\pgfqpoint{0.590551in}{0.590551in}}{\pgfqpoint{2.165354in}{2.165354in}}%
\pgfusepath{clip}%
\pgfsetrectcap%
\pgfsetroundjoin%
\pgfsetlinewidth{0.803000pt}%
\definecolor{currentstroke}{rgb}{0.690196,0.690196,0.690196}%
\pgfsetstrokecolor{currentstroke}%
\pgfsetstrokeopacity{0.250000}%
\pgfsetdash{}{0pt}%
\pgfpathmoveto{\pgfqpoint{1.196835in}{0.590551in}}%
\pgfpathlineto{\pgfqpoint{1.196835in}{2.755906in}}%
\pgfusepath{stroke}%
\end{pgfscope}%
\begin{pgfscope}%
\pgfsetbuttcap%
\pgfsetroundjoin%
\definecolor{currentfill}{rgb}{0.000000,0.000000,0.000000}%
\pgfsetfillcolor{currentfill}%
\pgfsetlinewidth{0.803000pt}%
\definecolor{currentstroke}{rgb}{0.000000,0.000000,0.000000}%
\pgfsetstrokecolor{currentstroke}%
\pgfsetdash{}{0pt}%
\pgfsys@defobject{currentmarker}{\pgfqpoint{0.000000in}{-0.048611in}}{\pgfqpoint{0.000000in}{0.000000in}}{%
\pgfpathmoveto{\pgfqpoint{0.000000in}{0.000000in}}%
\pgfpathlineto{\pgfqpoint{0.000000in}{-0.048611in}}%
\pgfusepath{stroke,fill}%
}%
\begin{pgfscope}%
\pgfsys@transformshift{1.196835in}{0.590551in}%
\pgfsys@useobject{currentmarker}{}%
\end{pgfscope}%
\end{pgfscope}%
\begin{pgfscope}%
\definecolor{textcolor}{rgb}{0.000000,0.000000,0.000000}%
\pgfsetstrokecolor{textcolor}%
\pgfsetfillcolor{textcolor}%
\pgftext[x=1.196835in,y=0.493329in,,top]{\color{textcolor}{\rmfamily\fontsize{8.330000}{9.996000}\selectfont\catcode`\^=\active\def^{\ifmmode\sp\else\^{}\fi}\catcode`\%=\active\def%{\%}$\mathdefault{10^{1}}$}}%
\end{pgfscope}%
\begin{pgfscope}%
\pgfpathrectangle{\pgfqpoint{0.590551in}{0.590551in}}{\pgfqpoint{2.165354in}{2.165354in}}%
\pgfusepath{clip}%
\pgfsetrectcap%
\pgfsetroundjoin%
\pgfsetlinewidth{0.803000pt}%
\definecolor{currentstroke}{rgb}{0.690196,0.690196,0.690196}%
\pgfsetstrokecolor{currentstroke}%
\pgfsetstrokeopacity{0.250000}%
\pgfsetdash{}{0pt}%
\pgfpathmoveto{\pgfqpoint{1.923416in}{0.590551in}}%
\pgfpathlineto{\pgfqpoint{1.923416in}{2.755906in}}%
\pgfusepath{stroke}%
\end{pgfscope}%
\begin{pgfscope}%
\pgfsetbuttcap%
\pgfsetroundjoin%
\definecolor{currentfill}{rgb}{0.000000,0.000000,0.000000}%
\pgfsetfillcolor{currentfill}%
\pgfsetlinewidth{0.803000pt}%
\definecolor{currentstroke}{rgb}{0.000000,0.000000,0.000000}%
\pgfsetstrokecolor{currentstroke}%
\pgfsetdash{}{0pt}%
\pgfsys@defobject{currentmarker}{\pgfqpoint{0.000000in}{-0.048611in}}{\pgfqpoint{0.000000in}{0.000000in}}{%
\pgfpathmoveto{\pgfqpoint{0.000000in}{0.000000in}}%
\pgfpathlineto{\pgfqpoint{0.000000in}{-0.048611in}}%
\pgfusepath{stroke,fill}%
}%
\begin{pgfscope}%
\pgfsys@transformshift{1.923416in}{0.590551in}%
\pgfsys@useobject{currentmarker}{}%
\end{pgfscope}%
\end{pgfscope}%
\begin{pgfscope}%
\definecolor{textcolor}{rgb}{0.000000,0.000000,0.000000}%
\pgfsetstrokecolor{textcolor}%
\pgfsetfillcolor{textcolor}%
\pgftext[x=1.923416in,y=0.493329in,,top]{\color{textcolor}{\rmfamily\fontsize{8.330000}{9.996000}\selectfont\catcode`\^=\active\def^{\ifmmode\sp\else\^{}\fi}\catcode`\%=\active\def%{\%}$\mathdefault{10^{2}}$}}%
\end{pgfscope}%
\begin{pgfscope}%
\pgfpathrectangle{\pgfqpoint{0.590551in}{0.590551in}}{\pgfqpoint{2.165354in}{2.165354in}}%
\pgfusepath{clip}%
\pgfsetrectcap%
\pgfsetroundjoin%
\pgfsetlinewidth{0.803000pt}%
\definecolor{currentstroke}{rgb}{0.690196,0.690196,0.690196}%
\pgfsetstrokecolor{currentstroke}%
\pgfsetstrokeopacity{0.250000}%
\pgfsetdash{}{0pt}%
\pgfpathmoveto{\pgfqpoint{2.649997in}{0.590551in}}%
\pgfpathlineto{\pgfqpoint{2.649997in}{2.755906in}}%
\pgfusepath{stroke}%
\end{pgfscope}%
\begin{pgfscope}%
\pgfsetbuttcap%
\pgfsetroundjoin%
\definecolor{currentfill}{rgb}{0.000000,0.000000,0.000000}%
\pgfsetfillcolor{currentfill}%
\pgfsetlinewidth{0.803000pt}%
\definecolor{currentstroke}{rgb}{0.000000,0.000000,0.000000}%
\pgfsetstrokecolor{currentstroke}%
\pgfsetdash{}{0pt}%
\pgfsys@defobject{currentmarker}{\pgfqpoint{0.000000in}{-0.048611in}}{\pgfqpoint{0.000000in}{0.000000in}}{%
\pgfpathmoveto{\pgfqpoint{0.000000in}{0.000000in}}%
\pgfpathlineto{\pgfqpoint{0.000000in}{-0.048611in}}%
\pgfusepath{stroke,fill}%
}%
\begin{pgfscope}%
\pgfsys@transformshift{2.649997in}{0.590551in}%
\pgfsys@useobject{currentmarker}{}%
\end{pgfscope}%
\end{pgfscope}%
\begin{pgfscope}%
\definecolor{textcolor}{rgb}{0.000000,0.000000,0.000000}%
\pgfsetstrokecolor{textcolor}%
\pgfsetfillcolor{textcolor}%
\pgftext[x=2.649997in,y=0.493329in,,top]{\color{textcolor}{\rmfamily\fontsize{8.330000}{9.996000}\selectfont\catcode`\^=\active\def^{\ifmmode\sp\else\^{}\fi}\catcode`\%=\active\def%{\%}$\mathdefault{10^{3}}$}}%
\end{pgfscope}%
\begin{pgfscope}%
\pgfsetbuttcap%
\pgfsetroundjoin%
\definecolor{currentfill}{rgb}{0.000000,0.000000,0.000000}%
\pgfsetfillcolor{currentfill}%
\pgfsetlinewidth{0.602250pt}%
\definecolor{currentstroke}{rgb}{0.000000,0.000000,0.000000}%
\pgfsetstrokecolor{currentstroke}%
\pgfsetdash{}{0pt}%
\pgfsys@defobject{currentmarker}{\pgfqpoint{0.000000in}{-0.027778in}}{\pgfqpoint{0.000000in}{0.000000in}}{%
\pgfpathmoveto{\pgfqpoint{0.000000in}{0.000000in}}%
\pgfpathlineto{\pgfqpoint{0.000000in}{-0.027778in}}%
\pgfusepath{stroke,fill}%
}%
\begin{pgfscope}%
\pgfsys@transformshift{0.688976in}{0.590551in}%
\pgfsys@useobject{currentmarker}{}%
\end{pgfscope}%
\end{pgfscope}%
\begin{pgfscope}%
\pgfsetbuttcap%
\pgfsetroundjoin%
\definecolor{currentfill}{rgb}{0.000000,0.000000,0.000000}%
\pgfsetfillcolor{currentfill}%
\pgfsetlinewidth{0.602250pt}%
\definecolor{currentstroke}{rgb}{0.000000,0.000000,0.000000}%
\pgfsetstrokecolor{currentstroke}%
\pgfsetdash{}{0pt}%
\pgfsys@defobject{currentmarker}{\pgfqpoint{0.000000in}{-0.027778in}}{\pgfqpoint{0.000000in}{0.000000in}}{%
\pgfpathmoveto{\pgfqpoint{0.000000in}{0.000000in}}%
\pgfpathlineto{\pgfqpoint{0.000000in}{-0.027778in}}%
\pgfusepath{stroke,fill}%
}%
\begin{pgfscope}%
\pgfsys@transformshift{0.816921in}{0.590551in}%
\pgfsys@useobject{currentmarker}{}%
\end{pgfscope}%
\end{pgfscope}%
\begin{pgfscope}%
\pgfsetbuttcap%
\pgfsetroundjoin%
\definecolor{currentfill}{rgb}{0.000000,0.000000,0.000000}%
\pgfsetfillcolor{currentfill}%
\pgfsetlinewidth{0.602250pt}%
\definecolor{currentstroke}{rgb}{0.000000,0.000000,0.000000}%
\pgfsetstrokecolor{currentstroke}%
\pgfsetdash{}{0pt}%
\pgfsys@defobject{currentmarker}{\pgfqpoint{0.000000in}{-0.027778in}}{\pgfqpoint{0.000000in}{0.000000in}}{%
\pgfpathmoveto{\pgfqpoint{0.000000in}{0.000000in}}%
\pgfpathlineto{\pgfqpoint{0.000000in}{-0.027778in}}%
\pgfusepath{stroke,fill}%
}%
\begin{pgfscope}%
\pgfsys@transformshift{0.907699in}{0.590551in}%
\pgfsys@useobject{currentmarker}{}%
\end{pgfscope}%
\end{pgfscope}%
\begin{pgfscope}%
\pgfsetbuttcap%
\pgfsetroundjoin%
\definecolor{currentfill}{rgb}{0.000000,0.000000,0.000000}%
\pgfsetfillcolor{currentfill}%
\pgfsetlinewidth{0.602250pt}%
\definecolor{currentstroke}{rgb}{0.000000,0.000000,0.000000}%
\pgfsetstrokecolor{currentstroke}%
\pgfsetdash{}{0pt}%
\pgfsys@defobject{currentmarker}{\pgfqpoint{0.000000in}{-0.027778in}}{\pgfqpoint{0.000000in}{0.000000in}}{%
\pgfpathmoveto{\pgfqpoint{0.000000in}{0.000000in}}%
\pgfpathlineto{\pgfqpoint{0.000000in}{-0.027778in}}%
\pgfusepath{stroke,fill}%
}%
\begin{pgfscope}%
\pgfsys@transformshift{0.978112in}{0.590551in}%
\pgfsys@useobject{currentmarker}{}%
\end{pgfscope}%
\end{pgfscope}%
\begin{pgfscope}%
\pgfsetbuttcap%
\pgfsetroundjoin%
\definecolor{currentfill}{rgb}{0.000000,0.000000,0.000000}%
\pgfsetfillcolor{currentfill}%
\pgfsetlinewidth{0.602250pt}%
\definecolor{currentstroke}{rgb}{0.000000,0.000000,0.000000}%
\pgfsetstrokecolor{currentstroke}%
\pgfsetdash{}{0pt}%
\pgfsys@defobject{currentmarker}{\pgfqpoint{0.000000in}{-0.027778in}}{\pgfqpoint{0.000000in}{0.000000in}}{%
\pgfpathmoveto{\pgfqpoint{0.000000in}{0.000000in}}%
\pgfpathlineto{\pgfqpoint{0.000000in}{-0.027778in}}%
\pgfusepath{stroke,fill}%
}%
\begin{pgfscope}%
\pgfsys@transformshift{1.035644in}{0.590551in}%
\pgfsys@useobject{currentmarker}{}%
\end{pgfscope}%
\end{pgfscope}%
\begin{pgfscope}%
\pgfsetbuttcap%
\pgfsetroundjoin%
\definecolor{currentfill}{rgb}{0.000000,0.000000,0.000000}%
\pgfsetfillcolor{currentfill}%
\pgfsetlinewidth{0.602250pt}%
\definecolor{currentstroke}{rgb}{0.000000,0.000000,0.000000}%
\pgfsetstrokecolor{currentstroke}%
\pgfsetdash{}{0pt}%
\pgfsys@defobject{currentmarker}{\pgfqpoint{0.000000in}{-0.027778in}}{\pgfqpoint{0.000000in}{0.000000in}}{%
\pgfpathmoveto{\pgfqpoint{0.000000in}{0.000000in}}%
\pgfpathlineto{\pgfqpoint{0.000000in}{-0.027778in}}%
\pgfusepath{stroke,fill}%
}%
\begin{pgfscope}%
\pgfsys@transformshift{1.084286in}{0.590551in}%
\pgfsys@useobject{currentmarker}{}%
\end{pgfscope}%
\end{pgfscope}%
\begin{pgfscope}%
\pgfsetbuttcap%
\pgfsetroundjoin%
\definecolor{currentfill}{rgb}{0.000000,0.000000,0.000000}%
\pgfsetfillcolor{currentfill}%
\pgfsetlinewidth{0.602250pt}%
\definecolor{currentstroke}{rgb}{0.000000,0.000000,0.000000}%
\pgfsetstrokecolor{currentstroke}%
\pgfsetdash{}{0pt}%
\pgfsys@defobject{currentmarker}{\pgfqpoint{0.000000in}{-0.027778in}}{\pgfqpoint{0.000000in}{0.000000in}}{%
\pgfpathmoveto{\pgfqpoint{0.000000in}{0.000000in}}%
\pgfpathlineto{\pgfqpoint{0.000000in}{-0.027778in}}%
\pgfusepath{stroke,fill}%
}%
\begin{pgfscope}%
\pgfsys@transformshift{1.126422in}{0.590551in}%
\pgfsys@useobject{currentmarker}{}%
\end{pgfscope}%
\end{pgfscope}%
\begin{pgfscope}%
\pgfsetbuttcap%
\pgfsetroundjoin%
\definecolor{currentfill}{rgb}{0.000000,0.000000,0.000000}%
\pgfsetfillcolor{currentfill}%
\pgfsetlinewidth{0.602250pt}%
\definecolor{currentstroke}{rgb}{0.000000,0.000000,0.000000}%
\pgfsetstrokecolor{currentstroke}%
\pgfsetdash{}{0pt}%
\pgfsys@defobject{currentmarker}{\pgfqpoint{0.000000in}{-0.027778in}}{\pgfqpoint{0.000000in}{0.000000in}}{%
\pgfpathmoveto{\pgfqpoint{0.000000in}{0.000000in}}%
\pgfpathlineto{\pgfqpoint{0.000000in}{-0.027778in}}%
\pgfusepath{stroke,fill}%
}%
\begin{pgfscope}%
\pgfsys@transformshift{1.163588in}{0.590551in}%
\pgfsys@useobject{currentmarker}{}%
\end{pgfscope}%
\end{pgfscope}%
\begin{pgfscope}%
\pgfsetbuttcap%
\pgfsetroundjoin%
\definecolor{currentfill}{rgb}{0.000000,0.000000,0.000000}%
\pgfsetfillcolor{currentfill}%
\pgfsetlinewidth{0.602250pt}%
\definecolor{currentstroke}{rgb}{0.000000,0.000000,0.000000}%
\pgfsetstrokecolor{currentstroke}%
\pgfsetdash{}{0pt}%
\pgfsys@defobject{currentmarker}{\pgfqpoint{0.000000in}{-0.027778in}}{\pgfqpoint{0.000000in}{0.000000in}}{%
\pgfpathmoveto{\pgfqpoint{0.000000in}{0.000000in}}%
\pgfpathlineto{\pgfqpoint{0.000000in}{-0.027778in}}%
\pgfusepath{stroke,fill}%
}%
\begin{pgfscope}%
\pgfsys@transformshift{1.415557in}{0.590551in}%
\pgfsys@useobject{currentmarker}{}%
\end{pgfscope}%
\end{pgfscope}%
\begin{pgfscope}%
\pgfsetbuttcap%
\pgfsetroundjoin%
\definecolor{currentfill}{rgb}{0.000000,0.000000,0.000000}%
\pgfsetfillcolor{currentfill}%
\pgfsetlinewidth{0.602250pt}%
\definecolor{currentstroke}{rgb}{0.000000,0.000000,0.000000}%
\pgfsetstrokecolor{currentstroke}%
\pgfsetdash{}{0pt}%
\pgfsys@defobject{currentmarker}{\pgfqpoint{0.000000in}{-0.027778in}}{\pgfqpoint{0.000000in}{0.000000in}}{%
\pgfpathmoveto{\pgfqpoint{0.000000in}{0.000000in}}%
\pgfpathlineto{\pgfqpoint{0.000000in}{-0.027778in}}%
\pgfusepath{stroke,fill}%
}%
\begin{pgfscope}%
\pgfsys@transformshift{1.543502in}{0.590551in}%
\pgfsys@useobject{currentmarker}{}%
\end{pgfscope}%
\end{pgfscope}%
\begin{pgfscope}%
\pgfsetbuttcap%
\pgfsetroundjoin%
\definecolor{currentfill}{rgb}{0.000000,0.000000,0.000000}%
\pgfsetfillcolor{currentfill}%
\pgfsetlinewidth{0.602250pt}%
\definecolor{currentstroke}{rgb}{0.000000,0.000000,0.000000}%
\pgfsetstrokecolor{currentstroke}%
\pgfsetdash{}{0pt}%
\pgfsys@defobject{currentmarker}{\pgfqpoint{0.000000in}{-0.027778in}}{\pgfqpoint{0.000000in}{0.000000in}}{%
\pgfpathmoveto{\pgfqpoint{0.000000in}{0.000000in}}%
\pgfpathlineto{\pgfqpoint{0.000000in}{-0.027778in}}%
\pgfusepath{stroke,fill}%
}%
\begin{pgfscope}%
\pgfsys@transformshift{1.634280in}{0.590551in}%
\pgfsys@useobject{currentmarker}{}%
\end{pgfscope}%
\end{pgfscope}%
\begin{pgfscope}%
\pgfsetbuttcap%
\pgfsetroundjoin%
\definecolor{currentfill}{rgb}{0.000000,0.000000,0.000000}%
\pgfsetfillcolor{currentfill}%
\pgfsetlinewidth{0.602250pt}%
\definecolor{currentstroke}{rgb}{0.000000,0.000000,0.000000}%
\pgfsetstrokecolor{currentstroke}%
\pgfsetdash{}{0pt}%
\pgfsys@defobject{currentmarker}{\pgfqpoint{0.000000in}{-0.027778in}}{\pgfqpoint{0.000000in}{0.000000in}}{%
\pgfpathmoveto{\pgfqpoint{0.000000in}{0.000000in}}%
\pgfpathlineto{\pgfqpoint{0.000000in}{-0.027778in}}%
\pgfusepath{stroke,fill}%
}%
\begin{pgfscope}%
\pgfsys@transformshift{1.704693in}{0.590551in}%
\pgfsys@useobject{currentmarker}{}%
\end{pgfscope}%
\end{pgfscope}%
\begin{pgfscope}%
\pgfsetbuttcap%
\pgfsetroundjoin%
\definecolor{currentfill}{rgb}{0.000000,0.000000,0.000000}%
\pgfsetfillcolor{currentfill}%
\pgfsetlinewidth{0.602250pt}%
\definecolor{currentstroke}{rgb}{0.000000,0.000000,0.000000}%
\pgfsetstrokecolor{currentstroke}%
\pgfsetdash{}{0pt}%
\pgfsys@defobject{currentmarker}{\pgfqpoint{0.000000in}{-0.027778in}}{\pgfqpoint{0.000000in}{0.000000in}}{%
\pgfpathmoveto{\pgfqpoint{0.000000in}{0.000000in}}%
\pgfpathlineto{\pgfqpoint{0.000000in}{-0.027778in}}%
\pgfusepath{stroke,fill}%
}%
\begin{pgfscope}%
\pgfsys@transformshift{1.762225in}{0.590551in}%
\pgfsys@useobject{currentmarker}{}%
\end{pgfscope}%
\end{pgfscope}%
\begin{pgfscope}%
\pgfsetbuttcap%
\pgfsetroundjoin%
\definecolor{currentfill}{rgb}{0.000000,0.000000,0.000000}%
\pgfsetfillcolor{currentfill}%
\pgfsetlinewidth{0.602250pt}%
\definecolor{currentstroke}{rgb}{0.000000,0.000000,0.000000}%
\pgfsetstrokecolor{currentstroke}%
\pgfsetdash{}{0pt}%
\pgfsys@defobject{currentmarker}{\pgfqpoint{0.000000in}{-0.027778in}}{\pgfqpoint{0.000000in}{0.000000in}}{%
\pgfpathmoveto{\pgfqpoint{0.000000in}{0.000000in}}%
\pgfpathlineto{\pgfqpoint{0.000000in}{-0.027778in}}%
\pgfusepath{stroke,fill}%
}%
\begin{pgfscope}%
\pgfsys@transformshift{1.810867in}{0.590551in}%
\pgfsys@useobject{currentmarker}{}%
\end{pgfscope}%
\end{pgfscope}%
\begin{pgfscope}%
\pgfsetbuttcap%
\pgfsetroundjoin%
\definecolor{currentfill}{rgb}{0.000000,0.000000,0.000000}%
\pgfsetfillcolor{currentfill}%
\pgfsetlinewidth{0.602250pt}%
\definecolor{currentstroke}{rgb}{0.000000,0.000000,0.000000}%
\pgfsetstrokecolor{currentstroke}%
\pgfsetdash{}{0pt}%
\pgfsys@defobject{currentmarker}{\pgfqpoint{0.000000in}{-0.027778in}}{\pgfqpoint{0.000000in}{0.000000in}}{%
\pgfpathmoveto{\pgfqpoint{0.000000in}{0.000000in}}%
\pgfpathlineto{\pgfqpoint{0.000000in}{-0.027778in}}%
\pgfusepath{stroke,fill}%
}%
\begin{pgfscope}%
\pgfsys@transformshift{1.853003in}{0.590551in}%
\pgfsys@useobject{currentmarker}{}%
\end{pgfscope}%
\end{pgfscope}%
\begin{pgfscope}%
\pgfsetbuttcap%
\pgfsetroundjoin%
\definecolor{currentfill}{rgb}{0.000000,0.000000,0.000000}%
\pgfsetfillcolor{currentfill}%
\pgfsetlinewidth{0.602250pt}%
\definecolor{currentstroke}{rgb}{0.000000,0.000000,0.000000}%
\pgfsetstrokecolor{currentstroke}%
\pgfsetdash{}{0pt}%
\pgfsys@defobject{currentmarker}{\pgfqpoint{0.000000in}{-0.027778in}}{\pgfqpoint{0.000000in}{0.000000in}}{%
\pgfpathmoveto{\pgfqpoint{0.000000in}{0.000000in}}%
\pgfpathlineto{\pgfqpoint{0.000000in}{-0.027778in}}%
\pgfusepath{stroke,fill}%
}%
\begin{pgfscope}%
\pgfsys@transformshift{1.890169in}{0.590551in}%
\pgfsys@useobject{currentmarker}{}%
\end{pgfscope}%
\end{pgfscope}%
\begin{pgfscope}%
\pgfsetbuttcap%
\pgfsetroundjoin%
\definecolor{currentfill}{rgb}{0.000000,0.000000,0.000000}%
\pgfsetfillcolor{currentfill}%
\pgfsetlinewidth{0.602250pt}%
\definecolor{currentstroke}{rgb}{0.000000,0.000000,0.000000}%
\pgfsetstrokecolor{currentstroke}%
\pgfsetdash{}{0pt}%
\pgfsys@defobject{currentmarker}{\pgfqpoint{0.000000in}{-0.027778in}}{\pgfqpoint{0.000000in}{0.000000in}}{%
\pgfpathmoveto{\pgfqpoint{0.000000in}{0.000000in}}%
\pgfpathlineto{\pgfqpoint{0.000000in}{-0.027778in}}%
\pgfusepath{stroke,fill}%
}%
\begin{pgfscope}%
\pgfsys@transformshift{2.142138in}{0.590551in}%
\pgfsys@useobject{currentmarker}{}%
\end{pgfscope}%
\end{pgfscope}%
\begin{pgfscope}%
\pgfsetbuttcap%
\pgfsetroundjoin%
\definecolor{currentfill}{rgb}{0.000000,0.000000,0.000000}%
\pgfsetfillcolor{currentfill}%
\pgfsetlinewidth{0.602250pt}%
\definecolor{currentstroke}{rgb}{0.000000,0.000000,0.000000}%
\pgfsetstrokecolor{currentstroke}%
\pgfsetdash{}{0pt}%
\pgfsys@defobject{currentmarker}{\pgfqpoint{0.000000in}{-0.027778in}}{\pgfqpoint{0.000000in}{0.000000in}}{%
\pgfpathmoveto{\pgfqpoint{0.000000in}{0.000000in}}%
\pgfpathlineto{\pgfqpoint{0.000000in}{-0.027778in}}%
\pgfusepath{stroke,fill}%
}%
\begin{pgfscope}%
\pgfsys@transformshift{2.270083in}{0.590551in}%
\pgfsys@useobject{currentmarker}{}%
\end{pgfscope}%
\end{pgfscope}%
\begin{pgfscope}%
\pgfsetbuttcap%
\pgfsetroundjoin%
\definecolor{currentfill}{rgb}{0.000000,0.000000,0.000000}%
\pgfsetfillcolor{currentfill}%
\pgfsetlinewidth{0.602250pt}%
\definecolor{currentstroke}{rgb}{0.000000,0.000000,0.000000}%
\pgfsetstrokecolor{currentstroke}%
\pgfsetdash{}{0pt}%
\pgfsys@defobject{currentmarker}{\pgfqpoint{0.000000in}{-0.027778in}}{\pgfqpoint{0.000000in}{0.000000in}}{%
\pgfpathmoveto{\pgfqpoint{0.000000in}{0.000000in}}%
\pgfpathlineto{\pgfqpoint{0.000000in}{-0.027778in}}%
\pgfusepath{stroke,fill}%
}%
\begin{pgfscope}%
\pgfsys@transformshift{2.360861in}{0.590551in}%
\pgfsys@useobject{currentmarker}{}%
\end{pgfscope}%
\end{pgfscope}%
\begin{pgfscope}%
\pgfsetbuttcap%
\pgfsetroundjoin%
\definecolor{currentfill}{rgb}{0.000000,0.000000,0.000000}%
\pgfsetfillcolor{currentfill}%
\pgfsetlinewidth{0.602250pt}%
\definecolor{currentstroke}{rgb}{0.000000,0.000000,0.000000}%
\pgfsetstrokecolor{currentstroke}%
\pgfsetdash{}{0pt}%
\pgfsys@defobject{currentmarker}{\pgfqpoint{0.000000in}{-0.027778in}}{\pgfqpoint{0.000000in}{0.000000in}}{%
\pgfpathmoveto{\pgfqpoint{0.000000in}{0.000000in}}%
\pgfpathlineto{\pgfqpoint{0.000000in}{-0.027778in}}%
\pgfusepath{stroke,fill}%
}%
\begin{pgfscope}%
\pgfsys@transformshift{2.431274in}{0.590551in}%
\pgfsys@useobject{currentmarker}{}%
\end{pgfscope}%
\end{pgfscope}%
\begin{pgfscope}%
\pgfsetbuttcap%
\pgfsetroundjoin%
\definecolor{currentfill}{rgb}{0.000000,0.000000,0.000000}%
\pgfsetfillcolor{currentfill}%
\pgfsetlinewidth{0.602250pt}%
\definecolor{currentstroke}{rgb}{0.000000,0.000000,0.000000}%
\pgfsetstrokecolor{currentstroke}%
\pgfsetdash{}{0pt}%
\pgfsys@defobject{currentmarker}{\pgfqpoint{0.000000in}{-0.027778in}}{\pgfqpoint{0.000000in}{0.000000in}}{%
\pgfpathmoveto{\pgfqpoint{0.000000in}{0.000000in}}%
\pgfpathlineto{\pgfqpoint{0.000000in}{-0.027778in}}%
\pgfusepath{stroke,fill}%
}%
\begin{pgfscope}%
\pgfsys@transformshift{2.488805in}{0.590551in}%
\pgfsys@useobject{currentmarker}{}%
\end{pgfscope}%
\end{pgfscope}%
\begin{pgfscope}%
\pgfsetbuttcap%
\pgfsetroundjoin%
\definecolor{currentfill}{rgb}{0.000000,0.000000,0.000000}%
\pgfsetfillcolor{currentfill}%
\pgfsetlinewidth{0.602250pt}%
\definecolor{currentstroke}{rgb}{0.000000,0.000000,0.000000}%
\pgfsetstrokecolor{currentstroke}%
\pgfsetdash{}{0pt}%
\pgfsys@defobject{currentmarker}{\pgfqpoint{0.000000in}{-0.027778in}}{\pgfqpoint{0.000000in}{0.000000in}}{%
\pgfpathmoveto{\pgfqpoint{0.000000in}{0.000000in}}%
\pgfpathlineto{\pgfqpoint{0.000000in}{-0.027778in}}%
\pgfusepath{stroke,fill}%
}%
\begin{pgfscope}%
\pgfsys@transformshift{2.537448in}{0.590551in}%
\pgfsys@useobject{currentmarker}{}%
\end{pgfscope}%
\end{pgfscope}%
\begin{pgfscope}%
\pgfsetbuttcap%
\pgfsetroundjoin%
\definecolor{currentfill}{rgb}{0.000000,0.000000,0.000000}%
\pgfsetfillcolor{currentfill}%
\pgfsetlinewidth{0.602250pt}%
\definecolor{currentstroke}{rgb}{0.000000,0.000000,0.000000}%
\pgfsetstrokecolor{currentstroke}%
\pgfsetdash{}{0pt}%
\pgfsys@defobject{currentmarker}{\pgfqpoint{0.000000in}{-0.027778in}}{\pgfqpoint{0.000000in}{0.000000in}}{%
\pgfpathmoveto{\pgfqpoint{0.000000in}{0.000000in}}%
\pgfpathlineto{\pgfqpoint{0.000000in}{-0.027778in}}%
\pgfusepath{stroke,fill}%
}%
\begin{pgfscope}%
\pgfsys@transformshift{2.579584in}{0.590551in}%
\pgfsys@useobject{currentmarker}{}%
\end{pgfscope}%
\end{pgfscope}%
\begin{pgfscope}%
\pgfsetbuttcap%
\pgfsetroundjoin%
\definecolor{currentfill}{rgb}{0.000000,0.000000,0.000000}%
\pgfsetfillcolor{currentfill}%
\pgfsetlinewidth{0.602250pt}%
\definecolor{currentstroke}{rgb}{0.000000,0.000000,0.000000}%
\pgfsetstrokecolor{currentstroke}%
\pgfsetdash{}{0pt}%
\pgfsys@defobject{currentmarker}{\pgfqpoint{0.000000in}{-0.027778in}}{\pgfqpoint{0.000000in}{0.000000in}}{%
\pgfpathmoveto{\pgfqpoint{0.000000in}{0.000000in}}%
\pgfpathlineto{\pgfqpoint{0.000000in}{-0.027778in}}%
\pgfusepath{stroke,fill}%
}%
\begin{pgfscope}%
\pgfsys@transformshift{2.616750in}{0.590551in}%
\pgfsys@useobject{currentmarker}{}%
\end{pgfscope}%
\end{pgfscope}%
\begin{pgfscope}%
\definecolor{textcolor}{rgb}{0.000000,0.000000,0.000000}%
\pgfsetstrokecolor{textcolor}%
\pgfsetfillcolor{textcolor}%
\pgftext[x=1.673228in,y=0.337912in,,top]{\color{textcolor}{\rmfamily\fontsize{10.000000}{12.000000}\selectfont\catcode`\^=\active\def^{\ifmmode\sp\else\^{}\fi}\catcode`\%=\active\def%{\%}JL dimension $k$}}%
\end{pgfscope}%
\begin{pgfscope}%
\pgfpathrectangle{\pgfqpoint{0.590551in}{0.590551in}}{\pgfqpoint{2.165354in}{2.165354in}}%
\pgfusepath{clip}%
\pgfsetrectcap%
\pgfsetroundjoin%
\pgfsetlinewidth{0.803000pt}%
\definecolor{currentstroke}{rgb}{0.690196,0.690196,0.690196}%
\pgfsetstrokecolor{currentstroke}%
\pgfsetstrokeopacity{0.250000}%
\pgfsetdash{}{0pt}%
\pgfpathmoveto{\pgfqpoint{0.590551in}{0.688442in}}%
\pgfpathlineto{\pgfqpoint{2.755906in}{0.688442in}}%
\pgfusepath{stroke}%
\end{pgfscope}%
\begin{pgfscope}%
\pgfsetbuttcap%
\pgfsetroundjoin%
\definecolor{currentfill}{rgb}{0.000000,0.000000,0.000000}%
\pgfsetfillcolor{currentfill}%
\pgfsetlinewidth{0.803000pt}%
\definecolor{currentstroke}{rgb}{0.000000,0.000000,0.000000}%
\pgfsetstrokecolor{currentstroke}%
\pgfsetdash{}{0pt}%
\pgfsys@defobject{currentmarker}{\pgfqpoint{-0.048611in}{0.000000in}}{\pgfqpoint{-0.000000in}{0.000000in}}{%
\pgfpathmoveto{\pgfqpoint{-0.000000in}{0.000000in}}%
\pgfpathlineto{\pgfqpoint{-0.048611in}{0.000000in}}%
\pgfusepath{stroke,fill}%
}%
\begin{pgfscope}%
\pgfsys@transformshift{0.590551in}{0.688442in}%
\pgfsys@useobject{currentmarker}{}%
\end{pgfscope}%
\end{pgfscope}%
\begin{pgfscope}%
\definecolor{textcolor}{rgb}{0.000000,0.000000,0.000000}%
\pgfsetstrokecolor{textcolor}%
\pgfsetfillcolor{textcolor}%
\pgftext[x=0.283449in, y=0.649861in, left, base]{\color{textcolor}{\rmfamily\fontsize{8.330000}{9.996000}\selectfont\catcode`\^=\active\def^{\ifmmode\sp\else\^{}\fi}\catcode`\%=\active\def%{\%}$\mathdefault{0.00}$}}%
\end{pgfscope}%
\begin{pgfscope}%
\pgfpathrectangle{\pgfqpoint{0.590551in}{0.590551in}}{\pgfqpoint{2.165354in}{2.165354in}}%
\pgfusepath{clip}%
\pgfsetrectcap%
\pgfsetroundjoin%
\pgfsetlinewidth{0.803000pt}%
\definecolor{currentstroke}{rgb}{0.690196,0.690196,0.690196}%
\pgfsetstrokecolor{currentstroke}%
\pgfsetstrokeopacity{0.250000}%
\pgfsetdash{}{0pt}%
\pgfpathmoveto{\pgfqpoint{0.590551in}{0.955829in}}%
\pgfpathlineto{\pgfqpoint{2.755906in}{0.955829in}}%
\pgfusepath{stroke}%
\end{pgfscope}%
\begin{pgfscope}%
\pgfsetbuttcap%
\pgfsetroundjoin%
\definecolor{currentfill}{rgb}{0.000000,0.000000,0.000000}%
\pgfsetfillcolor{currentfill}%
\pgfsetlinewidth{0.803000pt}%
\definecolor{currentstroke}{rgb}{0.000000,0.000000,0.000000}%
\pgfsetstrokecolor{currentstroke}%
\pgfsetdash{}{0pt}%
\pgfsys@defobject{currentmarker}{\pgfqpoint{-0.048611in}{0.000000in}}{\pgfqpoint{-0.000000in}{0.000000in}}{%
\pgfpathmoveto{\pgfqpoint{-0.000000in}{0.000000in}}%
\pgfpathlineto{\pgfqpoint{-0.048611in}{0.000000in}}%
\pgfusepath{stroke,fill}%
}%
\begin{pgfscope}%
\pgfsys@transformshift{0.590551in}{0.955829in}%
\pgfsys@useobject{currentmarker}{}%
\end{pgfscope}%
\end{pgfscope}%
\begin{pgfscope}%
\definecolor{textcolor}{rgb}{0.000000,0.000000,0.000000}%
\pgfsetstrokecolor{textcolor}%
\pgfsetfillcolor{textcolor}%
\pgftext[x=0.283449in, y=0.917248in, left, base]{\color{textcolor}{\rmfamily\fontsize{8.330000}{9.996000}\selectfont\catcode`\^=\active\def^{\ifmmode\sp\else\^{}\fi}\catcode`\%=\active\def%{\%}$\mathdefault{0.05}$}}%
\end{pgfscope}%
\begin{pgfscope}%
\pgfpathrectangle{\pgfqpoint{0.590551in}{0.590551in}}{\pgfqpoint{2.165354in}{2.165354in}}%
\pgfusepath{clip}%
\pgfsetrectcap%
\pgfsetroundjoin%
\pgfsetlinewidth{0.803000pt}%
\definecolor{currentstroke}{rgb}{0.690196,0.690196,0.690196}%
\pgfsetstrokecolor{currentstroke}%
\pgfsetstrokeopacity{0.250000}%
\pgfsetdash{}{0pt}%
\pgfpathmoveto{\pgfqpoint{0.590551in}{1.223216in}}%
\pgfpathlineto{\pgfqpoint{2.755906in}{1.223216in}}%
\pgfusepath{stroke}%
\end{pgfscope}%
\begin{pgfscope}%
\pgfsetbuttcap%
\pgfsetroundjoin%
\definecolor{currentfill}{rgb}{0.000000,0.000000,0.000000}%
\pgfsetfillcolor{currentfill}%
\pgfsetlinewidth{0.803000pt}%
\definecolor{currentstroke}{rgb}{0.000000,0.000000,0.000000}%
\pgfsetstrokecolor{currentstroke}%
\pgfsetdash{}{0pt}%
\pgfsys@defobject{currentmarker}{\pgfqpoint{-0.048611in}{0.000000in}}{\pgfqpoint{-0.000000in}{0.000000in}}{%
\pgfpathmoveto{\pgfqpoint{-0.000000in}{0.000000in}}%
\pgfpathlineto{\pgfqpoint{-0.048611in}{0.000000in}}%
\pgfusepath{stroke,fill}%
}%
\begin{pgfscope}%
\pgfsys@transformshift{0.590551in}{1.223216in}%
\pgfsys@useobject{currentmarker}{}%
\end{pgfscope}%
\end{pgfscope}%
\begin{pgfscope}%
\definecolor{textcolor}{rgb}{0.000000,0.000000,0.000000}%
\pgfsetstrokecolor{textcolor}%
\pgfsetfillcolor{textcolor}%
\pgftext[x=0.283449in, y=1.184636in, left, base]{\color{textcolor}{\rmfamily\fontsize{8.330000}{9.996000}\selectfont\catcode`\^=\active\def^{\ifmmode\sp\else\^{}\fi}\catcode`\%=\active\def%{\%}$\mathdefault{0.10}$}}%
\end{pgfscope}%
\begin{pgfscope}%
\pgfpathrectangle{\pgfqpoint{0.590551in}{0.590551in}}{\pgfqpoint{2.165354in}{2.165354in}}%
\pgfusepath{clip}%
\pgfsetrectcap%
\pgfsetroundjoin%
\pgfsetlinewidth{0.803000pt}%
\definecolor{currentstroke}{rgb}{0.690196,0.690196,0.690196}%
\pgfsetstrokecolor{currentstroke}%
\pgfsetstrokeopacity{0.250000}%
\pgfsetdash{}{0pt}%
\pgfpathmoveto{\pgfqpoint{0.590551in}{1.490603in}}%
\pgfpathlineto{\pgfqpoint{2.755906in}{1.490603in}}%
\pgfusepath{stroke}%
\end{pgfscope}%
\begin{pgfscope}%
\pgfsetbuttcap%
\pgfsetroundjoin%
\definecolor{currentfill}{rgb}{0.000000,0.000000,0.000000}%
\pgfsetfillcolor{currentfill}%
\pgfsetlinewidth{0.803000pt}%
\definecolor{currentstroke}{rgb}{0.000000,0.000000,0.000000}%
\pgfsetstrokecolor{currentstroke}%
\pgfsetdash{}{0pt}%
\pgfsys@defobject{currentmarker}{\pgfqpoint{-0.048611in}{0.000000in}}{\pgfqpoint{-0.000000in}{0.000000in}}{%
\pgfpathmoveto{\pgfqpoint{-0.000000in}{0.000000in}}%
\pgfpathlineto{\pgfqpoint{-0.048611in}{0.000000in}}%
\pgfusepath{stroke,fill}%
}%
\begin{pgfscope}%
\pgfsys@transformshift{0.590551in}{1.490603in}%
\pgfsys@useobject{currentmarker}{}%
\end{pgfscope}%
\end{pgfscope}%
\begin{pgfscope}%
\definecolor{textcolor}{rgb}{0.000000,0.000000,0.000000}%
\pgfsetstrokecolor{textcolor}%
\pgfsetfillcolor{textcolor}%
\pgftext[x=0.283449in, y=1.452023in, left, base]{\color{textcolor}{\rmfamily\fontsize{8.330000}{9.996000}\selectfont\catcode`\^=\active\def^{\ifmmode\sp\else\^{}\fi}\catcode`\%=\active\def%{\%}$\mathdefault{0.15}$}}%
\end{pgfscope}%
\begin{pgfscope}%
\pgfpathrectangle{\pgfqpoint{0.590551in}{0.590551in}}{\pgfqpoint{2.165354in}{2.165354in}}%
\pgfusepath{clip}%
\pgfsetrectcap%
\pgfsetroundjoin%
\pgfsetlinewidth{0.803000pt}%
\definecolor{currentstroke}{rgb}{0.690196,0.690196,0.690196}%
\pgfsetstrokecolor{currentstroke}%
\pgfsetstrokeopacity{0.250000}%
\pgfsetdash{}{0pt}%
\pgfpathmoveto{\pgfqpoint{0.590551in}{1.757990in}}%
\pgfpathlineto{\pgfqpoint{2.755906in}{1.757990in}}%
\pgfusepath{stroke}%
\end{pgfscope}%
\begin{pgfscope}%
\pgfsetbuttcap%
\pgfsetroundjoin%
\definecolor{currentfill}{rgb}{0.000000,0.000000,0.000000}%
\pgfsetfillcolor{currentfill}%
\pgfsetlinewidth{0.803000pt}%
\definecolor{currentstroke}{rgb}{0.000000,0.000000,0.000000}%
\pgfsetstrokecolor{currentstroke}%
\pgfsetdash{}{0pt}%
\pgfsys@defobject{currentmarker}{\pgfqpoint{-0.048611in}{0.000000in}}{\pgfqpoint{-0.000000in}{0.000000in}}{%
\pgfpathmoveto{\pgfqpoint{-0.000000in}{0.000000in}}%
\pgfpathlineto{\pgfqpoint{-0.048611in}{0.000000in}}%
\pgfusepath{stroke,fill}%
}%
\begin{pgfscope}%
\pgfsys@transformshift{0.590551in}{1.757990in}%
\pgfsys@useobject{currentmarker}{}%
\end{pgfscope}%
\end{pgfscope}%
\begin{pgfscope}%
\definecolor{textcolor}{rgb}{0.000000,0.000000,0.000000}%
\pgfsetstrokecolor{textcolor}%
\pgfsetfillcolor{textcolor}%
\pgftext[x=0.283449in, y=1.719410in, left, base]{\color{textcolor}{\rmfamily\fontsize{8.330000}{9.996000}\selectfont\catcode`\^=\active\def^{\ifmmode\sp\else\^{}\fi}\catcode`\%=\active\def%{\%}$\mathdefault{0.20}$}}%
\end{pgfscope}%
\begin{pgfscope}%
\pgfpathrectangle{\pgfqpoint{0.590551in}{0.590551in}}{\pgfqpoint{2.165354in}{2.165354in}}%
\pgfusepath{clip}%
\pgfsetrectcap%
\pgfsetroundjoin%
\pgfsetlinewidth{0.803000pt}%
\definecolor{currentstroke}{rgb}{0.690196,0.690196,0.690196}%
\pgfsetstrokecolor{currentstroke}%
\pgfsetstrokeopacity{0.250000}%
\pgfsetdash{}{0pt}%
\pgfpathmoveto{\pgfqpoint{0.590551in}{2.025377in}}%
\pgfpathlineto{\pgfqpoint{2.755906in}{2.025377in}}%
\pgfusepath{stroke}%
\end{pgfscope}%
\begin{pgfscope}%
\pgfsetbuttcap%
\pgfsetroundjoin%
\definecolor{currentfill}{rgb}{0.000000,0.000000,0.000000}%
\pgfsetfillcolor{currentfill}%
\pgfsetlinewidth{0.803000pt}%
\definecolor{currentstroke}{rgb}{0.000000,0.000000,0.000000}%
\pgfsetstrokecolor{currentstroke}%
\pgfsetdash{}{0pt}%
\pgfsys@defobject{currentmarker}{\pgfqpoint{-0.048611in}{0.000000in}}{\pgfqpoint{-0.000000in}{0.000000in}}{%
\pgfpathmoveto{\pgfqpoint{-0.000000in}{0.000000in}}%
\pgfpathlineto{\pgfqpoint{-0.048611in}{0.000000in}}%
\pgfusepath{stroke,fill}%
}%
\begin{pgfscope}%
\pgfsys@transformshift{0.590551in}{2.025377in}%
\pgfsys@useobject{currentmarker}{}%
\end{pgfscope}%
\end{pgfscope}%
\begin{pgfscope}%
\definecolor{textcolor}{rgb}{0.000000,0.000000,0.000000}%
\pgfsetstrokecolor{textcolor}%
\pgfsetfillcolor{textcolor}%
\pgftext[x=0.283449in, y=1.986797in, left, base]{\color{textcolor}{\rmfamily\fontsize{8.330000}{9.996000}\selectfont\catcode`\^=\active\def^{\ifmmode\sp\else\^{}\fi}\catcode`\%=\active\def%{\%}$\mathdefault{0.25}$}}%
\end{pgfscope}%
\begin{pgfscope}%
\pgfpathrectangle{\pgfqpoint{0.590551in}{0.590551in}}{\pgfqpoint{2.165354in}{2.165354in}}%
\pgfusepath{clip}%
\pgfsetrectcap%
\pgfsetroundjoin%
\pgfsetlinewidth{0.803000pt}%
\definecolor{currentstroke}{rgb}{0.690196,0.690196,0.690196}%
\pgfsetstrokecolor{currentstroke}%
\pgfsetstrokeopacity{0.250000}%
\pgfsetdash{}{0pt}%
\pgfpathmoveto{\pgfqpoint{0.590551in}{2.292764in}}%
\pgfpathlineto{\pgfqpoint{2.755906in}{2.292764in}}%
\pgfusepath{stroke}%
\end{pgfscope}%
\begin{pgfscope}%
\pgfsetbuttcap%
\pgfsetroundjoin%
\definecolor{currentfill}{rgb}{0.000000,0.000000,0.000000}%
\pgfsetfillcolor{currentfill}%
\pgfsetlinewidth{0.803000pt}%
\definecolor{currentstroke}{rgb}{0.000000,0.000000,0.000000}%
\pgfsetstrokecolor{currentstroke}%
\pgfsetdash{}{0pt}%
\pgfsys@defobject{currentmarker}{\pgfqpoint{-0.048611in}{0.000000in}}{\pgfqpoint{-0.000000in}{0.000000in}}{%
\pgfpathmoveto{\pgfqpoint{-0.000000in}{0.000000in}}%
\pgfpathlineto{\pgfqpoint{-0.048611in}{0.000000in}}%
\pgfusepath{stroke,fill}%
}%
\begin{pgfscope}%
\pgfsys@transformshift{0.590551in}{2.292764in}%
\pgfsys@useobject{currentmarker}{}%
\end{pgfscope}%
\end{pgfscope}%
\begin{pgfscope}%
\definecolor{textcolor}{rgb}{0.000000,0.000000,0.000000}%
\pgfsetstrokecolor{textcolor}%
\pgfsetfillcolor{textcolor}%
\pgftext[x=0.283449in, y=2.254184in, left, base]{\color{textcolor}{\rmfamily\fontsize{8.330000}{9.996000}\selectfont\catcode`\^=\active\def^{\ifmmode\sp\else\^{}\fi}\catcode`\%=\active\def%{\%}$\mathdefault{0.30}$}}%
\end{pgfscope}%
\begin{pgfscope}%
\pgfpathrectangle{\pgfqpoint{0.590551in}{0.590551in}}{\pgfqpoint{2.165354in}{2.165354in}}%
\pgfusepath{clip}%
\pgfsetrectcap%
\pgfsetroundjoin%
\pgfsetlinewidth{0.803000pt}%
\definecolor{currentstroke}{rgb}{0.690196,0.690196,0.690196}%
\pgfsetstrokecolor{currentstroke}%
\pgfsetstrokeopacity{0.250000}%
\pgfsetdash{}{0pt}%
\pgfpathmoveto{\pgfqpoint{0.590551in}{2.560151in}}%
\pgfpathlineto{\pgfqpoint{2.755906in}{2.560151in}}%
\pgfusepath{stroke}%
\end{pgfscope}%
\begin{pgfscope}%
\pgfsetbuttcap%
\pgfsetroundjoin%
\definecolor{currentfill}{rgb}{0.000000,0.000000,0.000000}%
\pgfsetfillcolor{currentfill}%
\pgfsetlinewidth{0.803000pt}%
\definecolor{currentstroke}{rgb}{0.000000,0.000000,0.000000}%
\pgfsetstrokecolor{currentstroke}%
\pgfsetdash{}{0pt}%
\pgfsys@defobject{currentmarker}{\pgfqpoint{-0.048611in}{0.000000in}}{\pgfqpoint{-0.000000in}{0.000000in}}{%
\pgfpathmoveto{\pgfqpoint{-0.000000in}{0.000000in}}%
\pgfpathlineto{\pgfqpoint{-0.048611in}{0.000000in}}%
\pgfusepath{stroke,fill}%
}%
\begin{pgfscope}%
\pgfsys@transformshift{0.590551in}{2.560151in}%
\pgfsys@useobject{currentmarker}{}%
\end{pgfscope}%
\end{pgfscope}%
\begin{pgfscope}%
\definecolor{textcolor}{rgb}{0.000000,0.000000,0.000000}%
\pgfsetstrokecolor{textcolor}%
\pgfsetfillcolor{textcolor}%
\pgftext[x=0.283449in, y=2.521571in, left, base]{\color{textcolor}{\rmfamily\fontsize{8.330000}{9.996000}\selectfont\catcode`\^=\active\def^{\ifmmode\sp\else\^{}\fi}\catcode`\%=\active\def%{\%}$\mathdefault{0.35}$}}%
\end{pgfscope}%
\begin{pgfscope}%
\definecolor{textcolor}{rgb}{0.000000,0.000000,0.000000}%
\pgfsetstrokecolor{textcolor}%
\pgfsetfillcolor{textcolor}%
\pgftext[x=0.227894in,y=1.673228in,,bottom,rotate=90.000000]{\color{textcolor}{\rmfamily\fontsize{10.000000}{12.000000}\selectfont\catcode`\^=\active\def^{\ifmmode\sp\else\^{}\fi}\catcode`\%=\active\def%{\%}rank (normalized)}}%
\end{pgfscope}%
\begin{pgfscope}%
\pgfpathrectangle{\pgfqpoint{0.590551in}{0.590551in}}{\pgfqpoint{2.165354in}{2.165354in}}%
\pgfusepath{clip}%
\pgfsetrectcap%
\pgfsetroundjoin%
\pgfsetlinewidth{1.405250pt}%
\definecolor{currentstroke}{rgb}{0.235294,0.607843,0.352941}%
\pgfsetstrokecolor{currentstroke}%
\pgfsetdash{}{0pt}%
\pgfpathmoveto{\pgfqpoint{0.688976in}{1.828580in}}%
\pgfpathlineto{\pgfqpoint{0.907699in}{1.425628in}}%
\pgfpathlineto{\pgfqpoint{0.978112in}{1.173214in}}%
\pgfpathlineto{\pgfqpoint{1.126422in}{0.990589in}}%
\pgfpathlineto{\pgfqpoint{1.226910in}{0.846735in}}%
\pgfpathlineto{\pgfqpoint{1.345144in}{0.786038in}}%
\pgfpathlineto{\pgfqpoint{1.445632in}{0.777482in}}%
\pgfpathlineto{\pgfqpoint{1.563867in}{0.719459in}}%
\pgfpathlineto{\pgfqpoint{1.671446in}{0.706089in}}%
\pgfpathlineto{\pgfqpoint{1.782590in}{0.708228in}}%
\pgfpathlineto{\pgfqpoint{1.890169in}{0.703950in}}%
\pgfpathlineto{\pgfqpoint{2.001312in}{0.703950in}}%
\pgfpathlineto{\pgfqpoint{2.110640in}{0.693789in}}%
\pgfpathlineto{\pgfqpoint{2.220035in}{0.700474in}}%
\pgfpathlineto{\pgfqpoint{2.329363in}{0.692720in}}%
\pgfpathlineto{\pgfqpoint{2.438758in}{0.697800in}}%
\pgfpathlineto{\pgfqpoint{2.548085in}{0.693789in}}%
\pgfpathlineto{\pgfqpoint{2.657480in}{0.696998in}}%
\pgfusepath{stroke}%
\end{pgfscope}%
\begin{pgfscope}%
\pgfsetrectcap%
\pgfsetmiterjoin%
\pgfsetlinewidth{0.803000pt}%
\definecolor{currentstroke}{rgb}{0.000000,0.000000,0.000000}%
\pgfsetstrokecolor{currentstroke}%
\pgfsetdash{}{0pt}%
\pgfpathmoveto{\pgfqpoint{0.590551in}{0.590551in}}%
\pgfpathlineto{\pgfqpoint{0.590551in}{2.755906in}}%
\pgfusepath{stroke}%
\end{pgfscope}%
\begin{pgfscope}%
\pgfsetrectcap%
\pgfsetmiterjoin%
\pgfsetlinewidth{0.803000pt}%
\definecolor{currentstroke}{rgb}{0.000000,0.000000,0.000000}%
\pgfsetstrokecolor{currentstroke}%
\pgfsetdash{}{0pt}%
\pgfpathmoveto{\pgfqpoint{2.755906in}{0.590551in}}%
\pgfpathlineto{\pgfqpoint{2.755906in}{2.755906in}}%
\pgfusepath{stroke}%
\end{pgfscope}%
\begin{pgfscope}%
\pgfsetrectcap%
\pgfsetmiterjoin%
\pgfsetlinewidth{0.803000pt}%
\definecolor{currentstroke}{rgb}{0.000000,0.000000,0.000000}%
\pgfsetstrokecolor{currentstroke}%
\pgfsetdash{}{0pt}%
\pgfpathmoveto{\pgfqpoint{0.590551in}{0.590551in}}%
\pgfpathlineto{\pgfqpoint{2.755906in}{0.590551in}}%
\pgfusepath{stroke}%
\end{pgfscope}%
\begin{pgfscope}%
\pgfsetrectcap%
\pgfsetmiterjoin%
\pgfsetlinewidth{0.803000pt}%
\definecolor{currentstroke}{rgb}{0.000000,0.000000,0.000000}%
\pgfsetstrokecolor{currentstroke}%
\pgfsetdash{}{0pt}%
\pgfpathmoveto{\pgfqpoint{0.590551in}{2.755906in}}%
\pgfpathlineto{\pgfqpoint{2.755906in}{2.755906in}}%
\pgfusepath{stroke}%
\end{pgfscope}%
\end{pgfpicture}%
\makeatother%
\endgroup%